\documentclass[floatfix,twocolumn,showpacs,preprintnumbers,amsmath,amssymb,pra,superscriptaddress,longbibliography]{revtex4-1}
\usepackage{color}
\usepackage[usenames,dvipsnames,svgnames,table]{xcolor}
\usepackage[colorlinks=true,linkcolor=blue,urlcolor=blue,citecolor=blue]{hyperref}
\usepackage{mathtools}
\usepackage{graphicx}
\usepackage{dcolumn}
\usepackage{array}
\usepackage{lipsum}
\usepackage{bm}
\usepackage{subfigure}
\usepackage{amssymb}
\usepackage{multirow}
\usepackage{tabularx}
\usepackage{amsmath}
\usepackage{braket}
\graphicspath{{plots/}}
 \usepackage{lipsum}
\usepackage{mathrsfs}

\newcommand{\beq}{\begin{equation}}
\newcommand{\eeq}{\end{equation}}
\newcommand{\bea}{\begin{eqnarray}}
\newcommand{\eea}{\end{eqnarray}}

\newcommand{\kv}{{\bf k}}
\newcommand{\qv}{{\bf q}}

\renewcommand{\vec}[1]{\mathbf{#1}}

\begin{document}
\title{
Electronic Density Response of Warm Dense Matter
}

\author{Tobias Dornheim}
\email{t.dornheim@hzdr.de}

\affiliation{Center for Advanced Systems Understanding (CASUS), D-02826 G\"orlitz, Germany}
\affiliation{Helmholtz-Zentrum Dresden-Rossendorf (HZDR), D-01328 Dresden, Germany}

\author{Zhandos A.~Moldabekov}

\affiliation{Center for Advanced Systems Understanding (CASUS), D-02826 G\"orlitz, Germany}
\affiliation{Helmholtz-Zentrum Dresden-Rossendorf (HZDR), D-01328 Dresden, Germany}

\author{Kushal~Ramakrishna}

\affiliation{Center for Advanced Systems Understanding (CASUS), D-02826 G\"orlitz, Germany}
\affiliation{Helmholtz-Zentrum Dresden-Rossendorf (HZDR), D-01328 Dresden, Germany}

\author{Panagiotis Tolias}
\affiliation{Space and Plasma Physics, Royal Institute of Technology (KTH), Stockholm, SE-100 44, Sweden}

\author{Andrew~D.~Baczewski}
\affiliation{Center for Computing Research, Sandia National Laboratories, Albuquerque NM 87185 USA}

\author{Dominik~Kraus}
\affiliation{Institut f\"ur Physik, Universit\"at Rostock, D-18057 Rostock, Germany}
\affiliation{Helmholtz-Zentrum Dresden-Rossendorf (HZDR), D-01328 Dresden, Germany}

\author{Thomas~R.~Preston}
\affiliation{European XFEL, D-22869 Schenefeld, Germany}

\author{David A.~Chapman}
\affiliation{First Light Fusion, Yarnton, Oxfordshire, United Kingdom}

\author{Maximilian~P.~B\"ohme}

\affiliation{Center for Advanced Systems Understanding (CASUS), D-02826 G\"orlitz, Germany}
\affiliation{Helmholtz-Zentrum Dresden-Rossendorf (HZDR), D-01328 Dresden, Germany}
\affiliation{Technische  Universit\"at  Dresden,  D-01062  Dresden,  Germany}

\author{Tilo~D\"oppner}
\affiliation{Lawrence Livermore National Laboratory (LLNL), California 94550 Livermore, USA}

\author{Frank Graziani}
\affiliation{Lawrence Livermore National Laboratory (LLNL), California 94550 Livermore, USA}

\author{Michael Bonitz}

\affiliation{Institut f\"ur Theoretische Physik und Astrophysik, Christian-Albrechts-Universit\"at zu Kiel, D-24098 Kiel, Germany}

\author{Attila Cangi}

\affiliation{Center for Advanced Systems Understanding (CASUS), D-02826 G\"orlitz, Germany}
\affiliation{Helmholtz-Zentrum Dresden-Rossendorf (HZDR), D-01328 Dresden, Germany}

\author{Jan Vorberger}
\affiliation{Helmholtz-Zentrum Dresden-Rossendorf (HZDR), D-01328 Dresden, Germany}

\begin{abstract}
Matter at extreme temperatures and pressures --- commonly known as \emph{warm dense matter} (WDM) in the literature --- is ubiquitous throughout our Universe and occurs in a number of astrophysical objects such as giant planet interiors and brown dwarfs. Moreover, WDM is very important for technological applications such as inertial confinement fusion, and is realized in the laboratory using different techniques. A particularly important property for the understanding of WDM is given by its electronic density response to an external perturbation. Such response properties are routinely probed in x-ray Thomson scattering (XRTS) experiments, and, in addition, are central for the theoretical description of WDM. 
In this work, we give an overview of  a number of recent developments in this field. To this end, we summarize the relevant theoretical background, covering the regime of linear-response theory as well as nonlinear effects, the fully dynamic response and its static, time-independent limit, and the connection between density response properties and imaginary-time correlation functions (ITCF). In addition, we introduce the most important numerical simulation techniques including \emph{ab initio} path integral Monte Carlo (PIMC) simulations and different thermal density functional theory (DFT) approaches. 
From a practical perspective, we present a variety of simulation results for different density response properties, covering the archetypal model of the uniform electron gas and realistic WDM systems such as hydrogen. Moreover, we show how the concept of ITCFs can be used to infer the temperature from XRTS measurements of arbitrarily complex systems without the need for any models or approximations.
Finally, we outline a strategy for future developments based on the close interplay between simulations and experiments.
\end{abstract}

\maketitle

\tableofcontents

\section{Introduction\label{sec:introduction}}

\begin{figure}[b]\centering
\includegraphics[width=0.5\textwidth]{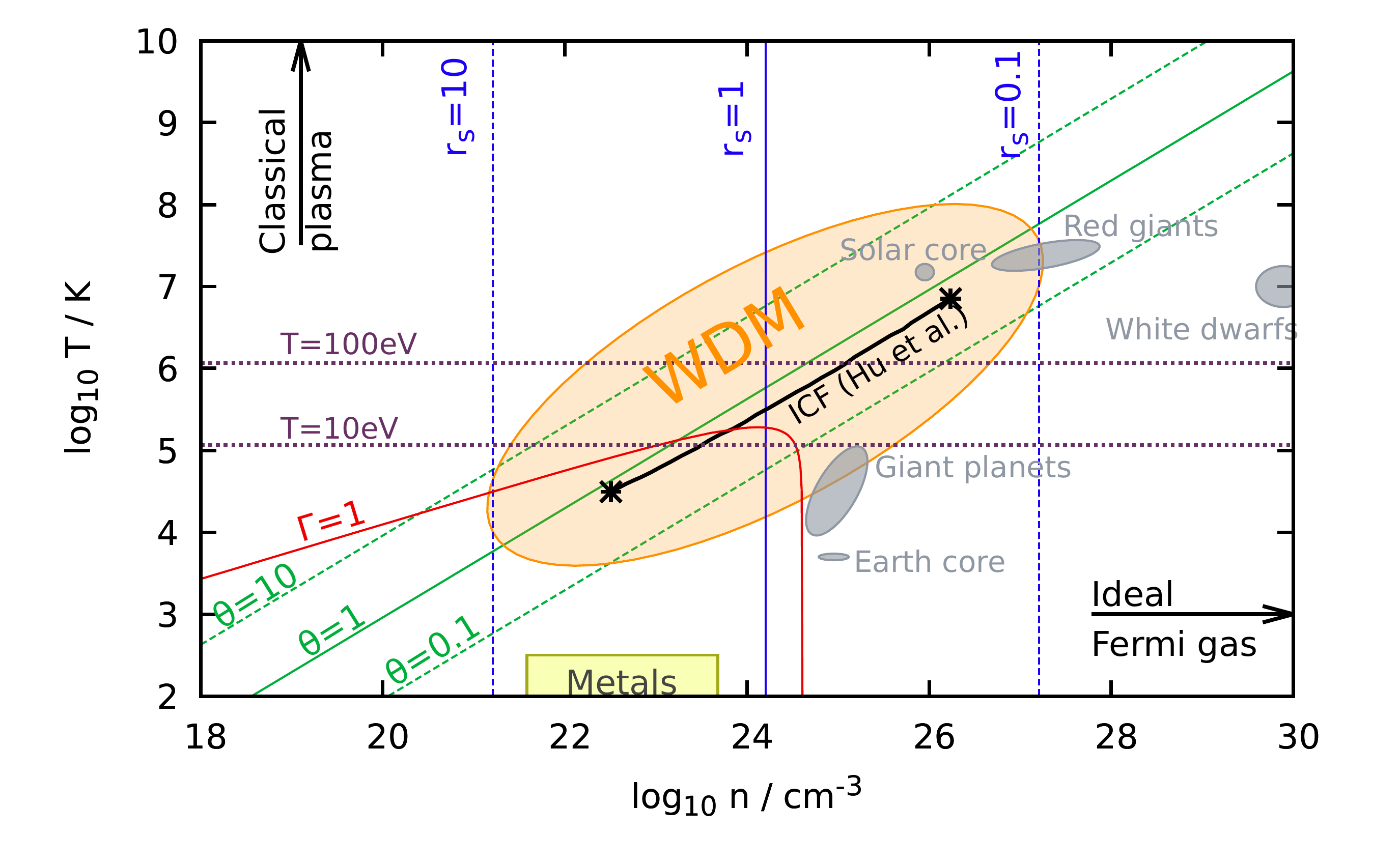}
\caption{\label{fig:overview} The warm-dense matter (WDM, orange bubble) regime is schematically illustrated in the density-temperature plane. Characteristic parameters are given by the Wigner-Seitz radius $r_s$ (vertical blue) and the degeneracy temperature $\theta=k_\textnormal{B}T/E_\textnormal{F}$ (diagonal green); also shown is the effective coupling parameter of the electrons $\Gamma$ (solid red), see Ref.~\cite{new_POP}. The grey bubbles indicate conditions of various astrophysical objects taken from Ref.~\cite{fortov_review}, and the black line the implosion path of a DT fuel capsule taken from Hu \emph{et al.}~\cite{hu_ICF}.
Adapted from Refs.~\cite{review,Dornheim_HEDP_2022}.
}
\end{figure}

Warm dense matter (WDM) is an extreme state that is characterized by high temperatures ($T\in 10^3-10^8\,$K), high pressures ($P\sim1-10^4\,$Gbar), and densities in the vicinity of and even partially exceeding solid state densities ($n\sim10^{21}-10^{28}\,$cm$^{-3}$). These conditions are ubiquitous throughout Nature and occur in a host of astrophysical objects~\cite{fortov_review,drake2018high}, see the overview in Fig.~\ref{fig:overview}. Prominent examples are giant planet interiors~\cite{Benuzzi_Mounaix_2014,SEJ14}, such as Jupiter in our solar system~\cite{militzer1,saumon1,Militzer_2008,Guillot2018,Brygoo2021}, but also exoplanets~\cite{Fortney_2007,wdm_book}.
Other natural realizations of WDM include brown dwarfs~\cite{saumon1,becker}, white dwarf atmospheres~\cite{lebedev2007high,Kritcher2020}, neutron star crusts~\cite{neutron_star_envelopes,Haensel}, and the remnants of meteor impacts~\cite{Kraus2016,Glenzer_2016}.
Another prominent example of WDM is the core of Earth. It consists primarily of iron at a temperature of $\sim$6000~K and pressure of $\sim$300~GPa, which is considered at the cold end of the WDM parameter space. Understanding the response properties, such as the electrical and thermal conductivity in warm dense iron~\cite{ohta2016experimental,konopkova2016direct} is tied to geophysical dynamics within Earth's interior that generate its magnetic field~\cite{dobson2016earth}. Experimental measurements and modeling are the subjects of very active investigations~\cite{Zhang_PRL_2020,ramakrishna2022electrical,ramakrishna2022electrical_dft_u,lobanov2022}.

In addition to its fundamental importance for stellar objects, WDM is also highly relevant for a number of technological applications, such as the discovery of novel materials~\cite{Kraus2016,Kraus2017,frydrych2020demonstration,Lazicki2021,Schuster_diamond_2022,He2022} and hot-electron chemistry~\cite{Brongersma2015,ernstorfer,ernstorfer2}. A particularly important application is given by inertial confinement fusion~\cite{Betti2016,Zylstra2022} as it is realized at the National Ignition Facility (NIF)~\cite{Moses_NIF}. On its pathway towards ignition, the fuel capsule traverses the WDM regime~\cite{hu_ICF} (see the black line in Fig.~\ref{fig:overview} that indicates the implosion path of a deuterium-tritium [DT] capsule~\cite{hu_ICF}).

Due to the high current interest in WDM, such extreme states are realized in large experimental facilities using different techniques; see the topical overview by Falk~\cite{falk_wdm}. Indeed, many ground-breaking results have been reported over the last years, including the formation of nanodiamonds at extreme pressure~\cite{Kraus2016,Kraus2017,Hamel_Diamond_arxiv} and the study of the liquid-liquid phase transition in hot dense hydrogen~\cite{Dzyabura_PNAS_2013,Knudson_Science_2015,Pierleoni_PNAS_2016,Knudson_PRL_2017,Celliers_Science_2018}. In addition, it can be expected that emerging capabilities at facilities such as NIF~\cite{Moses_NIF,Lutgert_POP_2022}, LCLS at SLAC~\cite{Glenzer_2016,LCLS_2016}, the Sandia Z Pulsed Power Facility~\cite{sinars2020review,Z_machine_1,Z_machine_2} in the USA, SACLA~\cite{Pile2011} in Japan, and the European XFEL~\cite{Tschentscher_2017} in Germany will open up new avenues for the study of matter at increasingly extreme densities and temperatures, with direct relevance to technological applications and laboratory astrophysics~\cite{lebedev2007high,Remington_2005,takabe_kuramitsu_2021}.

At the same time, it is important to note that a rigorous theoretical description of WDM is notoriously difficult~\cite{wdm_book,review,new_POP}. More specifically, WDM can be conveniently characterized in terms of three dimensionless parameters that are of the order of unity~\cite{review,Ott2018}. Firstly, the Wigner-Seitz radius $r_s=d/a_\textnormal{B}$ is given by the ratio of the average interparticle distance $d$ to the Bohr radius $a_\textnormal{B}$, see the vertical blue lines in Fig.~\ref{fig:overview}. Second, the degeneracy temperature $\theta=k_\textnormal{B}T/E_\textnormal{F}$ measures the thermal energy in units of the electronic Fermi energy $E_\textnormal{F}$~\cite{quantum_theory}, with $\theta\ll 1$ and $\theta\gg 1$ corresponding to the fully degenerate and semi-classical regime, respectively. Different values of $\theta$ are included as the green lines in Fig.~\ref{fig:overview}. A third useful parameter is the classical coupling parameter of the electrons $\Gamma$ that measures the ratio of the interaction to the kinetic energy on the level of a mean-field description~\cite{Ott2018}; it is depicted by the solid red line in Fig.~\ref{fig:overview}, and we also find $\Gamma\sim1$ in the WDM regime.

Consequently, there are no small parameters that can serve as the basis for a suitable expansion~\cite{wdm_book,new_POP}.
Nevertheless, the strong demand for theoretical models and simulations has sparked a surge of developments in the field. For example, \emph{ab initio} quantum Monte Carlo (QMC)~\cite{anderson2007quantum} methods are capable of providing exact results for the static properties of the uniform electron gas (UEG)~\cite{Schoof_CPP_2015,Dornheim_POP_2017,Yilmaz_JCP_2020,dornheim_prl,Malone_JCP_2015,Malone_PRL_2016,Joonho_JCP_2021,Schoof_PRL_2015,groth_prl,Groth_PRB_2016,Dornheim_PRB_2016,Dornheim_PRB_nk_2021,Dornheim_PRE_2021,Filinov_CPP_2021,Hunger_PRE_2021} --- the quantum version of the classical one-component plasma (OCP)~\cite{review,quantum_theory,loos} --- and light elements such as hydrogen~\cite{Bohme_PRL_2022,Bohme_PRE_2022} over substantial parts of the WDM regime. Moreover, path integral Monte Carlo (PIMC) simulations within the fixed-node approximation~\cite{Ceperley1991} --- also known as restricted PIMC (RPIMC) in the literature --- have allowed Militzer and co-workers~\cite{Driver_PRL_2012,Driver_PRB_2016,Driver_PRE_2018} to go to heavier elements and material mixtures~\cite{Zhang_JCP_2018}. These simulations constitute the basis for an extensive equation-of-state (EOS) database~\cite{Militzer_PRE_2021} that can be used for a host of practical applications.

A second important tool for the \emph{ab initio} simulation of WDM is thermal density functional theory (DFT)~\cite{Mermin_DFT_1965,Pittalis_PRL_2011,Pribram_Thermal_Context_2014,Dharma-wardana_Computation_2016,Smith2018}, which combines a computational cost that is generally less demanding compared to QMC methods with an often acceptable accuracy~\cite{Clay_PRB_2014,Clay_PRB_2016,Moldabekov_JCP_2021,Moldabekov_PRB_2022,Moldabekov_PRL_2022}. In particular, first accurate parametrizations of the exchange-correlation (XC) free energy $f_\textnormal{xc}$ of the warm dense UEG~\cite{ksdt,groth_prl,review,status} allow for thermal DFT simulations of WDM on the level of the local density approximation (LDA) that take consistently into account the dependence of the XC functional on the temperature parameter $\theta$. Such temperature effects can have a substantial impact on the DFT results for different observables~\cite{Sjostrom_PRB_2014,karasiev_importance,kushal}, and the development of improved XC functionals for WDM calculations constitutes an active area of research~\cite{pribram,Karasiev_PRL_2018,PhysRevB.101.245141,Karasiev_PRB_2022}. This is complemented by the development of efficient algorithms that allow one to extend DFT to higher temperatures~\cite{Cangi_PRB_2015,Zhang_POP_2016,Ding_PRL_2018,Cytter_PRB_2019,moussa2019assessment,Bethkenhagen_high_T_2021,White_2022,Fiedler_PRR_2022,BLANCHET2022108215}, which are unfeasible for standard Kohn-Sham implementations~\cite{Jones_RMP_2015}. Moreover, there has been a surge of research activities in utilizing machine-learning techniques in the context of DFT~\cite{Pederson2022}. The most promising approaches for improving calculations in the WDM regime include machine-learning interatomic potentials that enable large-scale \emph{ab initio} molecular dynamics calculations and actually replacing DFT calculations with machine-learning surrogate models~\cite{Fiedler_PRM_2022}. The development of accurate machine-learning interatomic potentials has become a broad field of research~\cite{Ceriotti2022}. They enable covering a large parameter space in temperature and pressure than is feasible with conventional \emph{ab initio} molecular dynamics as has recently been demonstrated in several works~\cite{Schoerner_PRB_2022,Schoerner_PRB_2022_2,Willman_PRB_2022,Hamel_Diamond_arxiv}. Replacing DFT all along in terms of machine learning models is an avenue of research with great potential. Pioneering efforts on models~\cite{Snyder_PRL_2012,Brockherde2017,Grisafi2019} have matured into production-level codes that accelerate DFT calculations for realistic materials at finite temperature and pressure~\cite{Ellis_PRB_2021,Fiedler_MLST_2022} and have recently enabled simulations at unprecedented system size~\cite{Fiedler_big}.

In combination with complementary methods such as quantum hydrodynamics~\cite{Diaw2017,zhandos_pop18,Larder_SciAdvance_2019,phdthesis,Moldabekov_SciPost_2022,Graziani_CPP_2022}, quantum scattering theory \cite{PhysRevB.102.195127,  PhysRevLett.119.045002}, and kinetic models~\cite{PhysRevE.103.063206, doi:10.1063/1.5095655} 
these developments have given new insights into the physics of WDM, including the EOS~\cite{Benedict_PRB_2014,Harbour_PRE_2017,Militzer_PRE_2021}, effective potentials~\cite{Dharma-wardana_PRE_2012,zhandos_cpp17,zhandos_cpp21,Dornheim_JCP_2022}, and a number of transport properties~\cite{GRABOWSKI2020100905} such as stopping power \cite{Magyar_CPP_2016,Hayes2020, White_2022, PhysRevE.101.053203} and
electrical conductivity \cite{Witte_PRL_2017, PhysRevE.103.063206}.


Many of the properties of WDM are encoded in its response to an external perturbation.
Of particular interest are perturbations that couple to the electronic charge density, e.g., an incident electromagnetic field or the Coulomb field of an incident charged particle. 
The response to these probes contains information about a wide range of excitations, many of which are sufficiently sensitive to the thermodynamic state of the WDM that they can be used to infer conditions like temperature~\cite{Dornheim_T_2022} or density~\cite{Harbour_PRE_2017,siegfried_review,Gregori_PRE_2003}.
X-ray Thomson Scattering (XRTS)~\cite{siegfried_review,sheffield2010plasma,Crowley_2013,kraus_xrts} is a widely used diagnostic that enables these inferences (among others) through measurements of the intensity of hard x-rays that are inelastically scattered from WDM samples.
The fact that hard x-rays are required is a reflection of the fact that the densities of WDM are sufficiently high that they are opaque to lower energy probes.
Those same hard x-rays can be used to probe the state of these systems, while isochorically heating it on femtosecond time scales~\cite{Sperling_PRL_2015,Witte_PRL_2017}.

More specifically, XRTS experiments rely on measurements of the angle- and energy-resolved  scattering intensity to probe the dynamic structure factor (DSF), $S(\mathbf{q},\omega)$, where $\mathbf{q}$ and $\omega$ are a wave number and frequency, respectively.
For sufficiently weak probes in the linear response regime, the DSF is related to the linear density response function $\chi({\bf q},\omega)$ through the fluctuation-dissipation theorem~\cite{quantum_theory} (FDT), Eq.~(\ref{eq:FDT}).
Kinematic constraints from the first Born approximation dictate that the scattering angle is related to the momentum transferred $\mathbf{q}$ to or from the sample in the scattering process.
Similarly, the shift in energy upon scattering $\omega$ is the quantity of interest, rather than the absolute energy of the detected light, as this encodes information about the excitation or de-excitation of the sample.
Thus it is typical to parametrize the XRTS scattering intensity $I(\mathbf{q},\omega)$ in terms of the same momentum and energy transfer as those parametrizing $S(\mathbf{q},\omega)$ and $\chi({\bf q},\omega)$, even though the salient experimental quantities (detection angle and photon energy) are different.

In fact, because the XRTS probe beam is only approximately monochromatic and the detector has energy-dependent sensitivity, the actually detected intensity is a convolution of the DSF and the combined source-instrument function $R(\omega)$,
\begin{eqnarray}\label{eq:convolution}
I(\mathbf{q},\omega) = S(\mathbf{q},\omega) \circledast R(\omega).
\end{eqnarray}
Nevertheless with careful characterization of $R(\omega)$, information about $S(\mathbf{q},\omega)$ and $\chi({\bf q},\omega)$ can be obtained from the measured intensity.
Thus in addition to its practical utility as a multi-purpose diagnostic, XRTS is ultimately a powerful technique to directly access the density response function.

From a theoretical perspective, it is very convenient to express $\chi(\mathbf{q},\omega)$ as~\cite{kugler1}
\begin{eqnarray}\label{eq:LFC}
\chi(\mathbf{q},\omega) = \frac{\chi_0(\mathbf{q},\omega)}{1-\frac{4\pi}{q^2}\left[1-G(\mathbf{q},\omega)\right]\chi_0(\mathbf{q},\omega)}\ ,
\end{eqnarray}
with $\chi_0(\mathbf{q},\omega)$ being a known reference function, such as the density response of an ideal Fermi gas (also known as Lindhard function in the literature) in the case that $\chi(\mathbf{q},\omega)$ describes the density response of the UEG. In this case, the complete wave-vector- and frequency-resolved information about electronic XC effects is contained in the local field correction (LFC) $G(\mathbf{q},\omega)$. Indeed, setting $G(\mathbf{q},\omega)\equiv0$ in Eq.~(\ref{eq:LFC}) corresponds to the well-known random phase approximation (RPA), which describes the density response on a mean-field level~\cite{bonitz_book}. Consequently, the LFC constitutes key input for a host of applications, such as the construction of advanced XC functionals for DFT via the adiabatic connection formula~\cite{pribram,Lu_JCP_2014,Patrick_JCP_2015}, the incorporation of electron-electron correlations into quantum hydrodynamics~\cite{zhandos_pop18,Diaw2017}, or the interpretation of XRTS experiments~\cite{Fortmann_PRE_2010,Preston_APL_2019,Dornheim_PRL_2020_ESA,Zan_PRE_2021}. In addition, the LFC is formally equivalent to the XC kernel from linear-response time-dependent DFT (LR-TDDFT)~\cite{dynamic1,marques2012fundamentals,IIT}; they are related by $K_\textnormal{xc}(\mathbf{q},\omega)=-4\pi/q^2G(\mathbf{q},\omega)$. Indeed, replacing $\chi_0(\mathbf{q},\omega)$ by the dynamic Kohn-Sham response function $\chi_\textnormal{S}(\mathbf{q},\omega)$ in Eq.~(\ref{eq:LFC}) constitutes the basis for the LR-TDDFT simulation of real materials~\cite{marques2012fundamentals,Mo_PRL_2018,Ramakrishna_2020,Ramakrishna_PRB_2021}, see Sec.~\ref{sec:DFT} below.

While the assumption of a \emph{linear response} is fairly ubiquitous throughout quantum many-body theory, it holds, strictly speaking, only in the limit of an infinitesimal perturbation amplitude $A$. Recently, Dornheim \emph{et al.}~\cite{Dornheim_PRL_2020} have presented the first \emph{ab initio} PIMC simulation results for the nonlinear density response of the warm dense UEG. Such nonlinear effects~\cite{golden_pra_85,mukamel_86,bergara1997,Bergara1999,Kalman,Mikhailov_Annalen,Mikhailov_PRL,Dornheim_PRR_2021,Dornheim_JPSJ_2021,Dornheim_CPP_2022} become important for example for XRTS experiments with ultra-high intensities that can be realized at X-ray free-electron laser (XFEL) facilities via the novel seeding technique~\cite{Fletcher2015}. Moreover, they are an important ingredient to the construction of effective pair potentials at small distances~\cite{Rasolt_PRB_1975,Gravel,zhandos_cpp21} and stopping power calculations for slow and/or heavy particles~\cite{PhysRevB.37.9268,Echenique_PRB_1986,Nagy_PRA_1989}. Finally, it has been shown that nonlinear effects depend much more sensitively on system parameters like the temperature $T$~\cite{Dornheim_PRL_2020,Dornheim_PRR_2021}, which makes their experimental observation a potentially interesting new method of diagnostics~\cite{Moldabekov_SciRep_2022}.
Describing the dynamics of a quantum many-body system due to a time-dependent external perturbation beyond the linear response regime is achieved by the real-time formulation of time-dependent DFT (RT-TDDFT)~\cite{RG84} which will be introduced below.

Lastly, we note that even the interpretation of XRTS experiments with comparably low scattering intensity such that any nonlinear effects can safely be neglected generally relies on a number of model assumptions. More specifically, the most widely used model is the Chihara decomposition~\cite{Chihara_1987,Gregori_PRE_2003,siegfried_review,kraus_xrts}, 
which is a chemical model assuming a clear distinction between \emph{bound} and \emph{free} electrons. This assumption is questionable in the WDM regime~\cite{Bohme_PRL_2022} and constitutes a de facto uncontrolled approximation in practice. Therefore, previous EOS measurements~\cite{Falk_HEDP_2012,Falk_PRL_2014,Kritcher2020} can substantially depend on the employed model and do not necessarily constitute a reliable benchmark for the development of new theoretical approaches such as TDDFT~\cite{dynamic2}, or reliable input for other applications. To overcome this unsatisfactory situation, Dornheim \emph{et al.}~\cite{Dornheim_T_2022,Dornheim_insight_2022,Dornheim_T_follow_up,Dornheim_PTR_2022} have very recently suggested switching from the usual frequency-representation to the imaginary-time domain. This gives one direct access to the physical properties of a given system of interest and allows one to extract for example the temperature without any models, approximations, or simulations. In addition, such imaginary-time correlation functions (ITCF) naturally emerge within Feynman's imaginary-time path integral picture of statistical mechanics~\cite{Berne_JCP_1983} and constitute an important link between simulations~\cite{dornheim_ML,Dornheim_JCP_ITCF_2021}, physical insights~\cite{Dornheim_insight_2022,Dornheim_PTR_2022}, and experimental measurements~\cite{Dornheim_T_2022,Dornheim_T_follow_up}.

\begin{table*}[t]
    \centering\hspace*{-0.35cm}
    \begin{tabular}{c|c|c|c|c|c}\hline
      &  $f(t)$ & $A_0$ &  $\gamma$ & Regime & Approach \\
        \hline\hline
   1  & $ 1$   & small & 1 & dynamic linear response & LR-TDDFT (Sec.~\ref{sec:DFT}),  PIMC (Sec.~\ref{sec:PIMC}) + analytic continuation (Sec.~\ref{sec:ITCF}) \\
    2 & $ 1$   & large & 1 & dynamic nonlinear response & Dynamic nonlinear response theory (Ref.~\cite{Dornheim_PRR_2021}) \\
    \hline
   3  & $ 1$   & small & 0 & static linear response & PIMC (Sec.~\ref{sec:PIMC}), KS-DFT (Sec.~\ref{sec:DFT}), dielectric theories~(Refs.~\cite{IIT,stls,stls2,stls_original,Tanaka_CPP_2017,tanaka_hnc,Tolias_JCP_2021})  \\
    4 & $ 1$   & large & 0 & static nonlinear response & PIMC (Sec.~\ref{sec:PIMC}), KS-DFT (Sec.~\ref{sec:DFT}) \\
    \hline
    5 & $A_0 \delta(t)$   & small & 0 & weak kick & real-time TDDFT (RT-TDDFT) (Sec.~\ref{sec:RTTDDFT}), NEGF (Sec.~\ref{sec:NEGF})  \\
    6 & $A_0 \delta(t)$   & large & 0 & strong kick & RT-TDDFT (Sec.~\ref{sec:RTTDDFT}), NEGF (Sec.~\ref{sec:NEGF}) \\ \hline\hline
    \end{tabular}
    \caption{
    Overview of possible excitation scenarios contained in the Hamiltonian Eq.~\eqref{eq:Hamiltonian_general}. More details on linear-response theory (cases 1 and 3) can be found in Sec.~\ref{sec:LRT}; static nonlinear response theory (case 4) is discussed in Sec.~\ref{sec:nonlinear} and its generalization to the dynamic nonlinear regime (case 2) has been presented in Ref.~\cite{Dornheim_PRR_2021}. The theoretical formalism behind cases 5 and 6 is discussed in Sec.~\ref{sec:NEGF}.
    }
    \label{tab:response-overview}
\end{table*}

In this work, we attempt to give a coherent picture of the current state-of-the-art with respect to our understanding of the electronic density response of WDM. While we put more emphasis on a number of recent promising developments with respect to theoretical models and different simulation techniques, we also indicate the respective connection to relevant experiments. Indeed, although the notion of evaluating experimental data in the imaginary-time domain might seem to be somewhere between impractical to outright outlandish, we demonstrate the simplicity of the idea using an actual XRTS data set.

The paper is organized as follows: Sec.~\ref{sec:theory} contains the relevant theoretical background, starting with an introduction of the most relevant numerical methods in Sec.~\ref{sec:methods}, namely path integral Monte Carlo (\ref{sec:PIMC}), thermal DFT (\ref{sec:DFT}), and real-time time-dependent DFT (RT-TDDFT) (\ref{sec:RTTDDFT}). Moreover, we briefly touch upon non-equilibrium Green's functions (NEGF) in Sec.~\ref{sec:NEGF}. In addition, we discuss important relations within linear-response theory (\ref{sec:LRT}), its relationship to imaginary-time correlation functions (\ref{sec:ITCF}), and finally some basic concepts for a nonlinear theory of the electronic density response (\ref{sec:nonlinear}).
Sec.~\ref{sec:results} is devoted to the discussion of a gamut of simulation results both for a uniform electron gas, and for real WDM systems. In particular, we start our discussion with results for the linear density response in the static limit in Sec.~\ref{sec:static_LRT_results}, including a discussion of PIMC simulations of the UEG (\ref{sec:UEG_results}) and recent PIMC and thermal DFT simulations of warm dense hydrogen (\ref{sec:hydrogen_results}). In Sec.~\ref{sec:dynamic_results}, we extend these considerations to the fully dynamic case, again starting with a PIMC-based investigation of the UEG (\ref{sec:dynamic_UEG}) followed by TDDFT calculations for hydrogen, iron, and aluminium in Sec.~\ref{sec:dynamic_WDM}.
The discussion of simulation results is completed in Sec.~\ref{sec:nonlinear_results}, where we present new results for the nonlinear density response of the warm dense UEG.
In Sec.~\ref{sec:experiment}, we connect our theoretical framework to experimental observations by re-interpreting the XRTS measurement of plasmons in warm dense beryllium by Glenzer \emph{et al.}~\cite{Glenzer_PRL_2007} in the imaginary-time domain~\cite{Dornheim_T_2022,Dornheim_T_follow_up}.
The paper is concluded in Sec.~\ref{sec:summary}, where we discuss how our understanding of the electronic density response of WDM can be improved in future works by combining new developments in \emph{ab initio} simulations with new experimental set-ups.

\section{Theory\label{sec:theory}}

We consider an $N$-particle system that is described by the sum of an unperturbed Hamiltonian $\hat{H}_0$ and an external  single-particle perturbation that we give in the most general form, 
\begin{eqnarray}\label{eq:Hamiltonian_general}
 \hat{H} &=& \hat{H}_0 + \hat{V}_{\mathbf{q},\omega,A}
\,,\quad 
 \hat H_0 = \hat K + \hat W + \hat V_\textnormal{pot}\,,\\ \label{eq:Hamiltonian_general2}
\hat{V}_{\mathbf{q},\omega,A} &=& 2 A f(t) \sum_{l=1}^N \textnormal{cos}\left( \mathbf{q}\cdot\hat{\mathbf{r}}_l - \gamma\omega t \right)\ .
\end{eqnarray}
Here $\hat{K}$, $\hat{W}$, and $\hat{V}_\textnormal{pot}$ are the kinetic, interaction and external Coulomb energy (e.g. due to a configuration of ions) of the electrons, respectively. Moreover, $\hat{V}_{\mathbf{q},\omega,A}$ corresponds to an additional external perturbation
with $\mathbf{q}$ and $\omega$ being the corresponding wave vector and frequency, and $A(t)=Af(t)$ is a real perturbation amplitude that can explicitly depend on the time~\cite{kwong_prl-00}. We note that both $\hat{V}_{\mathbf{q},\omega,A}$ and $\hat{V}_\textnormal{pot}$ contribute to the total external single-electron potential energy $\hat{V}_\textnormal{ext}=\hat{V}_\textnormal{pot}+\hat{V}_{\mathbf{q},\omega,A}$.  In Eq.~\eqref{eq:Hamiltonian_general2}, we introduced a parameter $\gamma$ that is either $0$ or $1$ to distinguish between static and monochromatic excitation scenarios. An overview of the different physical situations that have been treated with established analytical and computational approaches is presented in Tab.~\ref{tab:response-overview}. Note that we assume Hartree atomic units throughout unless indicated otherwise.

In the limit of small $A$, linear response theory (LRT)~\cite{quantum_theory} becomes valid, which leads to a number of simplifications and useful relations that are discussed in Sec.~\ref{sec:LRT}. A more general nonlinear theory of the electronic density response is discussed in Sec.~\ref{sec:nonlinear}. While it is often convenient to restrict oneself to a static perturbation amplitude $A(t)\equiv A$ in Eq.~(\ref{eq:Hamiltonian_general}), the more general form is useful to study the response of a given system to a finite perturbation pulse with real-time methods, which is discussed in more detail in Sec.~\ref{sec:NEGF}.

From a practical perspective, we note that a gamut of numerical methods for the description of WDM systems has been presented in the literature, including molecular dynamics, classical Monte Carlo simulations, average-atom models~\cite{FMT49,SHWI07,FBCR10,STJZS14,MBVSM21,callow_first_principles}, as well as integral equation theory approaches within the hypernetted-chain approximation for effective quantum potentials~\cite{Jones_HEDP_2007,STARRETT201435,perrot,Golubnichy_CPP_2002}, dielectric formalism schemes~\cite{IIT,stls_original,stls,stls2,Holas_PRB_1987,schweng,stolzmann,vs_original,Tanaka_CPP_2017,tanaka_hnc,arora,castello2021classical,Tolias_JCP_2021}, and quantum hydrodynamics~\cite{pop_qhd_bonitz,Diaw2017,zhandos_pop18,Moldabekov_SciPost_2022,Graziani_CPP_2022}. 

In Sec.~\ref{sec:methods}, we focus on four particularly important methods, namely path integral Monte Carlo (\ref{sec:PIMC}), thermal DFT (\ref{sec:DFT}), real-time time-dependent DFT (\ref{sec:RTTDDFT}), and nonequilibrium Green functions (\ref{sec:NEGF}).

\subsection{Numerical methods\label{sec:methods}}


\subsubsection{Path integral Monte Carlo\label{sec:PIMC}}

The \emph{ab initio} path integral Monte Carlo (PIMC) approach~\cite{cep,Berne_JCP_1982,Takahashi_Imada_PIMC_1984,Pollock_PRB_1984} is one of the most successful methods for the description of quantum degenerate many-body systems. It is based on an evaluation of the partition function in coordinate representation, which, in the case of an electron gas in the canonical ensemble (inverse temperature $\beta=1/k_\textnormal{B}T$, volume $\mathcal{V}=L^3$, and number density $n=N/\mathcal{V}$ are constant) with $N^\uparrow$ and $N^\downarrow$ spin-up and spin-down electrons, is given by
\begin{widetext} 
\begin{eqnarray}\label{eq:Z}
Z_{\beta,N,\mathcal{V}} &=& \frac{1}{N^\uparrow! N^\downarrow!} \sum_{\sigma^\uparrow\in S_N^\uparrow} \sum_{\sigma^\downarrow\in S_N^\downarrow} \textnormal{sgn}(\sigma^\uparrow,\sigma^\downarrow) \int_\mathcal{V} d\mathbf{R} \bra{\mathbf{R}} e^{-\beta\hat H} \ket{\hat{\pi}_{\sigma^\uparrow}\hat{\pi}_{\sigma^\downarrow}\mathbf{R}}\ .
\end{eqnarray}\end{widetext}
Here the summation over all elements $\sigma^i$ of the respective permutation group $S_N^i$, with $\hat{\pi}_{\sigma^i}$ being the corresponding permutation operator, ensures the exact anti-symmetry with respect to the exchange of coordinates of identical fermions. Since a detailed introduction to the PIMC method has been presented elsewhere~\cite{cep}, we here restrict ourselves to a brief sketch of the main idea.
The central obstacle with respect to a direct evaluation of Eq.~(\ref{eq:Z}) is the non commutability of the kinetic ($\hat{K}$) and potential ($\hat{V}=\hat{W}+\hat{V}_\textnormal{ext}$) contributions to the full Hamiltonian $\hat{H}=\hat{K} + \hat{V}$, which renders the matrix elements of the density operator $\hat{\rho}=e^{-\beta\hat{H}}$ generally unfeasible. To overcome this issue, one can exploit the well-known semi-group property of the density operator, which eventually leads to a summation over $P$ particle coordinates $\mathbf{R}_i$. Crucially, each of the corresponding $P$ density matrices has to be evaluated at $P$-times the original temperature, which allows for the introduction of suitable factorization schemes~\cite{sakkos_JCP_2009,brualla_JCP_2004,Zillich_JCP_2010}; the associated factorization error can be estimated from the Baker-Campbell-Hausdorff formula~\cite{kleinert2009path}, and vanishes in the limit of large $P$.

In addition, we note that each high-temperature factor can be interpreted as a propagation in the imaginary time $t = -i\hbar\tau $ over a discrete time step $\tau=\beta/P$. Consequently, each quantum particle is represented as a \emph{path} along the imaginary time, which can be mapped onto an ensemble of classical ringpolymers; this is the origin of the celebrated \emph{classical isomorphism} by Chandler and Wolynes~\cite{Chandler_JCP_1981}. Moreover, PIMC gives one direct access to different imaginary-time correlation functions, such as the imaginary-time version of the intermediate scattering function defined in Eq.~(\ref{eq:ISF}) below.
The basic idea of the PIMC method is to stochastically generate all possible path configurations according to their respective contribution to the partition function $Z$ based on the Metropolis algorithm~\cite{metropolis}. For indistinguishable quantum particles --- bosons or fermions --- this also requires the sampling of all possible permutation topologies~\cite{Dornheim_permutation_cycles}, which can be achieved efficiently using different implementations~\cite{mezza,Dornheim_PRB_nk_2021} of the worm algorithm by Boninsegni \emph{et al.}~\cite{boninsegni1,boninsegni2}.

In the case of bosons (or hypothetical distinguishable quantum particles which are sometimes referred to as boltzmannons in the literature~\cite{Clark_PRL_2009,dornheim_cpp}), the sign function in Eq.~(\ref{eq:Z}) is always positive, and the PIMC method allows for the quasi-exact simulation of up to $N\sim10^4$ particles, which has given important insights into phenomena such as superfluidity~\cite{ultracold2,Kwon_PRB_2006,Dornheim_PRA_2020}.
For fermions, such as the electrons that are a key constituent of WDM, the sign function changes its sign for every pair exchange. The resulting cancellation of positive and negative terms is the root cause of the notorious fermion sign problem~\cite{Loh_PRB_1990,troyer,dornheim_sign_problem,Dornheim_JPA_2021}, which leads to an exponential increase in the required compute time upon increasing the system size $N$ or decreasing the temperature $T$. Therefore, the direct PIMC method that is used in the present work is limited to specific parts of the WDM parameter space ($\theta\gtrsim0.5$) and to light elements such as hydrogen~\cite{Bohme_PRE_2022}.

For completeness, we note that the sign problem can formally be avoided by imposing restrictions
on the nodal structure of the thermal density matrix; this restricted PIMC method developed by Ceperley~\cite{Ceperley1991} is exact if the true nodal structure of the quantum many-body system of interest was known. In practice, however, one has to introduce approximations, and often the nodal structure of an ideal Fermi gas at the same conditions is used~\cite{Brown_PRL_2013,Ceperley1991}. First, this \emph{fixed-node approximation} introduces an uncontrolled error that can be of the order of $10\%$ in the XC energy of a UEG at high density and low temperature~\cite{Schoof_PRL_2015}, whereas it is less pronounced in the momentum distribution function around $\theta=1$~\cite{Dornheim_PRB_nk_2021,Dornheim_PRE_2021}. Second, the nodal restrictions break the imaginary-time translation invariance, which prevents the direct estimation of imaginary-time correlation functions such as $F(\mathbf{q},\tau)$, see Sec.~\ref{sec:ITCF} below.

\subsubsection{Thermal density functional theory\label{sec:DFT}}
Let us consider a (time-independent) electronic Hamiltonian of the form
\begin{eqnarray}\label{eq:Hamiltonian}
 \hat{H} = \hat{K} + \hat{W} + \hat{V}_\textnormal{ext}\ ,
\end{eqnarray}
with $\hat{K}$ being the kinetic part, $\hat{W}$ containing the electron-electron interaction, and $\hat{V}_\textnormal{ext}$ an external single-particle potential.  In particular, $\hat{V}_\textnormal{ext}$ takes into account the electron-ion interaction when a snapshot of ions is considered, which is a common practice when DFT calculations are coupled to molecular dynamics (MD) simulations of classical ions within the Born-Oppenheimer approximation~\cite{wdm_book}. Further, $\hat{V}_\textnormal{ext}$ can also include the monochromatic external perturbation that is used to obtain the electronic density response, cf.~Eq.~(\ref{eq:Hamiltonian_general2}) in the limit of $\omega\to0$ and $f(t)\equiv1$.
The basic idea of DFT is to express the ground-state energy of a many-electron system defined by Eq.~(\ref{eq:Hamiltonian}) in terms of the many-electron density $n(\mathbf{r})$. At finite temperature, this is achieved by Mermin's generalization~\cite{Mermin_DFT_1965} of the Hohenberg-Kohn theorems~\cite{hohenberg-kohn}. This enables a thermodynamic description of many-electron systems within the context of DFT by the grand-canonical potential as a density functional 
\begin{equation}\label{eq:thermal-DFT}
\Omega[n] = F^T[n] + V_\textnormal{ext}[n]     
\end{equation}
where 
\begin{equation}
F^T[n]= \min_{\hat\Gamma \to n} \left( K[\hat\Gamma] + \mathcal{S}[\hat\Gamma] + W[\hat\Gamma] \right)    
\end{equation}
is the universal functional that includes contributions from the kinetic energy, entropy $\mathcal{S}$, and electron-electron interaction. Note that $F^T[n]$ itself is minimized over the statistical operator $\hat\Gamma$ defined by the many-particle states of the Hamiltonian in Eq.~(\ref{eq:Hamiltonian}).
Then the grand-canonical potential of a many-electron system in thermal equilibrium is found by minimization
\begin{equation}
\Omega = \min_n \Omega[n] = \min_n \left( F^T[n] + V_\textnormal{ext}[n] \right)\ .
\end{equation}
Minimizing $\Omega[n]$ over trial densities requires knowledge of all terms in Eq.~(\ref{eq:thermal-DFT}) as a functional of the density. This is, however, not straightforward, because explicit density functionals of these terms (besides the trivial term $V_\textnormal{ext}[n]$) are not known.

A practical solution to this problem is the KS approach~\cite{KS65}. Here one defines an auxiliary system of non-interacting particles described by a set of non-interacting single-particle Schr\"odinger equations 
\begin{eqnarray}\label{eq:SPSE}
 \left(-\frac{\nabla^2}{2} + v_\textnormal{S}[n](\mathbf{r}) \right)\phi_{\alpha, {\bf k}}(\mathbf{r}) = \epsilon_{\alpha, {\bf k}} \phi_{\alpha, {\bf k}} (\mathbf{r})\ ,
\end{eqnarray}
where $\phi_{\alpha, {\bf k}}(\mathbf{r})$ and $\epsilon_{\alpha, {\bf k}}$ denote the KS orbitals and corresponding eigenvalues with band index $\alpha$ and Bloch wave number ${\bf k}$.
The KS approach allows writing the grand-canonical potential as
\begin{equation}
\Omega[n] = K_\textnormal{S}[n] + T \mathcal{S}_\textnormal{S}[n] + U[n] + \Omega_\textnormal{XC}[n] + V_\textnormal{ext}[n]\,,
\end{equation}
where $K_\textnormal{S}[n]$ denotes the KS kinetic energy, $\mathcal{S}_\textnormal{S}[n]$ the KS entropy, $U[n]$ the Hartree energy, and $\Omega_\textnormal{XC}$ the XC free energy.
The KS system is constructed such that the density it yields
\begin{equation}\label{eq:KS-density}
    n(\mathbf{r}) =  \sum_{\alpha, {\bf k}} f_{\alpha, {\bf k}}(\beta,\mu) |\phi_{\alpha, {\bf k}}(\mathbf{r})|^2
\end{equation}
is identical to the many-electron density of the corresponding interacting many-electron system defined in Eq.~(\ref{eq:Hamiltonian}). Due to their nature as effective single-particle states, the KS orbitals $\{\phi_{\alpha, {\bf k}}\}$ are populated according to the Fermi-Dirac distribution
\begin{eqnarray}\label{eq:Fermi}
 f_{\alpha, {\bf k}}(\beta,\mu) = \left[1 + e^{\beta(\epsilon_{\alpha, {\bf k}}-\mu)} \right]^{-1}\ ,
\end{eqnarray}
with $\beta = 1/k_\textnormal{B}T$ and $\mu$ the chemical potential~\cite{quantum_theory}. In combination, Eqs.~(\ref{eq:SPSE}) and (\ref{eq:Fermi}) give one direct access to $K_\textnormal{S}[n] = -1/2 \sum_{\alpha, {\bf k}} \int d\mathbf{r}\ \phi_{\alpha, {\bf k}}^*(\mathbf{r}) \nabla^2 \phi_{\alpha, {\bf k}}(\mathbf{r})$ and $\mathcal{S}_\textnormal{S}[n] = -\sum_{\alpha, {\bf k}} \left( f_{\alpha, {\bf k}} \ln(f_{\alpha, {\bf k}}) + (1-f_{\alpha, {\bf k}}) \ln(1-f_{\alpha, {\bf k}}) \right)$, whereas all non-ideal contributions to the full kinetic energy $K$ and entropy $\mathcal{S}$ are contained in the XC functional.
The equality of the density in Eq.~(\ref{eq:KS-density}) with the true many-electron density is achieved by the KS potential defined as $v_\textnormal{S}[n] = v_\textnormal{ext}(\mathbf{r}) + v_\textnormal{H}[n](\mathbf{r}) + v_\textnormal{XC}[n](\mathbf{r})$ in terms of the external potential $v_\textnormal{ext}(\mathbf{r})$, the Hartree potential $v_\textnormal{H}[n](\mathbf{r})=\delta U[n]/\delta n(\mathbf{r})$, and the XC potential $v_\textnormal{XC}[n](\mathbf{r})=\delta \Omega_\textnormal{XC}[n]/\delta n(\mathbf{r})$.
In particular, the XC contribution $\Omega_\textnormal{XC}[n]$ contains the full information about many-body correlations and, therefore, would require the exact solution of the original many-electron problem, which is not feasible. In practice, $\Omega_\textnormal{XC}[n]$, therefore, has to be approximated. Specifically, the particular choice of the XC functional substantially influences the accuracy of a DFT calculation~\cite{Burke_JCP_2012}, which makes both the benchmarking of existing functionals~\cite{Clay_PRB_2014,Clay_PRB_2016,Moldabekov_JCP_2021,Moldabekov_PRB_2022,Moldabekov_PRL_2022} and the construction of novel, more sophisticated approximations to $\Omega_\textnormal{XC}[n]$ indispensable.


We note that the KS orbitals by themselves should be viewed not as physical quantities but as auxiliary properties that are connected to the energy and density of a system.
At the same time, the $\{\phi_\alpha\}$ constitute an important ingredient to a number of other applications, such as the incorporation of nonlocality into quantum hydrodynamics via an \emph{ab initio} Bohm potential term~\cite{Moldabekov_SciPost_2022, pop_qhd_bonitz}. A particularly important application of the KS orbitals that have been obtained from an equilibrium DFT calculation is given by linear-response time-dependent DFT (LR-TDDFT), which is based on the KS response function~\cite{ullrich2012time,marques2012fundamentals} given by
\begin{eqnarray}\label{eq:chi0}
 \chi_\textnormal{S}(\mathbf{q}, \omega) &=& \frac{1}{\mathcal{V}}
\sum_{\mathbf{k}, \alpha, \alpha^{\prime}}
\frac{f_{\alpha\mathbf{k}}-f_{\alpha^{\prime} \mathbf{k} + \mathbf{q} }}{\omega + \epsilon_{\alpha\mathbf{k}} - \epsilon_{\alpha^{\prime} \mathbf{k} + \mathbf{q} } + i\eta}\\
& & \times
 \langle \phi_{\alpha \mathbf{k}} | e^{-i\mathbf{q} \cdot \mathbf{r}} | \phi_{\alpha^{\prime} \mathbf{k} + \mathbf{q} } \rangle
 \langle \phi_{\alpha\mathbf{k}} | e^{i\mathbf{q} \cdot \mathbf{r}^{\prime}} | \phi_{\alpha^{\prime} \mathbf{k} + \mathbf{q} } \rangle\ . \nonumber
\end{eqnarray}
Here, the parameter $0<\eta\ll 1$ ensures the retardation, the sum runs over the different eigenvalues $\epsilon$ and Fermi functions $f$ at different momenta $\kv$ and $\kv+\qv$, $\omega$ is the energy of the excited mode.

The physically meaningful dynamic electronic linear density response of a system of interest is then given by
\begin{eqnarray}\label{eq:kernel}
 \chi(\mathbf{q},\omega) = \frac{\chi_\textnormal{S}(\mathbf{q},\omega)}{1 - \left[v(q)
 +K_\textnormal{xc}(\mathbf{q},\omega)\right]\chi_\textnormal{S}(\mathbf{q},\omega)}\ ,
\end{eqnarray}
with $K_\textnormal{xc}(\mathbf{q},\omega)$ being the XC kernel mentioned in the discussion of Eq.~(\ref{eq:LFC}) above. Since the LHS of Eq.~(\ref{eq:kernel}) is a well-defined physical observable, the kernel $K_\textnormal{xc}(\mathbf{q},\omega)$ strongly depends on the particular form of the KS response function $\chi_\textnormal{S}(\mathbf{q},\omega)$. In other words, the true XC kernel has to depend both on the material and on the XC functional that has been used for the computation of the $\{\phi_\alpha\}$.
From a theoretical perspective, the XC kernel is readily defined in terms of a double functional derivative of the XC free energy~\cite{IIT}; this is difficult to evaluate in practice and, to our knowledge, only possible for simple XC functionals in the case of bulk systems.
Consequently, accurate and dynamic XC kernels exist only for model systems such as the UEG~\cite{RG84,Pitarke_PRB_2007,Panholzer_PRL_2018,Ruzsinszky_PRB_2020,dornheim_dynamic}.

An alternative approach to the computation of the XC kernel from the second order functional derivatives is  based on the perturbation of the 
system by an external harmonic field to measure the density response and from that to extract the  XC kernel by inverting  Eq.~(\ref{eq:kernel}).
This method was used by Moroni \textit{et al.}~\cite{moroni} to calculate the static density response function and the LFC of the UEG in the ground state from quantum Monte Carlo simulations. Recently, it was used within KS-DFT by Moldabekov \textit{et al.}~to compute the static XC kernel of the UEG and warm dense hydrogen on the basis of various XC functionals across Jacob's Ladder~\cite{https://doi.org/10.48550/arxiv.2209.00928, hybrid_results}. Some illustrative results from these studies are presented in Sec.~\ref{sec:hydrogen_results}.

We note that setting $K_\textnormal{xc}(\mathbf{q},\omega)\equiv0$ in Eq.~(\ref{eq:kernel}) is often denoted as RPA in the DFT literature, but does not correspond to a mean-field description as the KS-orbitals automatically contain a certain amount of information about electronic XC effects due to the employed XC functional.
Finally, having the LR-TDDFT result for $\chi(\mathbf{q},\omega)$ gives one direct access to a number of material properties, such as the dynamic structure factor $S(\mathbf{q},\omega)$, see Eq.~(\ref{eq:FDT}) below.

\subsubsection{\label{sec:RTTDDFT}Real-time time-dependent DFT (RT-TDDFT)}

The real-time approach to TDDFT (RT-TDDFT)~\cite{RG84,PhysRevB.54.4484,ullrich2011time} is a computationally efficient method for studying non-equilibrium electronic dynamics in first principles models of materials with an explicit treatment of the electron-ion interaction (usually within the Born-Oppenheimer approximation or sometimes using Ehrenfest molecular dynamics) and large basis sets.
The equations of motion in RT-TDDFT are the time-dependent Kohn-Sham equations,
\begin{equation}
 i \frac{\partial}{\partial t} \phi_{\alpha,{\bf k}}({\bf r},t) = \left\{-\frac{\nabla^2}{2} + v_{S}\left[n\right]({\bf r},t)\right\} \phi_{\alpha,{\bf k}}({\bf r},t) ~\label{eq:tdks_eqn},
\end{equation}
where we have $v_{S}\left[n\right]({\bf r},t) = v_\textnormal{ext}({\bf r},t) + v_H\left[n\right]({\bf r},t) + v_{xc}\left[n\right]({\bf r},t)$. In what follows, we shall abbreviate the effective Hamiltonian as $\hat{H}_{S}(t) = -\nabla^2/2 + v_S(\mathbf{r,t})$.
This is typically posed as an initial value problem in which $\phi_{\alpha,{\bf k}}({\bf r},t)$ are known for some $t=t_i$, typically from the solution to a static Mermin-Kohn-Sham DFT problem.
The time-dependent density is 
\begin{equation}
 n({\bf r},t) = \sum \limits_{\alpha,{\bf k}} f_{\alpha,{\bf k}}(\beta, \mu) |\phi_{\alpha, {\bf k}}({\bf r},t)|^2,
\end{equation}
and these occupation factors are typically consistent with the initial condition and held constant over the course of the dynamics.
Among the three terms in $v_{S}$, $v_\textnormal{ext}$ describes the attractive electron-ion interaction as well as impressed perturbations that drive the dynamics, $v_H$ is the time-dependent Hartree potential, and $v_{xc}$ is the time-dependent XC potential that describes electronic exchange and correlation.
The creation of approximations to the exact time-dependent Kohn-Sham potential that is both accurate and computationally efficient is one of the central problems in RT-TDDFT. 
The practical trade-off between the cost and accuracy of available and prospective approximations is the central factor in determining whether RT-TDDFT is the right method for a particular calculation.

While RT-TDDFT is capable of approximating these dynamics owing to arbitrary excitations [i.e., forms of $v_\textnormal{ext}({\bf r},t)$] it can also capture linear response functions in the limit that $v_\textnormal{ext}({\bf r},t)$ includes a contribution of the form $\delta v_0 f(t)$, where $\delta v_0$ is a constant that determines the strength of the perturbation and $f(t)$ is an arbitrary smooth function whose Fourier transform has support on the energies over which the linear response function is desired. As in LR-TDDFT, the response function computed using RT-TDDFT captures collective effects~\cite{baczewski2021predictions, Kononov2022} that are absent in approaches in which the Kubo-Greenwood (KG) formula is directly applied to the Kohn-Sham orbitals~\cite{kubo1957statistical,greenwood1958boltzmann,desjarlais2002electrical}.
Furthermore, the computational details of these formulations differ in such a way that RT-TDDFT can be made asymptotically less expensive than LR-TDDFT. 
To understand when and how RT-TDDFT is most efficient, we first need to describe typical approaches to the numerical solution of Eq.~(\ref{eq:tdks_eqn}).

The time integration of Eq.~(\ref{eq:tdks_eqn}) requires the implementation of a numerical approximation to the exact time-ordered unitary propagator that translates any KS orbital in time,
\begin{subequations}
\begin{align}
 &\mathcal{U}(t_f,t_i) = \mathcal{T}\exp\left[-i \int\limits_{t_i}^{t_f} d\tau \hat{H}_{S}(\tau) \right]~\label{eq:exact_propagator},\\
 &\text{where}~\phi_{\alpha,{\bf k}}({\bf r},t_f) = \mathcal{U}(t_f,t_i) \phi_{\alpha,{\bf k}}({\bf r},t_i),
\end{align}
\end{subequations}
and $\mathcal{T}$ is the time-ordering operator.
One approach to approximating this is first-order Trotterization of Eq.~(\ref{eq:exact_propagator}) into $N$ time steps of width $\Delta t=(t_f-t_i)/N$ over which the effective Hamiltonian is approximated as a constant at the midpoint of each step, eliminating the need for time ordering,
\begin{subequations}
\begin{align}
 &\mathcal{U}(t_f,t_i) \approx \prod \limits_{k=0}^{N-1} \mathcal{U}_{ap}[t_i+(k+1)\Delta t,t_i+k\Delta t],\\
 &\text{where}~\mathcal{U}_{ap}(t+\Delta t,t) = \exp\left[-i \Delta t \hat{H}_{S}\left(t+\frac{\Delta t}{2}\right)\right].
\end{align}
\end{subequations}
While conceptually straightforward, the practical implementation of this approach requires the numerical approximation of a matrix exponential and thus the diagonalization of $\hat{H}_{S}$ at each time step or some other more efficient method for numerically effecting this operation.

\begin{figure*}\centering
\includegraphics[width=0.45\textwidth]{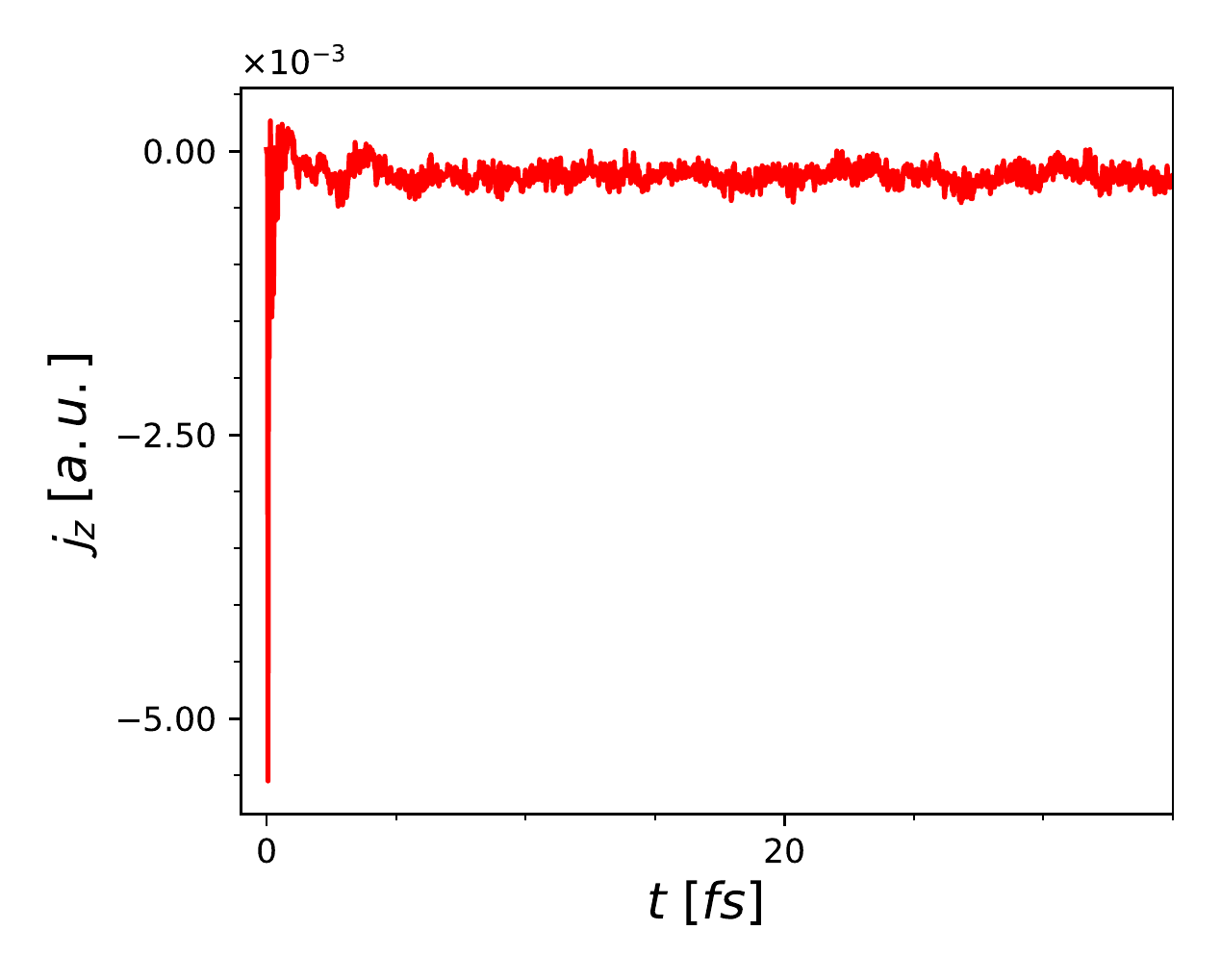}
\includegraphics[width=0.45\textwidth]{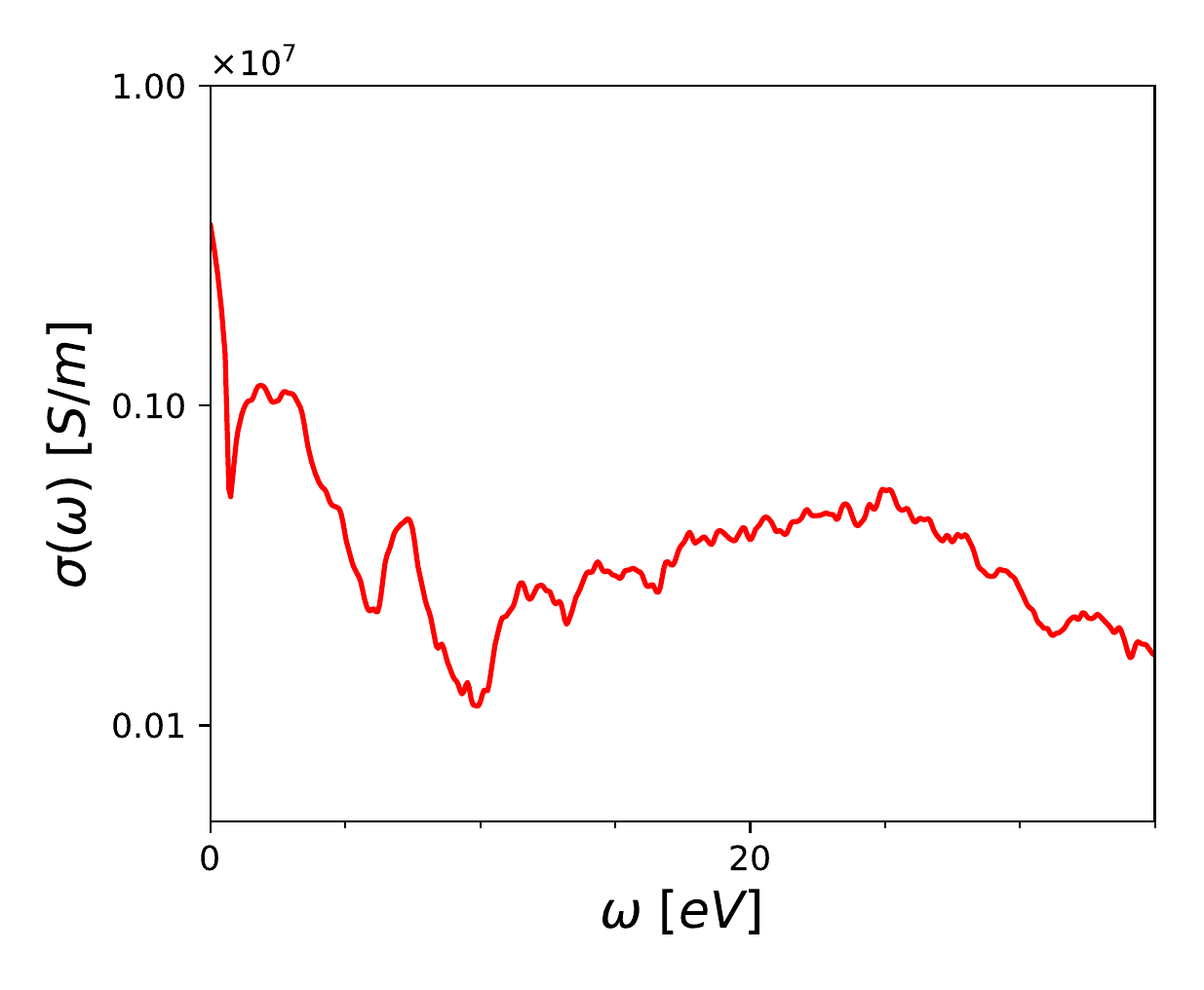}
\caption{
a) Induced current density in the $\vec{z}$ direction vs. time; b) frequency-dependent conductivity component $\boldsymbol{\sigma}_{zz}$.
}
\label{fig:cond_illustrate} 
\end{figure*}

Naive approaches to these operations have a cost that is cubic in the number of KS orbitals, which grows rapidly with temperature in and beyond the warm dense regime.
The parallel scalability of these approaches is also limited by a conflict between the optimal data distribution for effecting the product of $\hat{H}_{S}$ and a trial orbital, and orthogonalization of those orbitals, i.e., the optimal distribution for one is suboptimal for the other.
It turns out that it is possible to avoid both of these costs in RT-TDDFT and thus it can be made to be more efficient and scalable~\cite{enkovaara2010electronic,andrade2012time,baczewski2014numerical,Kononov2022} than implementations that rely directly on dense linear algebra, including many implementations of LR-TDDFT.
The precise conditions under which an RT-TDDFT linear response calculation is more efficient than an LR-TDDFT calculation is sensitive to many details of the calculation, but a large number of thermally occupied orbitals common in WDM calculations tend to favor RT-TDDFT as temperature increases.

While a scalar perturbation of the form $v_\textnormal{ext}({\bf r},t) = v_0\exp(i{\bf q}\cdot {\bf r})f(t)$ can be used to compute the DSF at non-zero ${\bf q}$~\cite{sakkos_JCP_2009,dynamic2}, calculation of the conductivity in the ${\bf q} \rightarrow 0$ limit requires the extension of RT-TDDFT to vector perturbations. 
More specifically, these vector perturbations are used to simulate a spatially uniform electric field, $\boldsymbol{E}(t) = -(1/c) (\partial \boldsymbol{A}/\partial t)$, impressed over the entire supercell~\cite{bertsch2000real,andrade2018negative}.
While the strength of this perturbation need not be weak in RT-TDDFT, as with the DSF one can implement an LR-TDDFT calculation with a sufficiently weak ${\bf E}(t)$ to compute the conductivity $\boldsymbol{\sigma}(\omega)$.
Then, rather than the time-dependent electronic charge density, the spatially uniform component of the time-dependent current density is related to the response function of interest.
Calculations of the conductivity are realized using a microscopic version of Ohm's law~\cite{ramakrishna2022electrical,ramakrishna2022electrical_dft_u},
\begin{equation}
\label{eq:ohms.law}
J_i(\omega)=\sigma_{ij}(\omega)\,{E}_j(\omega).
\end{equation}
Here the current density ($J$) and electric field ($E$) are vectors, while the conductivity ($\sigma$) is a tensor. The dielectric tensor is related to the conductivity tensor via
\begin{equation}
\epsilon_{ij}(\omega) = \delta_{ij} + i \frac{4\pi\sigma_{ij}(\omega) }{\omega}.     
\label{eps_sigma_relation}
\end{equation}

Fig.~\ref{fig:cond_illustrate} illustrates the methodology starting with a weak sigmoidal pulse applied in the $\vec{z}$ direction. The induced current density along $\vec{z}$  is shown in the left panel, and the frequency-dependent conductivity obtained using the Fourier transform of the induced current density is shown in the right panel. Note that the values for $j_{z}(t)$ in Fig. \ref{fig:cond_illustrate} are in atomic units.

\subsubsection{Non-equilibrium Green's functions}\label{sec:NEGF}

Non-equilibrium Green's functions (NEGF) theory is a powerful approach to treat electronic correlations, quantum and spin effects, as well as dynamical screening in non-ideal quantum systems, e.g.~Refs.~\cite{kremp2005quantum,bonitz_book}. There exist many applications to the warm dense uniform electron gas and to dense fully ionized plasmas \cite{kremp2005quantum,transfer1,Vorberger_PRE_2018}. Being computationally very demanding, many studies of real materials have been restricted to ground state properties, e.g.~Refs.~\cite{onida_rmp_02,romaniello_bprb_12}, providing important benchmarks for DFT, where QMC results are not available. On the other hand, an accurate analysis of the behavior of electrons out of equilibrium, including strong excitation and the short-time dynamics, is possible, but has been performed mostly for 
model systems, such as the UEG or lattice systems~\cite{hermanns_prb14,schluenzen_prb17}. Thus, NEGF are complementary to both PIMC and DFT.

The NEGF theory is formulated in second quantization (Fock space), starting from a complete set of single-particle orbitals, $\{\phi_i\}$ and the associated creation and annihilation operators, $\hat c^\dagger_i$ and $\hat c_j$. The central quantity is the single-particle Green's function, that is an ensemble average of the Heisenberg form of the field operators
\begin{align}
    g_{ij}(t,t') = -\frac{i}{\hbar}\left\langle \mathcal{T}_C\hat c_i(t)\hat c^\dagger_j(t')\right\rangle\,,
    \label{eq:1pnegf-def}
\end{align}
where $\mathcal{T}_C$ is the time ordering operator that is analogous to the one in  Eq.~\eqref{eq:exact_propagator}. Here the subscript $C$ indicates the use of the Keldysh time contour $\mathcal{C}$~\cite{keldysh_contour} that allows the extension of the power of Feynman diagrams to arbitrary nonequilibrium situations (for details, see the text books~\cite{stefanucci2013nonequilibrium,balzer2013book}). The angular brackets denote the expectation value that is computed either with the $N$-particle wave function (pure state) or N-particle density operator (mixed state) of the non-excited system.

The Green's function (\ref{eq:1pnegf-def}) gives access to all one-particle observables, such as the density matrix, which follows from the time-diagonal Green's function, $n_{ji}(t)= \frac{\hbar}{i}g_{ij}(t,t+\epsilon)$, where $\epsilon$ is an infinitesimal positive constant assuring the proper ordering of the two field operators.
A particular advantage of NEGF theory arises from the two-time structure of the function $g$: its values for different time arguments give direct access to the spectral function, the density of states, and the interaction energy of the system.

A special case is the coordinate representation which will be used in Sec.~\ref{sec:nonlinear}. There, a special notation for the field operators is used, $c_i(t) \to \psi(1)$, where $1=\{\textbf{r},t_1\}$ (the spin index is suppressed). Then the single-particle space and time dependent density is obtained from the space and time diagonal elements of the Green's function (\ref{eq:1pnegf-def}),
\begin{align}
    n(1) = \frac{\hbar}{i}g(1,1+\epsilon)\,.
    \label{eq:n-def}
\end{align}
The equations of motion for the NEGF are the Keldysh-Kadanoff-Baym equation (KBE, to be supplemented by the adjoint equation),
\begin{align}\nonumber
 & \sum_k\left[i\hbar\partial_t\delta_{ik}
 -h^{\rm HF}_{ik\sigma}(t)\right]g_{kj\sigma}(t,t')
 =\delta_{\cal C}(t-t')\delta_{ij} +\\
 &\qquad \qquad \qquad \sum_{k}\int_{\cal C} ds\,\Sigma^{\rm cor}_{ik\sigma}(t,s)g_{kj\sigma}(s,t')\,,
\label{eq:kbe}
\end{align}
where $h^{\rm HF}$ contains kinetic, potential and mean-field (Hartree-Fock) energy contributions, whereas correlation effects are included in the self-energy $\Sigma^{\rm cor}[g]$ which is a functional of the one-particle NEGF.

The KBE are analogous to the time-dependent Kohn-Sham equations of TD-DFT, Eq.~(\ref{eq:tdks_eqn}). The main differences are: 1) the KBE are equations for a density matrix that is similar to a product of two KS orbitals. 2) the energy terms are grouped slightly differently than in TDDFT --- the Hartree mean field and exact exchange are contained in $h^{\rm HF}$. 3) the exchange-correlation potential $v_{xc}[n]$ is replaced by the correlation self-energy $\Sigma^{\rm cor}$ which depends on $g$ rather than on the density and, therefore, depends on two times. Thus, it includes memory effects. 4) Similar to $v_{xc}$ in the case of DFT, $\Sigma^{\rm cor}$ is the only approximation of NEGF theory. Many systematic approaches exist for the self energy, most importantly Feynman diagram methods that include perturbation theory
and partial resummations that allow the systematic inclusion of dynamical screening effects and bound states, e.g.~Refs.~\cite{Schluenzen_2020,Joost_PRB_2022}. A comparison of the known approximations for $v_{xc}$ and $\Sigma^{\rm cor}$ is a topic of active research as it allows for benchmarks between the two approaches and offers the derivation of novel approximations for both, DFT and NEGF.

For the electronic density response of warm dense matter, NEGF theory offers a variety of approaches. First, is the derivation of equilibrium density response functions which are obtained by functional derivation of $g$, as will be demonstrated in Sec.~\ref{sec:nonlinear}. This method allows one to systematically derive not only the linear density response but also nonlinear generalizations. Second, there exists a straightforward real-time approach to the linear and nonlinear density response that was developed by Kwong and Bonitz \cite{kwong_prl-00}. To this end, an external potential is added to the general hamiltonian (\ref{eq:Hamiltonian_general}) that is of the form $V_{\rm ext}(r,t)= U_0(t)\cos{(\textbf{q  r)}}$. This potential imposes a density modulation to the system, and the short temporal duration of the pulse (``kick'') translates into a broad range of excitation energies.
\begin{figure}\centering
\includegraphics[width=0.55\textwidth]{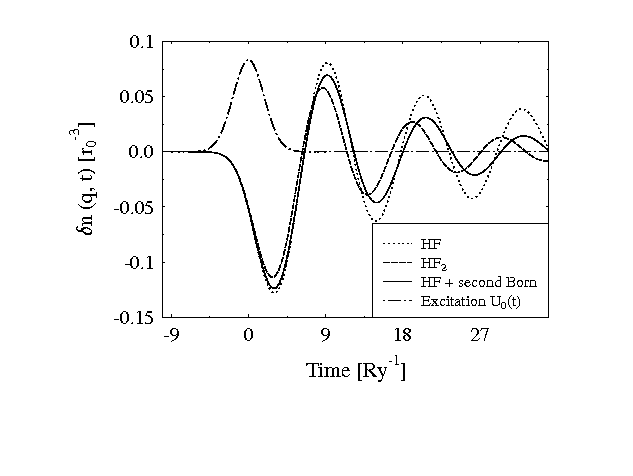}
\includegraphics[]{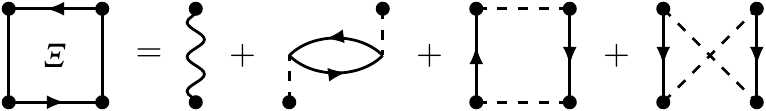}
\caption{Real-time NEGF simulation of the density response of the warm dense UEG following an excitation with a potential $V_{\rm ext}(r,t)= U_0(t)\cos{(\textbf{q  r)}}$, cf. dash-dotted line. Even after the excitation has vanished the electron density exhibits plasma oscillations where frequency and damping depend on the self-energy $\Sigma^{\rm cor}$ used in the solution of the KBE (\ref{eq:kbe}). HF (HF2): no correlation self-energy starting from an noninteracting (interacting) initial state --- this corresponds to the RPA result of equilibrium theory; HF+second Born:  $\Sigma^{\rm cor}$ within the second Born  approximation (2BA) showing correlation induced red shift and collisional damping in addition to Landau damping. Diagrams at the bottom show the relation to equilibrium approach: The use of the 2BA for $\Sigma^{\rm cor}$ is equivalent to solving the Bethe-Salpetr equation with a two-particle kernel $\Xi$ containing the indicated diagrams.
Fig. modified from Ref.~\cite{bonitz_book}.}
\label{fig:density_fluc_kbe} 
\end{figure} 
The solution of the KBE (\ref{eq:kbe}) for a UEG excited with this short pulse excites a monochromatic density  perturbation that decays with the plasmon frequency for that $q$ as shown in Fig.~\ref{fig:density_fluc_kbe}. The density fluctuation $\delta n(q,t)$ follows directly from the perturbation of the NEGF, via Eq.~(\ref{eq:n-def}),
\begin{align}
 \sum_p \delta g(p,t,t) = i \delta n(q,t) = \int_{-\infty}^{\infty} d{\bar t} \, \chi(q,t,{\bar t})U_0({\bar t}),
\label{eq:negf-density-response}
\end{align}
and is, to the lowest order (linear response), proportional to the excitation. In equilibrium, the density response function is stationary, $\chi(q,t,t')=\chi(q,t-t')$, and Eq.~(\ref{eq:negf-density-response}) can be solved in Fourier space for $\chi$: $\chi(q,\omega)=\delta n(q,\omega)/U_0(\omega)$ which gives access to all linear response functions, including the dielectric function and the dynamic structure factor, as discussed in Sec.~\ref{sec:LRT}.

A remarkable feature of real-time NEGF computations of the density response is that already fairly simple approximations for the self-energy 
$\Sigma^{\rm cor}$ yield high level results for the density response functions. The relation to the first (equilibrium) approach which requires to solve the Bethe-Salpeter equation (BSE) for the two-particle Green's function with a kernel $\Xi$ \cite{kremp2005quantum} is given  
by \cite{kwong_prl-00,bonitz_99_cpp2}
\begin{align}
\Xi(1,3;4,2) = \pm \delta\Sigma(1;3)/\delta g(4;2)\,.
\label{eq:xi0}
 \end{align}
The diagrams contributing to $\Xi$, if the KBE are solved on the level of second Born self-energies, are shown in the bottom of Fig.~\ref{fig:density_fluc_kbe}. Thereby, use of conserving self-energies automatically yields sum rule preserving approximations for $\Xi$. These attractive properties of NEGF  should also be fulfilled for other time-dependent approaches including real-time TDDFT offering interesting opportunities for the treatment of correlation effects in WDM.

Finally, we note that the real-time NEGF approach can be straightforwardly extended to the nonlinear response of the UEG as demonstrated in Ref.~\cite{kwong_prl-00}.

\subsection{Linear response theory\label{sec:LRT}}

At this point, let us set the perturbation amplitude in Eq.~(\ref{eq:Hamiltonian_general}) to be a constant (except for an infinitesimal damping that assures the vanishing of the perturbation for $t\to -\infty$), which leads to the harmonically perturbed Hamiltonian~\cite{moroni,moroni2,Dornheim_PRL_2020}
\begin{eqnarray}\label{eq:Hamiltonian_modified}
\hat H \to  \hat{H}_{\mathbf{q},\omega,A} = \hat{H}_0 + 2 A \sum_{l=1}^N \textnormal{cos}\left( \mathbf{q}\cdot\hat{\mathbf{r}}_l - \omega t \right)\ .
\end{eqnarray}
In the limit of small $A$, linear response theory (LRT)~\cite{quantum_theory} becomes valid, and the induced density is fully described by the simple linear relation
\begin{eqnarray}\label{eq:delta_n_omega}
 \delta n(\mathbf{q},\omega,A) = A  \chi(\mathbf{q},\omega) \phi_{\mathbf{q},\omega} \ ,
\end{eqnarray}
where $\phi_{\mathbf{q},\omega}$ is the Fourier transform of the cosine perturbation given by a sum of delta functions at positive and negative values of the wave vector $\mathbf{q}$ and the frequency $\omega$.
In other words, the system only exhibits a non-zero response at the wave vector and frequency of the original perturbation within LRT. Other effects such as the excitation of higher harmonics~\cite{Dornheim_PRR_2021} or mode-coupling between multiple perturbations~\cite{Dornheim_CPP_2022} are exclusively nonlinear effects, and are discussed in Sec.~\ref{sec:nonlinear} below.

For now, we shall postpone the discussion of the theoretical estimation of the linear density response function $\chi(\mathbf{q},\omega)$ and instead focus on its utility for the description of WDM and beyond. In this regard, a key relation is given by the
fluctuation-dissipation theorem (FDT)~\cite{quantum_theory}
\begin{eqnarray}\label{eq:FDT}
S(\mathbf{q},\omega) = - \frac{\textnormal{Im}\chi(\mathbf{q},\omega)}{\pi n (1-e^{-\beta\omega})}\ ,
\end{eqnarray}
which gives a direct and unique relation between the density response and the dynamic structure factor $S(\mathbf{q},\omega)$. In fact, $S(\mathbf{q},\omega)$ fully determines $\chi(\mathbf{q},\omega)$, as the real and imaginary parts of the latter are connected by the well-known Kramers-Kronig relations~\cite{quantum_theory}.
The DSF itself can be conveniently defined as the Fourier transform of the intermediate scattering function~\cite{siegfried_review}
\begin{eqnarray}\label{eq:DSF}
S(\mathbf{q},\omega) = \int_{-\infty}^\infty \textnormal{d}t\ e^{i\omega t} F(\mathbf{q},t)\ ,
\end{eqnarray}
which is defined as
\begin{eqnarray}\label{eq:ISF2}
F(\mathbf{q},t) = \braket{\hat n(\mathbf{q},t)\hat n(-\mathbf{q},0)}\ .
\end{eqnarray}

From a practical perspective, the DSF constitutes the central quantity in XRTS experiments with WDM~\cite{siegfried_review,kraus_xrts,sheffield2010plasma}; the measured scattering intensity $I(\mathbf{q},\omega)$ is given by the convolution of the DSF with the combined source and instrument function $R(\omega)$, see Eq.~(\ref{eq:convolution}),
and the wave vector $\mathbf{q}$ is determined by the scattering angle. Therefore, the availability of an accurate model for either $\chi(\mathbf{q},\omega)$ [subsequently evaluating Eq.~(\ref{eq:FDT})] or directly for $S(\mathbf{q},\omega)$ allows one to compare theoretical calculations with experimental measurements. Here LR-TDDFT (Sec.~\ref{sec:DFT}) is an example for the first route, whereas approximate Chihara models~\cite{Gregori_PRE_2003,Chihara_1987,siegfried_review,kraus_xrts} that decompose the full DSF into contributions from bound and free electrons and their respective transitions directly work with $S(\mathbf{q},\omega)$. In practice, these models can easily be convolved with $R(\omega)$, and are used to infer a-priori unknown system parameters such as the temperature. Evidently, the quality of a thus inferred EOS significantly depends on the level of accuracy of the employed approximation. Very recently, Dornheim \emph{et al.}~\cite{Dornheim_T_2022,Dornheim_insight_2022} have proposed to avoid this limitation by switching to the imaginary-time domain, which is discussed in Sec.~\ref{sec:ITCF} below.

A second highly important application of LRT stems from the combination of the FDT [Eq.~(\ref{eq:FDT})] with the adiabatic connection formula, which we describe in the following. As an initial step, we note that the DSF gives straightforward access to the static structure factor (SSF) $S(\mathbf{q})$ via
\begin{eqnarray}\label{SSFfreqint}
S(\mathbf{q}) = \int_{-\infty}^\infty \textnormal{d}\omega\ S(\mathbf{q},\omega)\ .
\end{eqnarray}
For completeness, we point out that the SSF contains the same information as the pair correlation function $g(\mathbf{r})$, and they are related by~\cite{hansen2013theory}
\begin{eqnarray}\label{eq:PCF}
g(\mathbf{r}) = 1 + \frac{1}{n}\int \frac{\textnormal{d}\mathbf{q}}{(2\pi)^3} \left(S(\mathbf{q})-1 \right) e^{i\mathbf{q}\cdot\mathbf{r}}\ .
\end{eqnarray}
As the next step, we can use either $S(\mathbf{q})$ or $g(\mathbf{r})$ to estimate the interaction energy $W$, for example~\cite{hansen2013theory}
\begin{eqnarray}\label{eq:interaction_energy}
W =\frac{1}{2}\int_\mathcal{V} \textnormal{d}\mathbf{r}_1\int_\mathcal{V}\textnormal{d}\mathbf{r}_2\ \frac{n^{(2)}(\mathbf{r}_1,\mathbf{r}_2)}{|\mathbf{r}_1 - \mathbf{r}_2|}\ .
\end{eqnarray}
where $n^{(2)}(\mathbf{r}_1,\mathbf{r}_2)=n(\mathbf{r}_1)n(\mathbf{r}_2)g(\mathbf{r}_1-\mathbf{r}_2)$ denotes the two-particle density. Note that, for the classical and quantum OCP, owing to the rigid charge neutralizing background, the substitution $n^{2}(\mathbf{r}_1,\mathbf{r}_2)\to{n}^{2}(\mathbf{r}_1,\mathbf{r}_2)-n(\mathbf{r}_1)n(\mathbf{r}_2)$ is implied, which is necessary for convergence~\cite{Baus_Hansen_OCP}. Finally, the adiabatic-connection formula relates Eq.~(\ref{eq:interaction_energy}) with the XC contribution $f_\textnormal{xc}$ to the free energy~\cite{quantum_theory}
\begin{eqnarray}\label{eq:adiabatic_connection}
f_\textnormal{xc} = \int_0^1 \textnormal{d}\lambda\ \braket{\hat{W}}_{\lambda}\lambda\ .
\end{eqnarray}
In other words, integrating over the interaction energy of a system where the interaction contribution to the Hamiltonian has been re-scaled by $\lambda\in[0,1]$, $\hat{H}_\lambda=\hat{H}_0+\lambda\hat{W}$, gives access to the free energy. 
From the $\lambda-$parametric form, it is rather evident that the adiabatic-connection formula constitutes the finite temperature version of the Hellmann-Feynmann theorem~\cite{pons_2020}, that is a very useful tool at zero temperature~\cite{quantum_theory}. In the context of the classical and quantum OCP, the coupling parameter forms of the adiabatic-connection formula are nearly exclusively employed~\cite{review,Baus_Hansen_OCP,plasma2};
\begin{subequations}
\begin{align}\label{eq:adiabatic_connectionOCP}
f_\textnormal{xc}(r_{\mathrm{s}},\theta)&=\frac{1}{r_{\mathrm{s}}^2}\int_0^{r_{\mathrm{s}}}r_{\mathrm{s}}^{\prime}W(r_{\mathrm{s}}^{\prime},\theta)dr_{\mathrm{s}}^{\prime},\,\mathrm{quantum\,\,OCP}\\
f_\textnormal{c}(\Gamma)&=\int_0^{\Gamma}\frac{W(\Gamma^{\prime})}{\Gamma^{\prime}}d\Gamma^{\prime},\,\mathrm{classical\,\,OCP}.
\end{align}
\end{subequations}
Consequently, knowledge of the response of a given system to all possible external perturbations, in the limit of a weak perturbation amplitude, gives access to all thermodynamic properties, and, thus, also to the free energy.

In practice, Eqs.~(\ref{eq:interaction_energy},\ref{eq:adiabatic_connection}) constitute the prime motivations for the field of the self-consistent dielectric formalism~\cite{stls_original,vs_original,IIT,schweng,farid,stolzmann,Dabrowski_PRB_1986,Holas_PRB_1987,arora,Tanaka_CPP_2017,tanaka_hnc,Tolias_JCP_2021,castello2021classical,stls,stls2}. 
Here the basic idea is to develop approximate expressions for the LFC $G(\mathbf{q},\omega)$, which give access to $\chi(\mathbf{q},\omega)$, see Eq.~(\ref{eq:LFC}) above, that can be employed via Eqs.~(\ref{eq:FDT},\ref{SSFfreqint},\ref{eq:interaction_energy},\ref{eq:adiabatic_connection}) to compute a host of dynamic, static and thermodynamic properties. It is noted that, in general, the LFC is a complicated functional of the DSF. In practice though, the LFC is a complicated functional of the SSF only, which still translates to a complex non-linear system of integral (in wavenumber and frequency) equations. Recent advances in this field have mainly focused on the accurate description of the strongly coupled electron liquid~\cite{dornheim_electron_liquid} with $r_s\gtrsim10$, with notable progress reported by Tanaka~\cite{tanaka_hnc} and Tolias \emph{et al.}~\cite{Tolias_JCP_2021}. These schemes are based on the combination of the non-perturbative integral equation theory of classical liquids with the quantum dielectric formalism. The latter scheme~\cite{Tolias_JCP_2021} simplifies to the former~\cite{tanaka_hnc} in well-defined limits and benefits from the availability of accurate OCP bridge functions extracted from MD simulations~\cite{YOCP_bridge_2021,OCP_bridge_2022}. Moreover, we stress that further development of dielectric theories remains an important goal also in the WDM regime, as the best available LFCs are frequency averaged, i.e., they neglect the dependence on $\omega$; $G(\mathbf{q},\omega)\equiv G(\mathbf{q})$. Having a reliable, and consistently frequency-dependent model for the LFC would allow a number of interesting investigations such as the determination of the effective mass within the concept of Fermi liquid theory~\cite{Eich_PRB_2017}, or 
the further analysis of a possible experimental detection of the \emph{roton feature}~\cite{dornheim_dynamic,Dornheim_Nature_2022} in the dilute electron gas.

Finally, we note that Eq.~(\ref{eq:adiabatic_connection}) has been used by a number of groups~\cite{IIT,stls2,ksdt,groth_prl,review} to construct representations of the XC free energy of the UEG based on different input data sets for the interaction energy $W$.

As an alternative route to obtain insights into the local field correction without any dielectric approximations, Moroni \emph{et al.}~\cite{moroni} have suggested to carry out QMC calculations based on the modified Hamiltonian of Eq.~(\ref{eq:Hamiltonian_modified}), but in the static limit of $\omega\to0$. 
In the ground state, the stiffness theorem~\cite{quantum_theory} then gives a straightforward relation between the induced change in the total energy and the static density response function $\chi(\mathbf{q})\equiv\chi(\mathbf{q},0)$. This approach has subsequently been used by different groups to estimate the static density response of a number of systems, including the UEG~\cite{moroni2,bowen2}, a charged Bose gas~\cite{Sugiyama_PRB_1994}, and neutron matter~\cite{PhysRevLett.116.152501}.

At finite temperature, it is more convenient to directly work with induced densities in reciprocal space~\cite{dornheim_pre,groth_jcp},
\begin{eqnarray}\label{eq:rho}
\braket{\hat\rho_\mathbf{k}}_{q,A} = \frac{1}{\mathcal{V}} \left< \sum_{l=1}^N e^{-i\mathbf{k}\cdot\hat{\mathbf{r}}_l} \right>_{q,A} \ , 
\end{eqnarray}
with $\braket{\dots}_{q,A}$ indicating that the expectation value is to be taken with respect to Eq.~(\ref{eq:Hamiltonian_modified}).
The sought-after static linear density response function $\chi(\mathbf{q})$ can then be obtained via the relation
\begin{eqnarray}\label{eq:LRT_rho}
\braket{\hat\rho_\mathbf{k}}_{q,A} = \delta_{\mathbf{k},\mathbf{q}} \chi(\mathbf{q}) A\ .
\end{eqnarray}
Furthermore, it is straightforward to invert Eq.~(\ref{eq:LFC}) to obtain the static local field correction $G(\mathbf{q})$. This has allowed Moroni \emph{et al.}~\cite{moroni2} to compute the first accurate results for $G(\mathbf{q})$ of the UEG in the zero-temperature limit, which have subsequently been used as input for the parametrization by Corradini \emph{et al.}~\cite{cdop}.

While being formally exact, the approach that is delineated by Eqs.~(\ref{eq:rho}) and (\ref{eq:LRT_rho}) requires, for any given combination of $(r_s,\theta)$, the execution of multiple independent QMC simulations for each individual $\mathbf{q}$ and for  different perturbation amplitudes $A$ (also to ensure non-linear effects are negligible). An elegant and computationally less demanding alternative is given by the PIMC estimation of the imaginary-time density-density correlation function $F(\mathbf{q},\tau)$ [Eq.~(\ref{eq:ISF})], which is discussed in the following section.

\subsection{Imaginary-time density-density correlation function\label{sec:ITCF}}

The imaginary-time density-density correlation function (ITCF) $F(\mathbf{q},\tau)$ is given by the intermediate scattering function defined in Eq.~(\ref{eq:ISF2}) above, but evaluated at the imaginary time $t=-i\hbar\tau$, with $0\leq\tau\leq\beta$; in the following, we will simply denote it as
\begin{eqnarray}\label{eq:ISF}
F(\mathbf{q},\tau) = \braket{\hat{n}(\mathbf{q},0)\hat{n}(-\mathbf{q},\tau)}\ .
\end{eqnarray}
From a theoretical perspective, the estimation of Eq.~(\ref{eq:ISF}) within a PIMC simulation is straightforward~\cite{Berne_JCP_1983,Dornheim_insight_2022}, and requires the correlated evaluation of two density operators in reciprocal space at two different imaginary-time slices. For completeness, note that QMC results for the ITCF have been reported for a gamut of systems beyond WDM, including ultracold helium~\cite{Boninsegni1996,Vitali_PRB_2010,Nava_PRB_2013,Ferre_PRB_2016,Dornheim_SciRep_2022}, exotic supersolids~\cite{Saccani_Supersolid_PRL_2012}, and the UEG in the zero-temperature limit~\cite{Motta_JCP_2015}.

A first important relation is given by the connection between $F(\mathbf{q},\tau)$ and the static density response function $\chi(\mathbf{q})$, which is known as the
imaginary-time version of the fluctuation-dissipation theorem~\cite{Dornheim_insight_2022,bowen2},
\begin{eqnarray}\label{eq:chi_static}
\chi(\mathbf{q}) = - n \int_0^\beta \textnormal{d}\tau\ F(\mathbf{q},\tau)\ .
\end{eqnarray}
In practice, Eq.~(\ref{eq:chi_static}) therefore allows one to estimate the full wave-vector dependence of the static linear density response function from a single simulation of the unperturbed system without worrying about non-linear effects. This approach was extensively used by Dornheim and co-workers over different parameter regimes~\cite{dynamic_folgepaper,dornheim_ML,dornheim_HEDP,dornheim_electron_liquid,Dornheim_HEDP_2022}.

A second important relation is given by the connection between $F(\mathbf{q},\tau)$ and the DSF, which simply comprises a two-sided Laplace transform,
\begin{eqnarray}\label{eq:Laplace}
F(\mathbf{q},\tau) &=& \mathcal{L}\left[S(\mathbf{q},\omega)\right] \\\nonumber &=& \int_{-\infty}^\infty \textnormal{d}\omega\ e^{-\omega\tau} S(\mathbf{q},\omega)\ .
\end{eqnarray}
Traditionally, Eq.~(\ref{eq:Laplace}) constitutes the starting point for a so-called \emph{analytic continuation}~\cite{JARRELL1996133}, i.e., a numerical inversion to obtain $S(\mathbf{q},\omega)$ from the QMC data for the ITCF. This is a well-known, though notoriously difficult problem. In fact, an inverse Laplace transform is ill-posed, and the inevitable statistical error bars in the QMC input data lead to different instabilities in practice~\cite{JARRELL1996133}. As a consequence, a host of different strategies to deal with the \emph{analytic continuation} have been suggested in the literature~\cite{Boninsegni1996,Mishchenko_PRB_2000,Vitali_PRB_2010,Sandvik_PRE_2016,Otsuki_PRE_2017,Goulko_PRB_2017,Boninsegni_maximum_entropy,PhysRevB.98.245101,dornheim_dynamic,dynamic_folgepaper,Fournier_PRL_2020,Nichols_PRE_2022}, but the quality of the thus reconstructed $S(\mathbf{q},\omega)$ generally remains unclear.
A notable exception is given by the warm dense UEG, for which a number of exact analytical constraints allow one to sufficiently reduce the space of possible $S(\mathbf{q},\omega)$ to render the numerical inversion of Eq.~(\ref{eq:Laplace}) tractable~\cite{dornheim_dynamic,dynamic_folgepaper,Dornheim_PRE_2020,Hamann_PRB_2020}. This has allowed Dornheim \emph{et al.}~\cite{dornheim_dynamic} to present the first accurate results for $S(\mathbf{q},\omega)$, which has given important insights into the nature of the XC induced red-shift in the dispersion of the UEG, cf.~Sec.~\ref{sec:dynamic_UEG} below.

On the other hand, it is well understood that the two-sided Laplace transform constitutes a unique mathematical transformation. Therefore, the ITCF contains by definition the same amount of information as the DSF, albeit in an at first unfamiliar representation~\cite{Dornheim_insight_2022,Dornheim_T_2022,Dornheim_PTR_2022}.
For example, consider the exact spectral representation of the DSF~\cite{quantum_theory}
\begin{eqnarray}\label{eq:spectral}
S(\mathbf{q},\omega) = \sum_{m,l} P_m \left\|{n}_{ml}(\mathbf{q}) \right\|^2 \delta(\omega - \omega_{lm})\ .
\end{eqnarray}
In other words, $S(\mathbf{q},\omega)$ is given by the sum over all possible transitions between the eigenstates $m$ and $l$ of the full $N$-body Hamiltonian induced by the density operator $\hat{n}(\mathbf{q})$, with $P_m$ being the probability to occupy the initial state $m$, $\left\|{n}_{ml}(\mathbf{q}) \right\|$ is the corresponding transition
matrix element, and $\omega_{lm}=(E_l-E_m)/\hbar$ denoting the energy difference.
Inserting Eq.~(\ref{eq:spectral}) into Eq.~(\ref{eq:Laplace}) then gives the analogous representation in the $\tau$-domain,
\begin{eqnarray}\label{eq:spectral_F}
F(\mathbf{q},\tau) &=& \sum_{m,l} P_m \left\|{n}_{ml}(\mathbf{q}) \right\|^2 e^{-\tau\omega_{lm}} \ .
\end{eqnarray}
Evidently, the $\tau$-decay of $F(\mathbf{q},\tau)$ for a given wave vector $\mathbf{q}$ is shaped by the characteristic frequencies in the corresponding $S(\mathbf{q},\omega)$. In particular, Eq.~(\ref{eq:spectral_F}) directly implies that energetically low-lying excitations such as the \emph{roton feature} in the dilute UEG~\cite{Dornheim_Nature_2022} will directly manifest as a less pronounced decay with $\tau$. Conversely, it holds that $\omega(q)\sim q^2$ in the non-collective single-particle regime with $q\gg q_\textnormal{F}$ (with $q_\textnormal{F}$ being the usual Fermi wave number~\cite{quantum_theory}), which will lead to a steeper decay of the ITCF.
Dornheim \emph{et al.}~\cite{Dornheim_insight_2022} have suggested characterizing this with a relative decay measure of the form
\begin{eqnarray}\label{eq:decay_measure}
\Delta F_\tau(\mathbf{q}) = \frac{F(\mathbf{q},0)-F(\mathbf{q},\tau)}{F(\mathbf{q},0)}\ ,
\end{eqnarray}
and corresponding results are shown in Fig.~\ref{fig:dispersion} below. 

In addition to being of great value for the interpretation of simulation results for $F(\mathbf{q},\tau)$, these considerations open up new avenues for the interpretation of XRTS experiments with WDM~\cite{Dornheim_T_2022}. In particular, the numerical deconvolution of the measured XRTS signal [Eq.~(\ref{eq:convolution}) above] is generally prevented by the inevitable experimental noise. Therefore, XRTS experiments do not give direct access to the DSF, which contains the relevant physical information about the given system of interest. 
Instead, one has to take the aforementioned detour over a model description of $S(\mathbf{q},\omega)$, which then has to be inserted into Eq.~(\ref{eq:convolution}) for a comparison to the experimental observation. Unfortunately, this forward-fitting approach introduces a bias to the interpretation of the experiment that depends on the employed approximation. To our knowledge, no exact simulation or theory is available for the description of $S(\mathbf{q},\omega)$ of real materials in the WDM regime.

In stark contrast, switching to the Laplace domain, i.e., inserting both the XRTS intensity $I(\mathbf{q},\omega)$ and the combined source and instrument function $R(\omega)$ into Eq.~(\ref{eq:Laplace}) makes the deconvolution trivial; this is due to the convolution theorem of $\mathcal{L}\left[\dots\right]$, which states that
\begin{eqnarray}\label{eq:convolution_theorem}
\mathcal{L}\left[S(\mathbf{q},\omega)\right]=\frac{\mathcal{L}\left[S(\mathbf{q},\omega) \circledast R(\omega)\right]}{\mathcal{L}\left[R(\omega)\right]}=\frac{\mathcal{L}\left[I(\mathbf{q},\omega)\right]}{\mathcal{L}\left[R(\omega)\right]}\,.
\end{eqnarray}
Indeed, the LHS of Eq.~(\ref{eq:convolution_theorem}) is the ITCF $F(\mathbf{q},\tau)$, which contains the same physical information as the DSF.
As we shall see, this allows for the straightforward interpretation of XRTS experiments without any models, simulations or approximations.

For a uniform system in thermodynamic equilibrium, the DSF fulfills the well-known \emph{detailed balance} between positive and negative frequencies~\cite{siegfried_review,quantum_theory,DOPPNER2009182},
\begin{eqnarray}\label{eq:detailed_balance}
S(\mathbf{q},-\omega) = S(\mathbf{q},\omega) e^{-\beta\omega}\ .
\end{eqnarray}
The detailed balance condition is a direct consequence of the exact spectral representation of the DSF, see Eq.~(\ref{eq:spectral}), and is central to the derivation of the FDT, see Eq.~(\ref{eq:FDT}). In essence, detailed balance is reflected by the FDT, given the odd frequency-parity of $\textnormal{Im}\chi(\mathbf{q},\omega)$. Inserting Eq.~(\ref{eq:detailed_balance}) into Eq.~(\ref{eq:Laplace}) gives the important
symmetry relation~\cite{Dornheim_T_2022,Dornheim_T_follow_up}
\begin{eqnarray}\label{eq:symmetry}
F(\mathbf{q},\tau) &=& \int_0^\infty \textnormal{d}\omega\ S(\mathbf{q},\omega)\left\{ e^{-\omega\tau} + e^{-\omega(\beta-\tau)} \right\}\\\nonumber
 &=& F(\mathbf{q},\beta-\tau)\ .
\end{eqnarray}
In particular, Eq.~(\ref{eq:symmetry}) implies that $F(\mathbf{q},\tau)$ is symmetric around $\tau=\beta/2$. In practice, one can thus directly diagnose the temperature of a given system from the Laplace transform of the XRTS signal; no forward-modelling or simulations are required.

As a final useful property of the ITCF, we mention its relation to the frequency-moments of the DSF, which are defined as~\cite{quantum_theory,kugler_bounds,kugler1}
\begin{eqnarray}\label{eq:moments}
M_\alpha^{S} = \braket{\omega^\alpha}_{S} = \int_{-\infty}^\infty \textnormal{d}\omega\ S(\mathbf{q},\omega)\ \omega^\alpha\ .
\end{eqnarray}
For example, the first moment (i.e., $\alpha=1$) is given by the exact f-sum rule~\cite{quantum_theory,pines_nozieresI_book}
\begin{eqnarray}\label{eq:f_sum_rule}
M_1^{S} = \braket{\omega^1}_{S} = \frac{\mathbf{q}^2}{2}\ .
\end{eqnarray}
The same information is encoded into the ITCF via its derivatives with respect to $\tau$ around $\tau=0$~\cite{Dornheim_insight_2022,Dornheim_moments_2022},
\begin{eqnarray}\label{eq:moments_derivative}
M_\alpha^{S}= \left( -1 \right)^\alpha \left. \frac{\partial^\alpha}{\partial\tau^\alpha} F(\mathbf{q},\tau) \right|_{\tau=0} \ .
\end{eqnarray}
In addition to being interesting in their own right, the frequency moments constitute exact constraints that assist in the analytic continuation that is necessary for the reconstruction of the DSF from the QMC ITCF data, see the inversion of the Laplace transform of Eq.~(\ref{eq:Laplace})~\cite{dornheim_dynamic,dynamic_folgepaper,Dornheim_PRE_2020,Hamann_PRB_2020}. They also
constitute the key input for the non-perturbative method of moments that is used by Tkachenko and co-workers to estimate $S(\mathbf{q},\omega)$ based on static structural properties~\cite{tkachenko_book,Vorberger_PRL_2012,Tkachenko_CPP_2018,Ara_POP_2021}.

\subsection{Nonlinear density response theory\label{sec:nonlinear}}

Even though the idea of LRT is prevalent and very useful in almost all areas of physics, an incomparably richer treasure trove of physics can be obtained by considering stronger deviations from equilibrium, i.e., the nonlinear response. Not only can the response of the investigated system to stronger external perturbations be described, but this response also gives direct access to higher order correlation functions~\cite{Dornheim_JPSJ_2021}.

The general definition of higher order response functions is given by an induced density $\delta n(\vec r,t)$ expansion
\begin{align}
\delta{n}(1)=\sum_{l=2}^{N}\int
\mathscr{W}^{(l-1)}(\{l\})
\prod_{i=2}^{l}V(i)d(i)\,,\label{eq:n_ind_re_compact}
\end{align}
where $1=\{\vec r_1,t_1\}$, $\{l\}=(1,2,...,l)$, $d1=\{d\vec r_1,dt_1\}$. After writing the first three terms explicitly and setting $\chi(1,2)\equiv\mathscr{W}^{(1)}(1,2)$, we have~\cite{PhysRevB.37.9268,Bergara1999}
\begin{align}
\delta n(1)\! &=\!\!\int\!\mathrm{d}2  \;\chi(1,2) V(2)+ \nonumber\\
&\quad\int\! \mathrm{d} 2\mathrm{d} 3\;
\mathscr{W}^{(2)}(1,2,3)
V(2)V(3)+ \nonumber \\
&\quad\int\! \mathrm{d}2 \mathrm{d}3 \mathrm{d}4\; \mathscr{W}^{(3)}(1,2,3,4)V(2)V(3)V(4)\cdots\label{eq:n_ind_re}
\end{align}
Here, all time integrations run from $-\infty$ to $t_1$, meaning that we need the retarded response functions. Still assuming a relatively weak perturbation, the series can be truncated after the second (quadratic) or third (cubic) term. Thus, the quadratic response function $\mathscr{W}^{(2)}$ and the cubic response function $\mathscr{W}^{(3)}$ are introduced. 

Mathematically, these response functions can be defined as functional derivatives of the one-particle Green's function $g(11^{\prime})=-i\langle T\{\psi^+(1^{\prime})\psi(1) \}\rangle$ with respect to the external perturbation. They are special cases of more general higher order correlation functions. For the linear response function, we have
\beq
\chi(1,2)=L(12,1^{\prime}2^{\prime})\left|_{\footnotesize
\begin{array}{l}
1^{\prime}=1^+\\
2^{\prime}=2
\end{array}
}\right.
=\pm i \frac{\delta g(11^{\prime})}{\delta V(2^{\prime}2)}
\left|_{\footnotesize
\begin{array}{l}
1^{\prime}=1^+\\
2^{\prime}=2
\end{array}
}\right.\,.
\label{eq:def_l}
\eeq
Here, $t_1^+=t_1+\varepsilon$. The function $L$ is in general a true four-point function describing, among other properties, the correlations of two density fluctuations and the scattering of two particles in a medium. The definition of the quadratic response function is
\beq
\mathscr{W}^{(2)}(1,2,3)=(\pm i)^2\frac{\delta^2 g(11^+)}{\delta V(33) \delta V(22)}\,.
\label{eq:def_y}
\eeq
Again, this is a special case of the three-particle correlator $Y(123,1^{\prime}2^{\prime}3^{\prime})$. Similar considerations are valid for the cubic response function. Eqs.~(\ref{eq:def_l}), (\ref{eq:def_y}) can be used to derive exact equations of motion for the linear and quadratic response functions and also to obtain RPA and other approximations for any of the higher order response functions.

It may sometimes be useful to introduce effective quantities like the polarization function and dielectric functions to describe in-matter effects. In the case of linear response, it holds
\beq
\chi(1,2)=\int d3\; \Pi(13)K(23)\,,
\eeq
where the polarization function $\Pi$ and the generalized linear dielectric function $K$ were introduced. In such a way, the effective response described by the polarization function is connected to the total response described by $\chi$. A similar relation for the quadratic response function reads
\bea
\mathscr{W}^{(2)}(1,2,3)&=&\int d4 d5\; \Pi(1,4,5)K(35)K(24)\nonumber\\
&&+\int d4\;\Pi(14)K(3,2,4)\,.
\eea
This defines the quadratic polarization $\Pi(123)$ and the quadratic dielectric function $K(123)$. Naturally, this can be extended to the cubic case.

For simplicity, we restrict ourselves to the quadratic response; the higher-order response has been investigated elsewhere~\cite{Dornheim_PRR_2021,Dornheim_JCP_ITCF_2021}. We also assume the system to be homogeneous and the external perturbation to be of harmonic nature. Note that the latter does not restrict the generality of the considerations. We shall only treat the static $\omega=0$ case herein. The induced density contains various combinations of higher harmonics of the perturbing potential. In wavenumber space, the result is, up to the second order,
\bea\label{eq:response}
\delta n(\boldsymbol{k})&=&A\chi(\mathbf{k})\Big\{\delta(\boldsymbol{k}-\qv)+\delta(\boldsymbol{k}+\qv)\Big\}+ \nonumber\\
&&A^2\Big\{ 
\mathscr{W}^{(2)}(\boldsymbol{k}-\qv,\qv)\left[\delta(\boldsymbol{k}-2\qv)+\delta(\boldsymbol{k)}\right]+\nonumber\\
&&\,\mathscr{W}^{(2)}(\boldsymbol{k}+\qv,-\qv)\left[\delta(\boldsymbol{k}+2\qv)+\delta(\boldsymbol{k})\right]
\Big\}
\,.
\eea
Here, $\chi(\mathbf{q})$ denotes the usual static ($\omega=0$) limit of the linear response function~(\ref{eq:def_l})~\cite{quantum_theory,nolting}, $\mathscr{W}^{(2)}(\mathbf{q}_1,\mathbf{q}_2)$ the static quadratic response function. A more in-depth derivation up to the third order is given in Ref.~\cite{Dornheim_PRR_2021}. 

In addition, we note that the quadratic density response function (see Ref.~\cite{Dornheim_JCP_ITCF_2021} for details) is connected to the imaginary-time structure of the system by the relation
\begin{widetext}
\begin{eqnarray}\label{eq:Y_imaginary}
\mathscr{W}^{(2)}(\qv_1,\qv_2)=\frac{1}{2\mathcal{V}}\int\limits_0^{\beta}d\tau_1\int\limits_0^{\beta}d\tau_2\,
\langle
\tilde \rho(\qv_1+\qv_2,0)\tilde \rho(-\qv_1,-\tau_1)\tilde \rho(-\qv_2,-\tau_2)
\rangle\,.
\end{eqnarray}
\end{widetext}
The density operator
\begin{eqnarray}\label{eq:n_tilde}
\tilde \rho(\mathbf{q},\tau) = \sum_{l=1}^N \textnormal{exp}\left(
-i\mathbf{q}\cdot\hat{\mathbf{r}}_{l,\tau}
\right)
\end{eqnarray}
is not normalized, and $\mathbf{r}_{l,\tau}$ denotes the position of particle $l$ at an imaginary time $\tau\in[0,\beta]$. Eq.~(\ref{eq:Y_imaginary}) thus directly implies that all quadratic terms of the nonlinear density response (including mode-coupling effects, see below) can be obtained from a single simulation of the unperturbed UEG.

One can determine all contributions of linear or non-linear origin to the induced density at any wavenumber or frequency from a higher order generalization of Eq.~(\ref{eq:response}). For instance, the induced density at the first harmonic, i.e. original perturbing wave vector, is given as a sum of linear and cubic contributions
\begin{eqnarray}\label{eq:rho1_fit}
\braket{\hat\rho_\mathbf{q}}_{q,A} &=& \chi(q) A + \chi^{(1,3)}(q) A^3+\ldots\,,
\end{eqnarray}
where we have introduced $\chi^{(1,3)}(q)=\mathscr{W}^{(3)}(-q,q,q)+\mathscr{W}^{(3)}(q,-q,q)+\mathscr{W}^{(3)}(q,q,-q)$.
The signal at the second harmonic is given as a sum of quadratic and quartic contributions
\begin{eqnarray}\label{eq:rho2_fit}
\braket{\hat{\rho}_{2\mathbf{q}}}_{q,A} = \chi^{(2,2)}(q)A^2 + \chi^{(2,4)}(q)A^4 + \dots\,.
\end{eqnarray}
The diagonal response function is again a special case of the general quadratic response function $\chi^{(2,2)}(q)=\mathscr{W}^{(2)}(q,q)$, and can therefore be estimated from PIMC results for the quadratic ITCF defined by
\begin{eqnarray}\label{eq:F2}
F^{(2)}(\mathbf{q},\tau_1,\tau_2) = \braket{\tilde \rho(2\mathbf{q},0)\tilde \rho(-\mathbf{q},\tau_1)\tilde \rho(-\mathbf{q},\tau_2)}\ ,
\end{eqnarray}
cf.~Eq.~(\ref{eq:Y_imaginary}).
At the third harmonic, the signal is given as a sum of cubic and quintic contributions
\begin{eqnarray}\label{eq:rho3_fit}
\braket{\hat{\rho}_{3\mathbf{q}}}_{q,A} = \chi^{(3,3)}(q)A^3 + \chi^{(3,5)}(q)A^5 + \dots\,.
\end{eqnarray}
The cubic response again determines the non-linear signal, but the cubic response at the third harmonic is a different realization of the general third order response function, $\chi^{(3,3)}(q)=\mathscr{W}^{(3)}(q,q,q)$, compared to the cubic response at the first harmonic.

The ideal quadratic response function can be derived analogously to the (finite-T) Lindhard formula known from LRT. Two terms are contributing. At the second harmonic, a recursion relation is known expressing the ideal quadratic response in terms of the ideal linear response evaluated at the first and second harmonics. This result is due to  Mikhailov~\cite{Mikhailov_Annalen,Mikhailov_PRL}
\begin{eqnarray}\label{eq:Mikhailov2}
\chi^{(2,2)}_0(q) = \frac{2}{q^2}\left[\chi^{(1,1)}_0(2q)-\chi^{(1,1)}_0(q)\right]\,.
\end{eqnarray}
The exact equation of motion for the quadratic response allows for an RPA-like approximation with two linear response functions in the denominator
\begin{equation}\label{eq:chi2_RPA}
    \chi^{(2,2)}_{\rm RPA}( q)= \frac{\chi^{(2,2)}_{0}( q)}{\left[1-v(q)\chi^{(1,1)}_{0}(q)\right]^{2} \left[1-v(2q)\chi^{(1,1)}_{0}( 2q)\right]}.
\end{equation}
There is a square of the Lindhard dielectric function at the first harmonic and a factor comprised of the Lindhard dielectric function at the second harmonic. 

The general structure of the equations of motion for the linear and quadratic responses is very similar. The idea of introducing LFCs into the RPA formula for the quadratic response is thus obvious
 \begin{eqnarray}\label{eq:quadratic_LFC}
     \chi^{(2,2)}_{\rm LFC}( q) &=&  \chi^{(2,2)}_{0}( q) \left[1-v(q)\left[1-G(q)\right]\chi^{(1,1)}_{0}(q)\right]^{-2}\nonumber\\  & & \times \left[1-v(2q)\left[1-G(2q)\right]\chi^{(1,1)}_{0}(2q)\right]^{-1}. \label{eq:chi2_LFC}
 \end{eqnarray}

\begin{figure*}\centering
\includegraphics[width=0.475\textwidth]{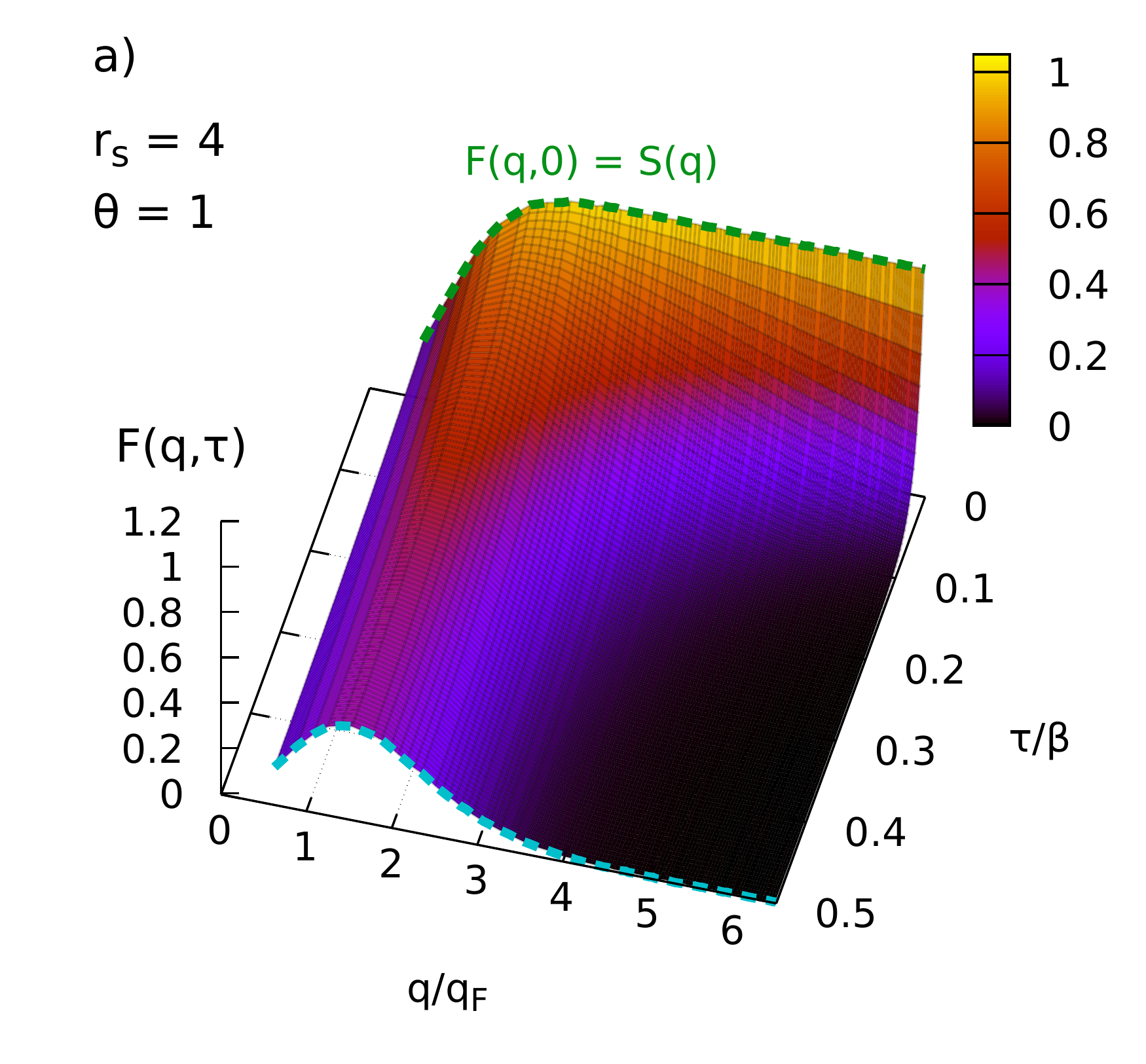}\includegraphics[width=0.45\textwidth]{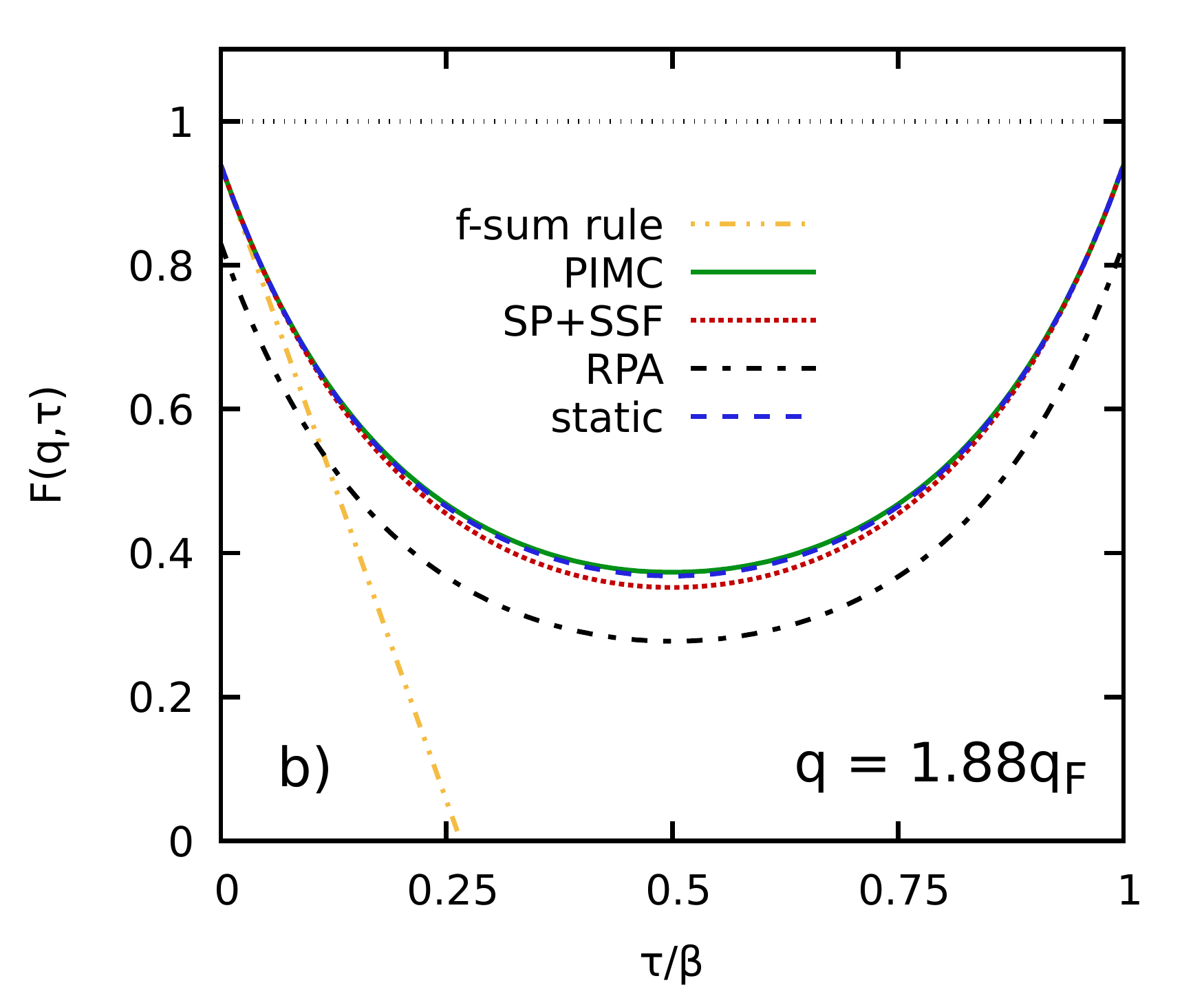}\\\vspace*{-0.02cm}\includegraphics[width=0.46\textwidth]{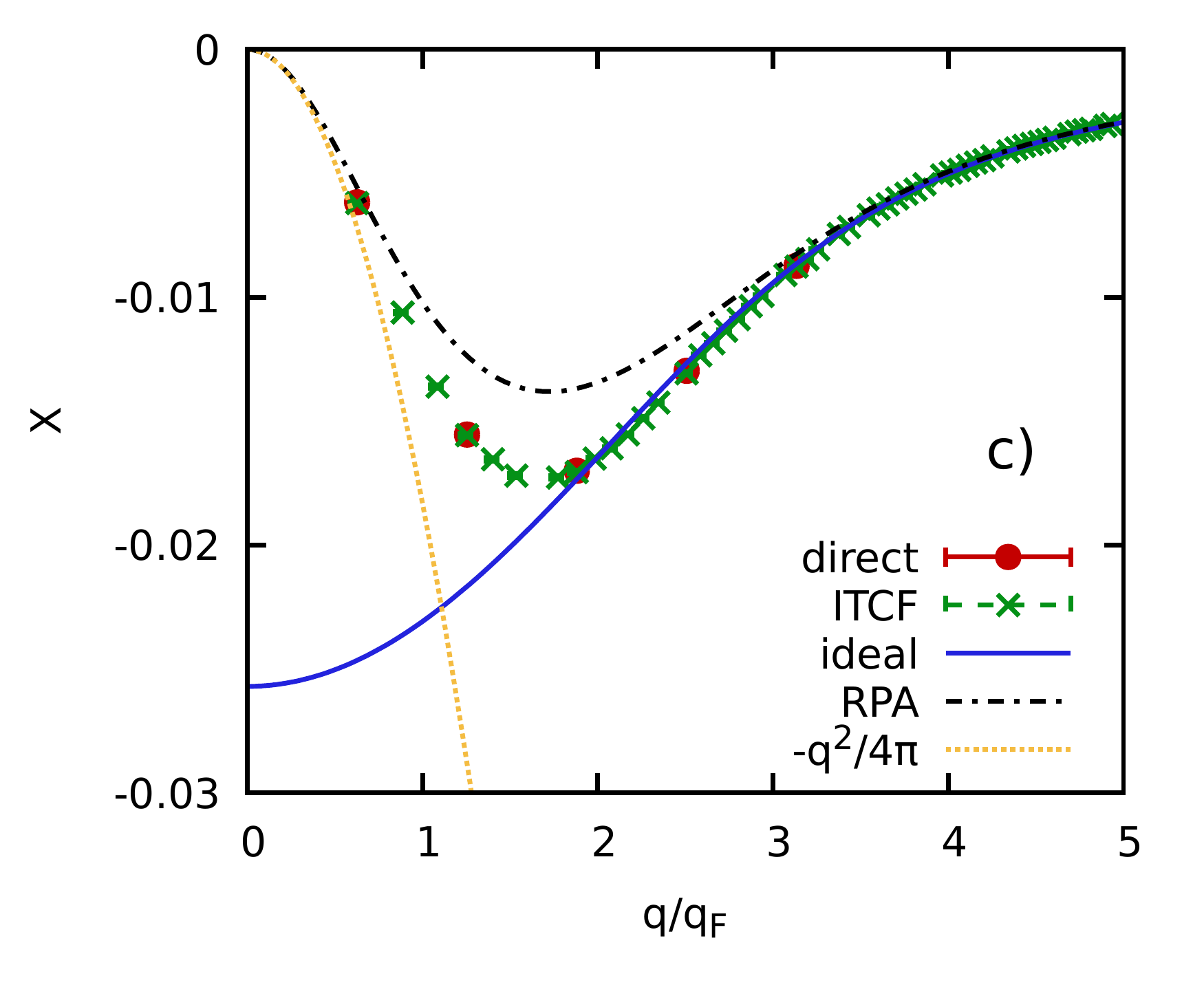}\includegraphics[width=0.46\textwidth]{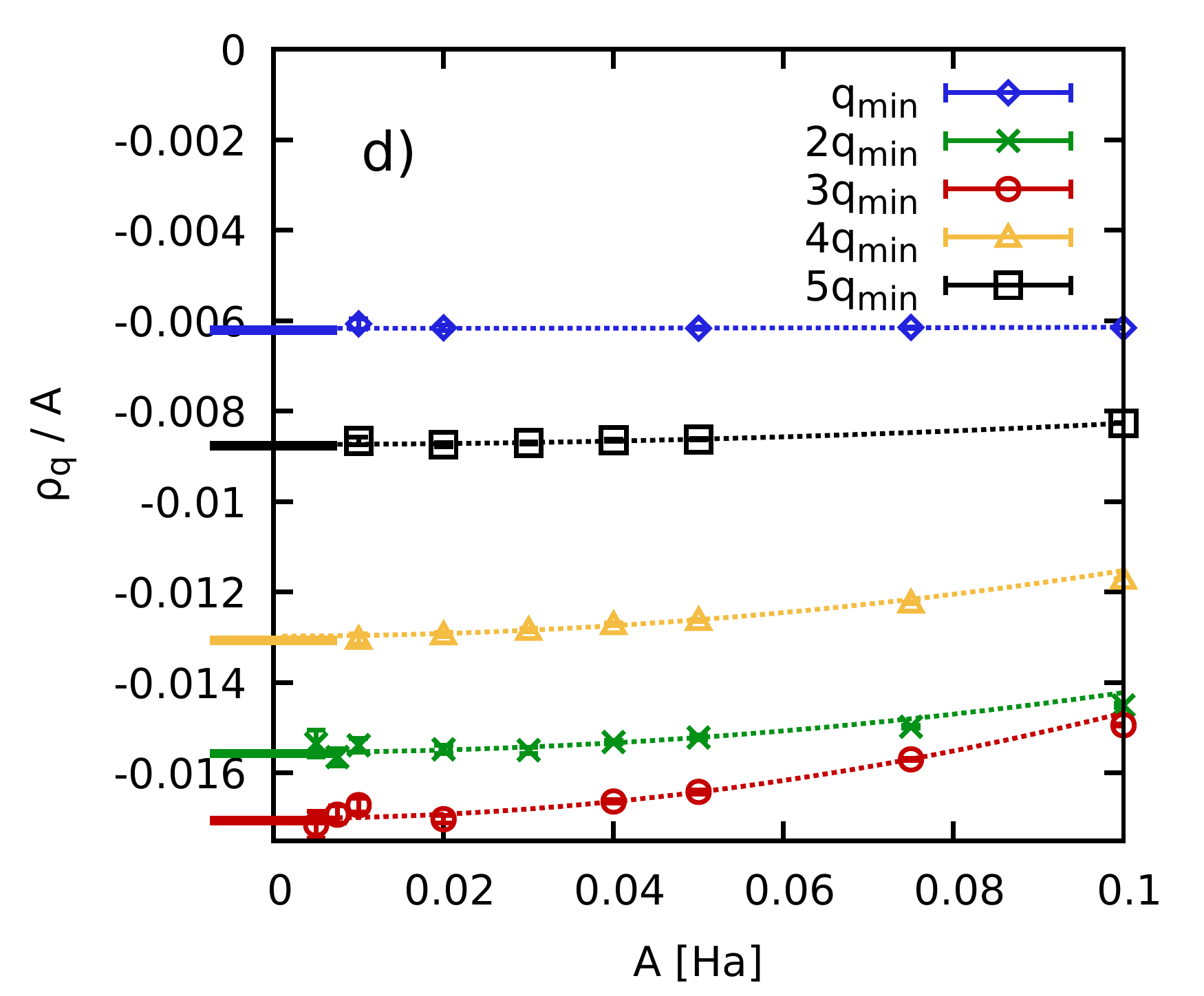}
\caption{\label{fig:LRT_N34_rs4_theta1_3D}
PIMC results for the ITCF and the static linear density response of the UEG for $N=34$, $r_s=4$, and $\theta=1$. a) ITCF in the $q$-$\tau$-plane in the interval $0\leq\tau\leq\beta/2$; the dashed green line indicates the limit of $\tau=0$ where the ITCF becomes equal to the static structure factor, $F(\mathbf{q},0)=S(\mathbf{q})$, and the light blue line the thermal structure factor $F(\mathbf{q},\beta/2)$. b) full $\tau$-dependence of $F(\mathbf{q},\tau)$ for a particular wave vector with $q=1.88q_\textnormal{F}$; solid green: PIMC data from Ref.~\cite{Dornheim_insight_2022}, dotted red: single-particle decay model from Ref.~\cite{Dornheim_PTR_2022}, dash-dotted black: RPA, dashed blue: \emph{static approximation}~\cite{dornheim_dynamic}, $G(\mathbf{q},\omega)\equiv G(\mathbf{q},0)$. c) Static linear density response function $\chi(\mathbf{q})$, green crosses: ITCF-based evaluation of Eq.~(\ref{eq:chi_static}); red dots: estimation from PIMC simulations of the harmonically perturbed system [cf.~Eq.~(\ref{eq:Hamiltonian_modified})] via Eq.~(\ref{eq:rho1_fit}); solid blue: ideal Lindhard function, $\chi_0(\mathbf{q})$; dash-dotted black: RPA; dotted yellow: long-wavelength expansion Eq.~(\ref{eq:perfect_screening}). d) Symbols: raw PIMC results for the induced density, Eq.~(\ref{eq:rho}) for different $q$ as a function of the perturbation amplitude $A$; dotted lines: corresponding fits according to Eq.~(\ref{eq:rho1_fit}).
}
\end{figure*}

\section{Simulation results\label{sec:results}}

Throughout this work, we limit ourselves to the discussion of simulation results for spin-unpolarized systems with an equal number of spin-up and spin-down electrons, $N^\uparrow=N^\downarrow=N/2$. We note that spin-effects are expected to play a particularly important role for WDM in magnetic fields, and have been studied extensively throughout the literature, e.g.~Refs.~\cite{Dornheim_CPP_2021,Brown_PRL_2013,ksdt,groth_prl,review,arora,Tanaka_CPP_2017,Dornheim_PRE_2021,Dornheim_PRR_2022}.

In addition, we note that both QMC and DFT simulations are restricted to a finite number of particles $N$ in a finite simulation cell. The corresponding finite-size effects have been analyzed in detail both for ambient conditions~\cite{Chiesa_PRL_2006,Fraser_PRB_1996,Drummond_PRB_2008,Holzmann_PRB_2016,Krakauer_PRL_2008} and in the WDM regime~\cite{dornheim_prl,Dornheim_JCP_2021,Dornheim_PRE_2020} and are not covered in the present work.

\subsection{Static linear density response\label{sec:static_LRT_results}}

\subsubsection{Uniform electron gas\label{sec:UEG_results}}

Let us begin the discussion of simulation results with an analysis of the static linear density response of the warm dense UEG~\cite{review,status}. In particular, the UEG constitutes one of the most fundamental model systems in physics, quantum chemistry and related disciplines. Often considered as the archetypal system of interacting electrons, the availability of accurate QMC simulations results~\cite{Ceperley_Alder_PRL_1980,Ortiz_PRB_1994,Ortiz_PRL_1999,Spink_PRB_2013} in the zero-temperature limit that were subsequently used as input for various parametrizations~\cite{vwn,Perdew_Zunger_PRB_1981,Perdew_Wang_PRB_1992,Gori-Giorgi_PRB_2000} has been of pivotal importance for the success of DFT at ambient conditions~\cite{Jones_RMP_2015}.
The first ground-state calculations for the static linear density response of the UEG have been presented by Moroni \emph{et al.}~\cite{moroni,moroni2}. They were parametrized by Corradini \emph{et al.}~\cite{cdop} and have been confirmed in the recent study by Chen and Haule~\cite{Chen2019}.

At finite temperature, a host of approximate results for the static linear density response has been reported based on e.g.~dielectric theories~\cite{IIT,stls,stls2,stolzmann,schweng,arora,Tanaka_CPP_2017,tanaka_hnc,Tolias_JCP_2021} and classical mappings~\cite{perrot,DuftyDutta_PRE_2013,LiuWu_JCP_2014}. To our knowledge, the first accurate QMC results for the linear response of the UEG at finite temperature have been presented in Refs.~\cite{dornheim_pre,groth_jcp,review} using the permutation blocking PIMC (PB-PIMC)~\cite{Dornheim_NJP_2015,Dornheim_JCP_2015,Dornheim_CPP_2019} and the configuration PIMC (CPIMC)~\cite{Schoof_CPP_2011,Schoof_CPP_2015,Schoof_PRL_2015,Groth_PRB_2016} methods.
More specifically, these results have been based on a direct simulation of a harmonically perturbed system, see Eqs.~(\ref{eq:Hamiltonian_modified}), (\ref{eq:rho}) and (\ref{eq:LRT_rho}) above. While formally being exact, these efforts have required to perform a considerable number of independent QMC simulations to acquire the density response for a single combination of $\mathbf{q}$ with $(r_s,\theta)$. Therefore, they have been limited to a few density-temperature combinations.

A more efficient strategy to investigate the static linear density response of any given system is offered by the imaginary-time version of the fluctuation-dissipation theorem [Eq.~(\ref{eq:chi_static})], see the discussion in Sec.~\ref{sec:ITCF} above. In Fig.~\ref{fig:LRT_N34_rs4_theta1_3D}, we show a corresponding investigation of the warm dense UEG at the electronic Fermi temperature $\theta=1$ and the metallic density of $r_s=4$, which is close to sodium~\cite{Huotari_PRL_2010,felde}. Panel a) shows the raw PIMC results for the ITCF $F(\mathbf{q},\tau)$ [Eq.~(\ref{eq:ISF})] in the $\tau$-$q$-plane.
For completeness, we note that the $q$-grid is defined by the system size~\cite{Chiesa_PRL_2006,Drummond_PRB_2008,dornheim_prl,Dornheim_CPP_2016,Dornheim_JCP_2021}, which limits us to discrete values with $q\geq 2\pi/L$. The $\tau$-grid, on the hand, directly follows from the number of high-temperature factors $P$ in our PIMC simulations (see Sec.~\ref{sec:PIMC} above), and can, in principle, be made arbitrarily fine.
In addition, we note that it is fully sufficient to restrict ourselves to the half-interval of $0\leq\tau\leq\beta/2$ due to the symmetry relation derived in Eq.~(\ref{eq:symmetry}). The $F(\mathbf{q},\tau)$ symmetry can also be discerned in panel b), where we show the full $\tau$-dependence of the ITCF for a particular wave vector. Returning to panel a), it becomes evident that the ITCF approaches the SSF $S(\mathbf{q})$ in the limit of $\tau\to0$, $F(\mathbf{q},0)=S(\mathbf{q})$; this is indicated by the bold-dashed green curve. 

While the SSF does not exhibit any maxima due to the absence of spatial order at these conditions, its \emph{thermal analogue} given by $F(\mathbf{q},\beta/2)$ (light blue dashed curve) exhibits a distinct structure with a pronounced peak around $q\sim2q_\textnormal{F}$. We note that the physical behavior of $F(\mathbf{q},\tau)$ has often been described as \emph{featureless} in the literature~\cite{Boninsegni1996,cep} and has been treated as a means for the estimation of other properties such as the DSF. Very recently, we have shown that $F(\mathbf{q},\tau)$ directly gives a number of physical insights into phenomena such as the roton feature of the UEG and the related XC induced red shift of $S(\mathbf{q},\omega)$ without any analytic continuation~\cite{Dornheim_insight_2022,Dornheim_PTR_2022}, see also the discussion of Fig.~\ref{fig:dispersion} below. For example, the single-particle dispersion $\omega(q)\sim q^2$ directly manifests in $F(\mathbf{q},\tau)$ as the steep exponential decay for large $q$; in this regime, the ITCF can indeed be modeled very accurately based on a simple, semi-analytical single-particle model for the imaginary-time diffusion process, see Ref.~\cite{Dornheim_PTR_2022} for details. Moreover, the well-known sum-rules of the DSF manifest in the ITCF as derivatives with respect to $\tau$ around $\tau=0$ [Eq.~(\ref{eq:moments_derivative})].

This is demonstrated in Fig.~\ref{fig:LRT_N34_rs4_theta1_3D} b), where we illustrate $F(\mathbf{q},\tau)$ for $q=1.88q_\textnormal{F}$. More specifically, the solid green curve shows our exact PIMC results, and the dash-double-dotted yellow line has been obtained from the exact f-sum rule [Eq.~(\ref{eq:f_sum_rule})] that describes the slope of $F(\mathbf{q},\tau)$ at the origin; we find perfect agreement between the two curves, as it is expected. The availability of exact PIMC benchmark data for the case of the UEG also allows us to unambiguously assess the accuracy of different approximations. The dash-dotted grey curve corresponds to the ubiquitous RPA, which provides the least accurate description of all curves shown in Fig.~\ref{fig:LRT_N34_rs4_theta1_3D} b). Evidently, the mean-field description of the dynamic density response function [i.e., setting $G(\mathbf{q},\omega)\equiv0$ in Eq.~(\ref{eq:LFC})] on which RPA is based is not appropriate in this regime, where electronic XC effects play an important role. For completeness, we note that the RPA still exactly fulfills the f-sum rule, which only describes the derivative at $\tau=0$, but is agnostic with respect to the particular value of $F(\mathbf{q},0)$.

The dashed blue curve has been computed within the \emph{static approximation}, i.e., by setting $G(\mathbf{q},\omega)\equiv G(\mathbf{q},0)$ in Eq.~(\ref{eq:LFC}). In particular, Dornheim \emph{et al.}~\cite{dornheim_dynamic} have reported that the \emph{static approximation} gives highly accurate results for the DSF $S(\mathbf{q},\omega)$ of the UEG for metallic densities $r_s\lesssim 4$. Evidently, the same trend manifests in the ITCF, which is in very good agreement with the exact PIMC results over the entire $\tau$-range. For completeness, we note that the \emph{static approximation} induces a slight overestimation of $F(\mathbf{q},0)=S(\mathbf{q})$ for $q\gtrsim 2q_\textnormal{F}$~\cite{Dornheim_PRL_2020_ESA,Dornheim_PRB_ESA_2021}. While almost being negligible for an individual $q$, this systematic error accumulates within integrated properties such as the interaction energy $W$ [Eq.~(\ref{eq:interaction_energy})]. This problem can be solved by combining the static local field correction of the UEG with a consistent limit for large wave numbers $q=|\mathbf{q}|\gg q_\textnormal{F}$ that is connected to the so-called on-top pair correlation function, $g(r=0)$. The corresponding \emph{effective static approximation}~\cite{Dornheim_PRL_2020_ESA} has been analytically parametrized in Ref.~\cite{Dornheim_PRB_ESA_2021}.

Finally, the dotted red curve in Fig.~\ref{fig:LRT_N34_rs4_theta1_3D} b) has been obtained by combining the SSF with the semi-analytical single-particle imaginary-time diffusion model presented in the recent Ref.~\cite{Dornheim_PTR_2022}. Remarkably, this simple model, too, fulfills the exact f-sum rule and accurately captures the $\tau$-dependence of the ITCF over the entire $\tau$-range.

\begin{figure*}\centering
\includegraphics[width=0.98\textwidth]{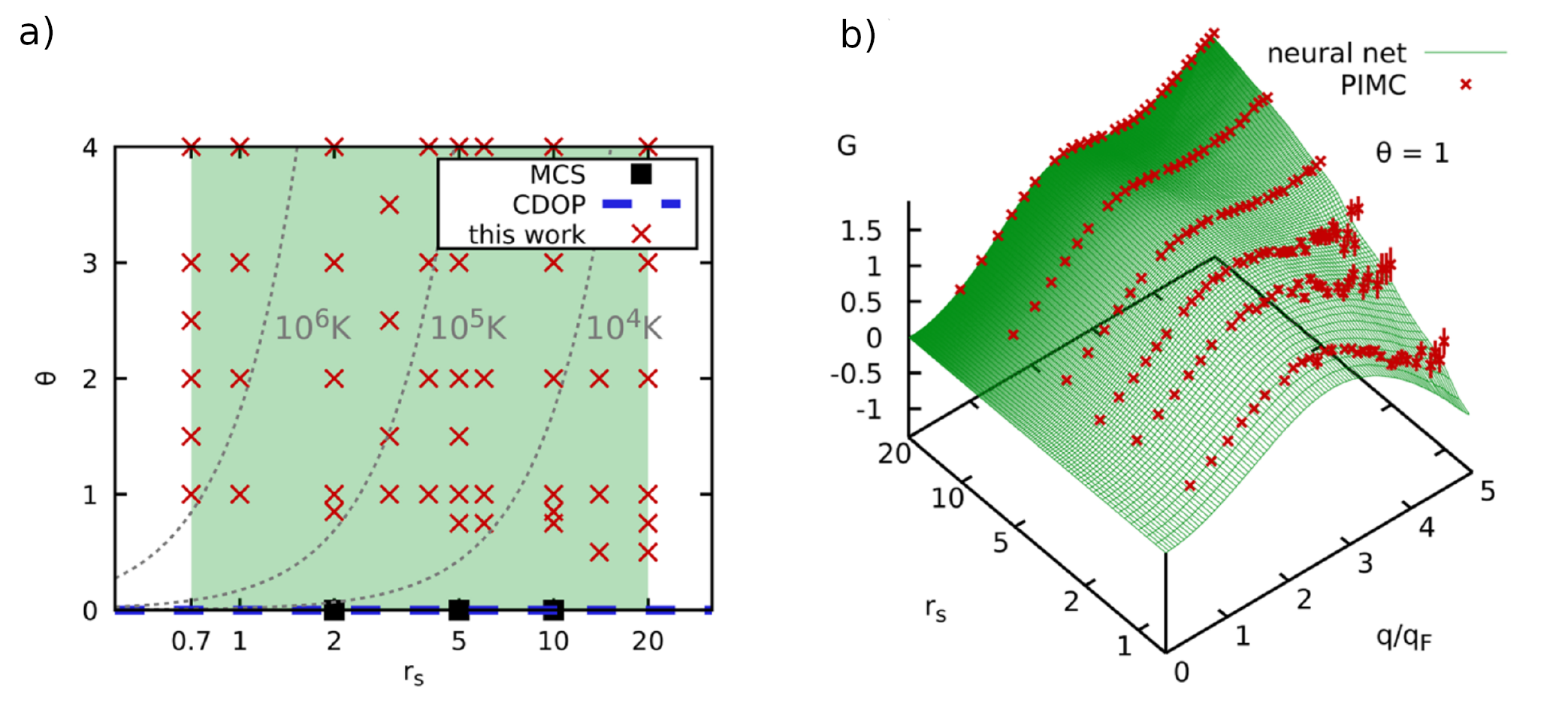}
\caption{\label{fig:ML} Neural network representation of the static local field correction $G(q;r_s,\theta)$. a) Available training data in the $r_s$-$\theta$-plane, red crossed: PIMC simulations based on the ITCF via Eq.~(\ref{eq:chi_static}); black squares (MCS): ground-state QMC calculations by Moroni \emph{et al.}~\cite{moroni2} based on simulations of the harmonically perturbed system, cf.~Eq.~(\ref{eq:Hamiltonian_modified}); dashed blue (CDOP): ground-state parametrization based on the MCS data by Corradini \emph{et al.}~\cite{cdop}. The shaded green area indicates the validity range of the neural network. b) Static local field correction at $\theta=1$ in the $r_s$-$q$-plane. The red crosses show PIMC results and the green surface is the prediction by the neural network.
 Reprinted from Dornheim \emph{et al.}~\cite{dornheim_ML}, \textit{J.~Chem.~Phys.}~\textbf{151}, 194104 (2019) with the permission of AIP Publishing.
}
\end{figure*} 

Let us next get to the task at hand, which is the estimation of the static linear density response function $\chi(\mathbf{q})$ from the PIMC data for $F(\mathbf{q},\tau)$ via Eq.~(\ref{eq:chi_static}). The results are shown in Fig.~\ref{fig:LRT_N34_rs4_theta1_3D} c), where we show the thus obtained PIMC results for $\chi(q)$ as the green crosses. In the limit of small $q$, the exact density response is given by~\cite{kugler_bounds} (dotted yellow curve)
\begin{eqnarray}\label{eq:perfect_screening}
\lim_{q\to0}\chi(\mathbf{q}) = - \frac{q^2}{4\pi}\ ;
\end{eqnarray}
this is a direct consequence of the perfect screening in the UEG and is not reproduced by the density response of the ideal Fermi gas (solid blue curve).

The density response of the UEG to an external static perturbation attains a maximum around $q\sim2q_\textnormal{F}$. From a physical perspective, this can be understood intuitively from the following considerations: for intermediate $q$, the wavelength $\lambda=2\pi/q$ of the perturbation is comparable to the average interparticle distance $d$. Consequently, such a perturbation induces a spatial pattern that automatically minimizes the interaction energy between the electrons, which maximizes the reaction of the system~\cite{Dornheim_alignment_2022}. A more technical explanation can be gleamed from the ITCF depicted in panel a). In particular, Eq.~(\ref{eq:chi_static}) states that $\chi(\mathbf{q})$ is proportional to the area under $F(\mathbf{q},\tau)$ for a given value of $\mathbf{q}$. Therefore, the peak in $\chi(\mathbf{q})$ is directly related to the maximum in $F(\mathbf{q},\beta/2)$. In other words, the existence of a hypothetical spatial pattern that minimizes the interaction energy --- but which is not actually manifested in the unperturbed UEG as there is no maximum in $S(\mathbf{q})$ at these conditions --- manifests itself as a reduced decay along $\tau$ for intermediate $q$; this is directly translated to the static density response function shown in Fig.~\ref{fig:LRT_N34_rs4_theta1_3D}.

The dash-dotted black curve shows the static linear density response function within the RPA, which systematically underestimates the true response of the UEG for $q\gtrsim q_\textnormal{F}$. This, too, can be attributed to the behaviour of the ITCF shown in Fig.~\ref{fig:LRT_N34_rs4_theta1_3D} b), which is too small over the entire $\tau$-range. At the same time, the RPA reproduces the correct perfect screening described by Eq.~(\ref{eq:perfect_screening}).

Finally, we have carried out extensive new PIMC simulations for the harmonically perturbed electron gas; the results for the dependence of the induced density [Eq.~(\ref{eq:rho})] on the perturbation amplitude $A$ are shown in Fig.~\ref{fig:LRT_N34_rs4_theta1_3D} d) for integer multiples of the minimum wave number $q_\textnormal{min}=2\pi/L$. Specifically, the data points show our raw PIMC simulation data (divided by $A$, such that the response attains a constant value in the LRT limit), and the corresponding dotted curves have been obtained from cubic fits according to Eq.~(\ref{eq:rho1_fit}). Evidently, the functional form nicely reproduces the PIMC results everywhere, as it is expected~\cite{Dornheim_PRL_2020,Dornheim_PRR_2021}. The horizontal bars show the LRT limit computed from the ITCF, which is consistent both to the fits and to the PIMC data on which they are based. The linear coefficient in Eq.~(\ref{eq:rho1_fit}) directly corresponds to the linear response function $\chi(\mathbf{q})$, which are included in Fig.~\ref{fig:LRT_N34_rs4_theta1_3D} c) as the red dots. For completeness, we note that the cubic coefficients correspond to the cubic response at the first harmonic~\cite{Dornheim_PRL_2020,Dornheim_PRR_2021}.
Evidently, these independent data for $\chi(\mathbf{q})$ that are based on the actual response of the UEG are in excellent agreement to the ITCF-based evaluation of Eq.~(\ref{eq:chi_static}) over the entire $q$-range. This demonstrates the high consistency of both our PIMC simulations, as well as of the underlying theoretical framework.

Based on the described estimation of $\chi(\mathbf{q})$ from the ITCF, Dornheim \emph{et al.}~\cite{dornheim_ML} have carried out extensive PIMC calculations for $\sim50$ density-temperature combinations; the corresponding parameters are depicted as the red crosses in the $r_s$-$\theta$-plane shown in Fig.~\ref{fig:ML} a). In addition, the black squares and dashed blue line show the ground-state QMC calculations by Moroni \emph{et al.}~\cite{moroni2} (MCS) and the corresponding parametrization by Corradini \emph{et al.}~\cite{cdop} (CDOP).  The task at hand was to obtain an accurate representation of the static local field correction $G(\mathbf{q})$ covering the entire relevant parameter range, i.e., the shaded green area in Fig.~\ref{fig:ML} a). 

On the one hand, the inversion of Eq.~(\ref{eq:LFC}) to compute $G(\mathbf{q})$ from PIMC data for $\chi(\mathbf{q})$ is straightforward. 
On the other hand, $G(\mathbf{q};r_s,\theta)$ exhibits a nontrivial dependence on $\mathbf{q}$ based on different combination of $r_s$ and $\theta$. This is illustrated in Fig.~\ref{fig:ML} b), where we show the static local field correction in the $r_s$-$q$-plane at the electronic Fermi temperature, $\theta=1$. Specifically, the red crosses show the $q$-dependence of our PIMC results for $G(\mathbf{q})$, and we observe the following trend. For low densities (i.e., large $r_s$), $G(\mathbf{q})$ exhibits a pronounced increase in the limit of large $q$. This increase becomes less pronounced with decreasing $r_s$, and actually becomes negative for high densities. From a physical perspective, this interesting behaviour can be traced back to the XC contribution to the kinetic energy~\cite{holas_limit,Hou_PRB_2022}, which is negative for some parameters at finite temperatures~\cite{Militzer_Pollock_PRL_2002,dynamic_folgepaper}.
To construct a reliable representation of $G(q;r_s,\theta)$ that is capable to capture these interesting trends, Dornheim \emph{et al.}~\cite{dornheim_ML} have trained a deep neural network to \emph{learn} the appropriate functional dependence. The results are shown as the green surface in Fig.~\ref{fig:ML}. Evidently, the neural net nicely reproduces the PIMC data where they are available, and smoothly interpolates in between. The high quality of this surrogate model~\cite{surrogate} was subsequently confirmed by the validation against independent data that had not been included into the training procedure, and by the very recent study of electronic exchange-correlation effects by Hou~\emph{et al.}~\cite{Hou_PRB_2022}.

Since its publication in 2019, the neural-net representation of $G(\mathbf{q};r_s,\theta)$ has been used for a number of applications, including the modelling of XRTS experiments~\cite{Dornheim_PRL_2020_ESA}, the estimation of ionization-potential depression~\cite{Zan_PRE_2021}, and as input for a theory of nonlinear effects in the UEG~\cite{Dornheim_PRR_2021}, which is covered in more detail in Sec.~\ref{sec:nonlinear_results} below.

\subsubsection{Warm dense hydrogen\label{sec:hydrogen_results}}

\begin{figure}\centering
\includegraphics[width=0.45\textwidth]{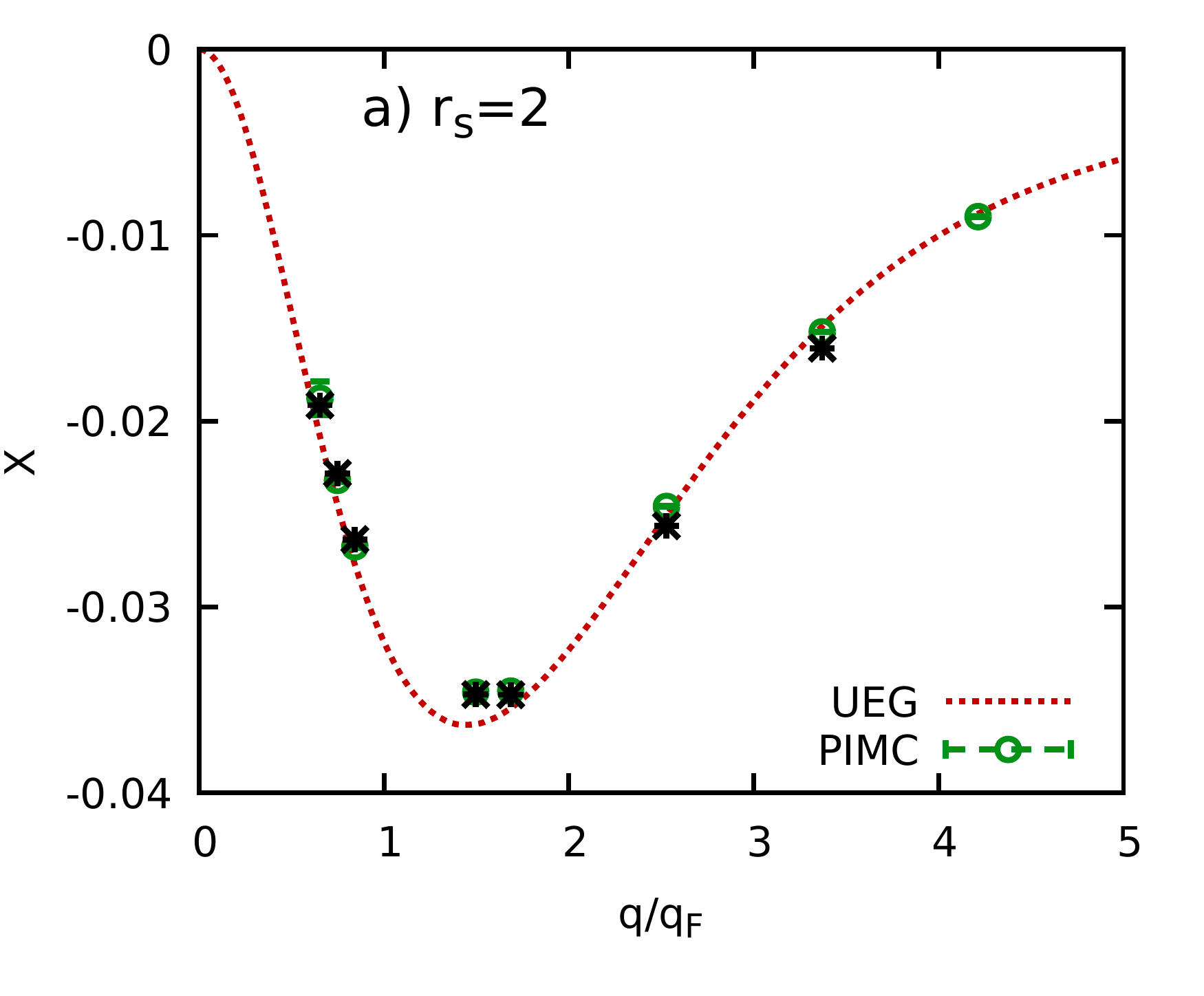}\\\vspace*{-1.19cm}\includegraphics[width=0.46\textwidth]{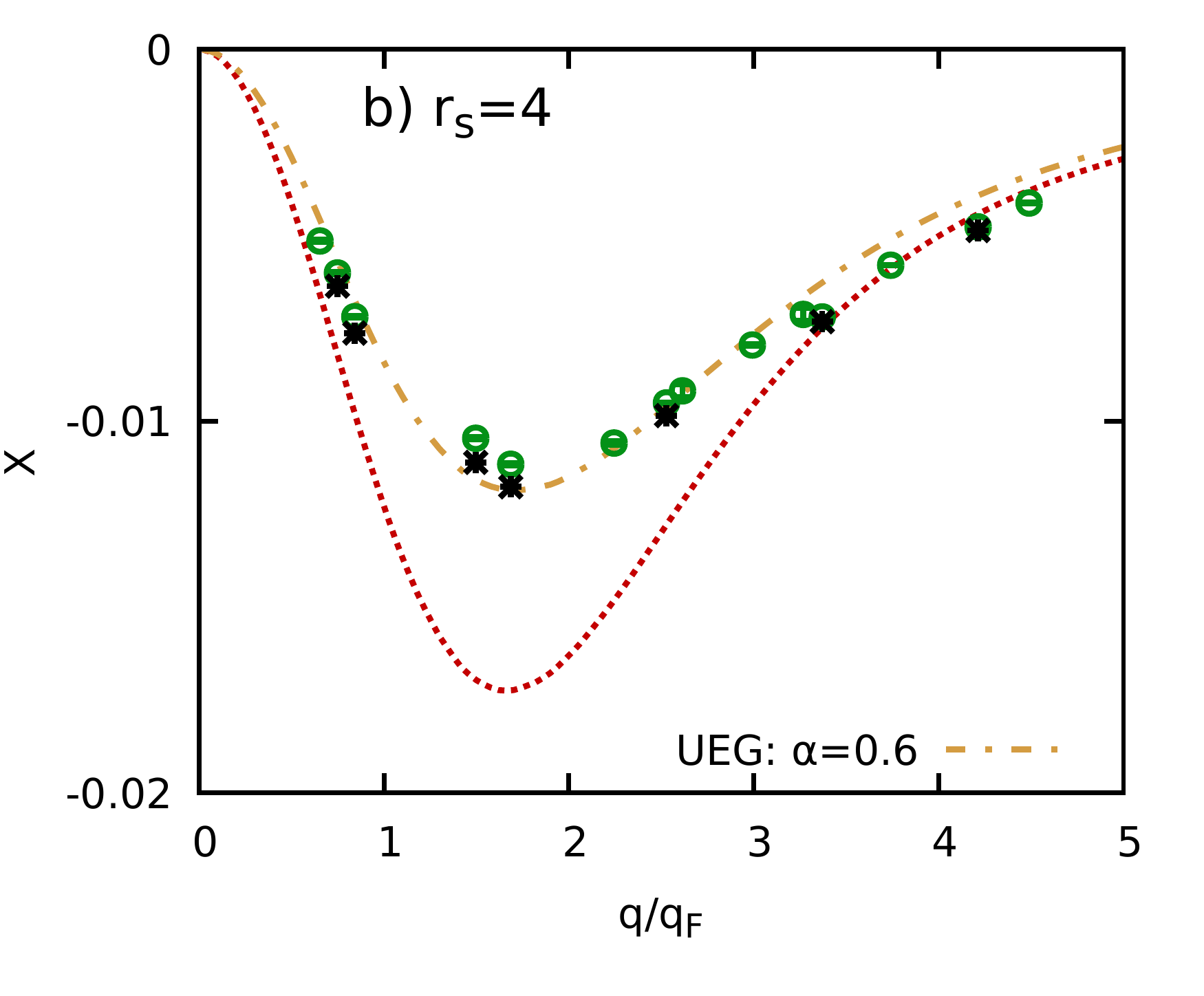}
\caption{\label{fig:H_chi} Static density response of hydrogen at the electronic Fermi temperature, $\theta=1$, for a) $r_s=2$ and b) $r_s=4$. Green circles: PIMC reference data by B\"ohme \emph{et al.}~\cite{Bohme_PRL_2022}; black stars: DFT results using the PBE XC functional by Moldabekov \emph{et al.}~\cite{Moldabekov_PRL_2022}; dotted red: density response of the UEG at the same conditions evaluated using the neural-net representation from Ref.~\cite{dornheim_ML}; dash-dotted yellow [only panel b)]: density response of a UEG with modified $r_s*$ and $\theta*$ corresponding to a free-electron fraction of $\alpha=0.6$.
}
\end{figure}

Let us next investigate the static linear density response of a real system. In Fig.~\ref{fig:H_chi}, we show the static density response function for hydrogen at the electronic Fermi temperature, i.e., $\theta=1$. Panel a) corresponds to the metallic density of $r_s=2$, and the green circles have been obtained by B\"ohme \emph{et al.}~\cite{Bohme_PRL_2022} by exactly solving the electronic problem in a fixed ion potential with PIMC. In particular, they have applied an external harmonic potential and subsequently measured the response in the electronic density; see also Ref.~\cite{Bohme_PRE_2022} for a more detailed description of the simulation setup. The dotted red line shows results for the density response of the UEG at the same conditions, and has been computed based on the neural-net representation of $G(q;r_s,\theta)$~\cite{dornheim_ML}. Evidently, the electronic density response of hydrogen at these conditions strongly resembles the behaviour of a free electron gas, as most of the electrons are delocalized here~\cite{Militzer_PRE_2001}.

Let us for now ignore the black stars and proceed to Fig.~\ref{fig:H_chi}b), where we present the same analysis for a lower density at $r_s=4$. Such conditions are expected to exhibit a number of interesting physical effects and can be realized experimentally for example in hydrogen jets~\cite{Zastrau}. Low-density systems constitute an interesting benchmark for the assessment of electronic XC effects~\cite{low_density1,low_density2}, which are more pronounced at larger values of the quantum coupling parameter $r_s$~\cite{Ott2018}. In fact, they constitute a realistic option for an experimental observation of the \emph{roton feature} in the dispersion of the warm dense electron gas that has been reported and analysed in earlier works~\cite{dornheim_dynamic,Dornheim_Nature_2022,Dornheim_insight_2022}. In addition, these conditions are characterized by a partial localization of the electrons around the ions~\cite{Militzer_PRE_2001}, which means that widely used concepts such the decomposition into bound and free electrons~\cite{Chihara_1987,Gregori_PRE_2003} are expected to break down. 
Indeed, the analysis by B\"ohme \emph{et al.}~\cite{Bohme_PRL_2022} has revealed that the actual electronic density response of hydrogen (green) is substantially lowered compared to the UEG model at the same conditions (dotted red). 
The dash-dotted yellow curve in Fig.~\ref{fig:H_chi} b) has also been computed based on the UEG, but with modified parameters $r_s*$ and $\theta*$ that correspond to an effective free-electronic fraction of $\alpha=0.6$, i.e., $60\%$; this is consistent with the value of $\alpha=0.54$ reported by Militzer and Ceperley~\cite{Militzer_PRE_2001}.
While this effective density response overall exhibits the correct magnitude around $q\sim2q_\textnormal{F}$, it deviates from the PIMC reference data in particular for large $q$. 
In fact, the true electronic density response of hydrogen even exceeds the red UEG curve for $q\gtrsim 4q_\textnormal{F}$, which has been attributed to an isotropy breaking due to the presence of the ions at small length scales $\lambda_q=2\pi/q$.
This is a strong indication that ionization models do not universally work over all relevant length scales.

While being important in their own right, the availability of highly accurate PIMC benchmark results also allows one to benchmark a number of previously used approximations. Of particular interest is the assessment of the accuracy of thermal DFT, which constitutes the workhorse of WDM theory~\cite{wdm_book,Holst_PRB_2008,Smith2018}. In Fig.~\ref{fig:H_chi}, we include DFT results that have been obtained by Moldabekov \emph{et al.}~\cite{Moldabekov_PRL_2022} based on the PBE XC functional~\cite{Perdew_PRL_1996} for the electronic density response of the same ion snapshot as the black stars. As mentioned above, hydrogen basically behaves as a free electron gas at $r_s=2$ and the DFT results are in good agreement with the PIMC reference data. 
For $r_s=4$, the agreement is less good, and the DFT calculations overestimate the actual response over the entire depicted $q$-range. This can be traced back to an underestimation of the localization of the electrons around the ions due to the PBE XC functional, which is known to be afflicted with self-interaction errors~\cite{Cohen_Science}. At the same time, we note that the overall agreement between PIMC and PBE-based DFT calculations is better compared to the effective UEG model with $\alpha=0.6$. In particular, the DFT simulations are capable of accurately reproducing the effect of the isotropy breaking at large $q$; this has important implications, as we shall discuss in the following.

Very recently, Moldabekov \emph{et al.}~\cite{Moldabekov_PRL_2022} have suggested that DFT calculations of a harmonically perturbed system that is governed by the $\omega\to0$ limit of the Hamiltonian Eq.~(\ref{eq:Hamiltonian_modified}) give direct access to the physical static linear density response function $\chi(\mathbf{q})$ by analysing the induced change in the single-electron density $n(\mathbf{r})$; these results are shown as the black stars in Fig.~\ref{fig:H_chi}. This procedure is formally exact, and only depends on the employed approximation to the XC functional in practice. Having found $\chi(\mathbf{q})$, it is straightforward to invert Eq.~(\ref{eq:kernel}) to obtain the static XC kernel $K_\textnormal{xc}(\mathbf{q})$ that is fully consistent with any KS-reference function $\chi_\textnormal{S}(\mathbf{q},\omega)$. This procedure is outlined in more detail in Fig.~\ref{fig:Zhandos} below, and allows one to generate the material-specific XC kernel for any XC functional; no evaluation of a functional derivative is required. Indeed, the latter point is notoriously difficult in practice and had hitherto to our knowledge prevented the computation of XC kernels of extended systems beyond adiabatic LDA (ALDA) or adiabatic generalized-gradient approximations (AGGA) for extended systems~\cite{Byun_2020}.

\begin{figure}\centering
\includegraphics[width=0.45\textwidth]{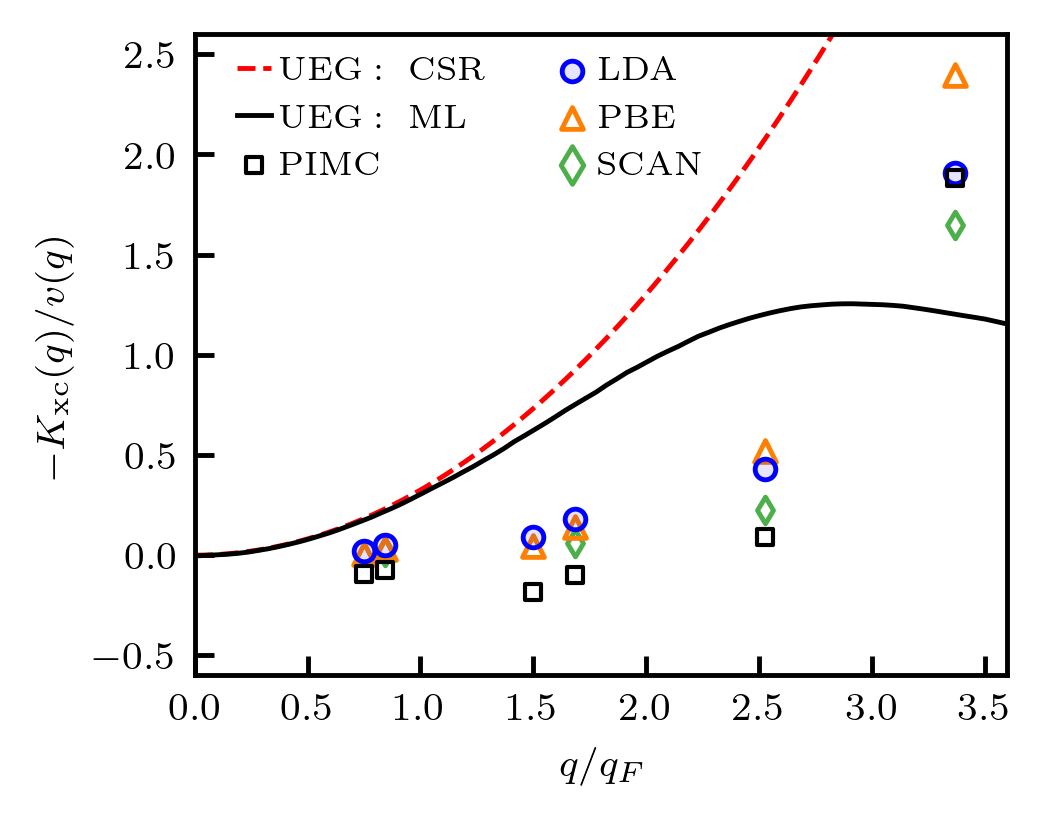}
\caption{\label{fig:H_kernel} Static XC kernel, plotted as the local field correction $G(q)=-K_\textnormal{xc}(q)/v(q)$, of hydrogen at $\theta=1$ and $r_s=4$. The solid black and dashed red line show the full $G(q)$ of the UEG at the same conditions evaluated from the neural-net representation from Ref.~\cite{dornheim_ML} and the corresponding exact long wavelength expansion given by the compressibility sum-rule (CSR) Eq.~(\ref{eq:CSR}). The black squares have been computed from the exact PIMC reference data by B\"ohme \emph{et al.}~\cite{Bohme_PRL_2022}, and the blue circles, orange triangles, and green diamonds from DFT results for $\chi(\mathbf{q})$ (see Fig.~\ref{fig:H_chi}) using the LDA~\cite{Perdew_Zunger_PRB_1981}, PBE~\cite{Perdew_PRL_1996}, and SCAN~\cite{SCAN} XC functionals. Adapted from Moldabekov \emph{et al.}~\cite{Moldabekov_PRL_2022}. 
}
\end{figure}

In Fig.~\ref{fig:H_kernel}, we show the corresponding results for the XC kernel of hydrogen for $r_s=4$ and $\theta=1$. As a reference function to obtain the kernel from the physical density response function $\chi(\mathbf{q})$, Moldabekov \emph{et al.}~\cite{Moldabekov_PRL_2022}
have used the same mean-field response function $\chi_0(\mathbf{q})$ that has been obtained without any XC functional  for all data sets. 
Specifically, the black squares have been computed from the PIMC data for $\chi(\mathbf{q})$ by B\"ohme \emph{et al.}~\cite{Bohme_PRL_2022}, and constitute a highly accurate benchmark for the other results. 
The dashed red curve corresponds to the popular ALDA kernel, which is computed from the parabolic compressibility sum rule expansion around $q=0$ for a UEG,
\begin{eqnarray}\label{eq:CSR}
\lim_{q\to0}G(q) = - \frac{q^2}{4\pi} \frac{\partial^2}{\partial n^2} \left( n F_\textnormal{xc} \right)\ ,
\end{eqnarray}
at the same density and temperature. Clearly, it does not capture the true behavior of the kernel, and substantially overestimates it over the entire depicted $q$-range. The solid dashed line has been obtained from the accurate neural-net representation of $G(q;r_s,\theta)$ of the UEG~\cite{dornheim_ML}; it, by definition, reproduces the ALDA expansion for small $q$, and does not reproduce the PIMC results for larger $q$ either. In fact, using the RPA expression for $\chi(\mathbf{q})$ in Eq.~(\ref{eq:kernel}), which corresponds to setting $K_\textnormal{xc}(\mathbf{q})\equiv0$, constitutes a superior approximation compared to either ALDA or the full UEG model, which lead to an actual deterioration of the attained accuracy. This has important implications for the LR-TDDFT simulation of WDM, as we discuss in more detail in Sec.~\ref{sec:dynamic_results} below.

In stark contrast to the ALDA model, the new DFT-based results for $K_\textnormal{xc}(\mathbf{q})$ from Ref.~\cite{Moldabekov_PRL_2022} that have been computed within the LDA (blue circles, using the functional by Perdew and Zunger~\cite{Perdew_Zunger_PRB_1981}), PBE~\cite{Perdew_PRL_1996} (orange triangles), and SCAN~\cite{SCAN} (green diamonds) qualitatively capture the $q$-dependence of the PIMC reference data over the entire depicted $q$-range. 
In particular, SCAN exhibits the best accuracy for intermediate wave numbers with $q\sim2q_\textnormal{F}$, although there remain significant differences to the exact PIMC reference data.

To summarise, the recent approach for the computation of $K_\textnormal{xc}(\mathbf{q})$ constitutes a promising route to study electron-electron correlations of real materials within the framework of KS-DFT. In combination with accurate reference results such as the PIMC data for hydrogen by B\"ohme \emph{et al.}~\cite{Bohme_PRL_2022}, it provides an ideally suited tool for the assessment of the accuracy of different XC functionals. Future efforts in this direction will include the rigorous benchmark of orbital-dependent hybrid functionals, which have already been successfully applied to the study of WDM~\cite{Witte_PRL_2017,Ravasio_PRL_2021}, and the development and benchmark of new functionals that are explicitly designed to meet the challenges of capturing the electronic density response of matter at extreme conditions.

\subsection{Dynamic linear density response\label{sec:dynamic_results}}

In the previous section, we have given an overview of some recent promising developments regarding the description of the electronic density response in the static limit of $\omega\to0$. While being an important step in the right direction, many important applications such as the modelling of XRTS experiments~\cite{siegfried_review,dynamic2,Mo_PRL_2018,Ramakrishna_PRB_2021} or the construction of advanced XC functionals within the adiabatic-connection fluctuation-dissipation formulation of DFT~\cite{pribram,Patrick_JCP_2015} require as input some information about the full frequency-dependence of the electronic density response of a given system.

Unfortunately, the accurate estimation of the dynamic electronic properties of WDM is even more difficult than in the static case, and only a handful of methods are serious contenders. Among these, the nonequilibrium Green's functions (NEGF) method~\cite{stefanucci2013nonequilibrium,Schluenzen_2020} deserves a special place, as it is, as the name suggests, capable of treating real nonequilibrium conditions as they occur, for example during the stopping of a projectile in a medium~\cite{Magyar_CPP_2016,Zeb_PRL_2012,Yost_PRB_2017,Schluenzen_CPP_2019}. While some remarkable methodological improvements have been reported over the last few years~\cite{Schluenzen_PRL_2020,Joost_PRB_2020}, it remains unclear if NEGF calculations of highly excited states as they occur in WDM are computationally feasible, and if common approximations to the self-energy are sufficient to capture the impact of electron-electron interactions. To our knowledge, no NEGF calculations of real WDM systems have been presented in the literature.

A second route towards dynamic electronic properties of WDM is RT-TDDFT~\cite{castro2006octopus,Schleife_JCP_2012,marques2012fundamentals,kolesov2016real,Kononov2022}. RT-TDDFT is a computationally efficient approach that allows for the use of large supercells and basis sets along with an accurate description of the electron-ion interaction. Among its key capabilities for WDM purposes are its ability to model XRTS experiments without the ubiquitous Chihara decomposition~\cite{dynamic2} and the possibility to estimate both nonequilibrium~\cite{correa2012nonadiabatic} and nonlinear effects~\cite{andrade2018negative}. For example, RT-TDDFT allows for a description of nonadiabatic electron-ion dynamics such as those that are relevant to electronic stopping power~\cite{Ehrenfest_JCP_2005,Magyar_CPP_2016,correa2018calculating,ding2018ab,white2018time,White_2022}, which is critical to self-heating in inertial fusion applications~\cite{lindl1995development,zylstra2019alpha}. On the downside, systematically improvable approximations to the true dynamic XC potential~\cite{maitra2002memory,elliott2012universal} --- which replaces the standard XC functional of equilibrium DFT --- have been difficult to develop and most calculations are done with adiabatic approximations to standard functionals. 
Nevertheless, RT-TDDFT has performed well when compared to experiments, and stopping power calculations have so far demonstrated reasonable agreement with WDM experiments~\cite{ding2018ab,malko2022proton} and databases of experimental results in less extreme conditions~\cite{schleife2015accurate,kang2019pushing,kononov2020pre,kononov2021anomalous}.

Currently, the arguably most widely used \emph{ab initio} method for the description of electron dynamics in WDM is given by linear-response TDDFT (LR-TDDFT)~\cite{dynamic1,Mo_PRL_2018,Ramakrishna_PRB_2021,marques2012fundamentals}. To be more specific, LR-TDDFT is based on a standard equilibrium KS-DFT simulation of a given system, which gives access to a set of KS-orbitals $\{\phi_\alpha\}$. The latter are then used to compute the dynamic KS-response function $\chi_\textnormal{S}(\mathbf{q},\omega)$ via Eq.~(\ref{eq:chi0}); the key ingredient is then given by the XC kernel $K_\textnormal{xc}(\mathbf{q},\omega)$, which relates $\chi_\textnormal{S}(\mathbf{q},\omega)$ to the actual physical dynamic linear density response function $\chi(\mathbf{q},\omega)$ of the system of interest via Eq.~(\ref{eq:kernel}). Formally, LR-TDDFT and RT-TDDFT should give equal results in the limit of weak perturbations; a practical investigation of this assumption for the density response of WDM constitutes an important topic for future research. The main bottleneck of LR-TDDFT is given by its dependence on a particular XC kernel, which has to be material-specific and must be consistent to the XC functional that has been used to determine $\chi_\textnormal{S}(\mathbf{q},\omega)$. In practice, both conditions are generally violated by the most commonly used model kernels \cite{Byun_2020, PhysRevB.35.5585}, and the impact of such inconsistencies on the calculated observables remains as of yet poorly understood. Here we stress that this is not a shortcoming of the general formulation of the LR-TDFT, but of the inconsistency in the ad hoc approximations used for the XC kernel.  
As we have elaborated in Sec.~\ref{sec:hydrogen_results} above, Moldabekov \emph{et al.}~\cite{Moldabekov_PRL_2022} have recently introduced a formally exact framework to estimate the consistent static kernel for any given XC functional and material, and its utilization for LR-TDDFT calculations within the \emph{static approximation} where $K_\textnormal{xc}(\mathbf{q},\omega)\equiv K_\textnormal{xc}(\mathbf{q})$ in Eq.~(\ref{eq:kernel}) is discussed in Sec.~\ref{sec:dynamic_WDM}.

As the final method for the computation of the dynamic density response, we mention PIMC. While, from a conceptual perspective, the real-time propagation of any observable in Feynman's path-integral picture of quantum mechanics~\cite{kleinert2009path} is straightforward, it leads to an oscillating complex exponential function in practice; this is the origin of the infamous \emph{dynamic phase problem}~\cite{Zhang_PRL_2003,Lothar_PRL_2008,Church_JCP_2021}. Still, PIMC methods give direct access to the imaginary-time dynamics of a given system~\cite{Berne_JCP_1983,Dornheim_JCP_ITCF_2021,Dornheim_insight_2022} in thermodynamic equilibrium; exact PIMC results for the ITCF $F(\mathbf{q},\tau)$ of the warm dense UEG are shown in Fig.~\ref{fig:LRT_N34_rs4_theta1_3D} above.
Indeed, the ITCF contains the same information as the DSF $S(\mathbf{q},\omega)$ --- albeit in an unfamiliar representation~\cite{Dornheim_insight_2022,Dornheim_PTR_2022} --- and is directly useful for the interpretation of XRTS experiments~\cite{Dornheim_T_2022}, see Sec.~\ref{sec:experiment} below.
Yet, the analytic continuation~\cite{JARRELL1996133} from the $\tau$-domain to real frequencies is notoriously difficult. It is, however, possible for the case of the UEG~\cite{dornheim_dynamic,dynamic_folgepaper,Hamann_PRB_2020} as we shall explain in the following section.

\subsubsection{Uniform electron gas\label{sec:dynamic_UEG}}

The accurate estimation of the dynamic density response and dynamic structure factor of the UEG has been an active topic of investigation over many decades~\cite{pines1999elementary,dynamic1,Dabrowski_PRB_1986,Holas_PRB_1987,Jongbae_Hong_1993,kwong_prl-00,Takada_PRL_2002,kremp2005quantum,Pitarke_PRB_2007,Takada_PRB_2016,Kas_PRL_2017,Panholzer_PRL_2018,dornheim_dynamic,Ruzsinszky_PRB_2020,Hamann_PRB_2020,Hamann_CPP_2020,Ara_POP_2021,Leblanc_arxiv_2022}. In the ground state, Takada and collaborators~\cite{Takada_PRL_2002,Takada_PRB_2016} have presented accurate data for $S(\mathbf{q},\omega)$ over a broad range of densities, with important implications for scattering experiments with aluminum~\cite{Takada_PRL_2002,felde}. Only very recently, LeBlanc \emph{et al.}~\cite{Leblanc_arxiv_2022} have presented accurate data for the UEG density response based on novel diagrammatic Monte Carlo calculations at low temperatures, $\theta=0.1$; it remains to be seen if these promising efforts can be extended to the WDM regime in future works.

\begin{figure}\centering
\includegraphics[width=0.5\textwidth]{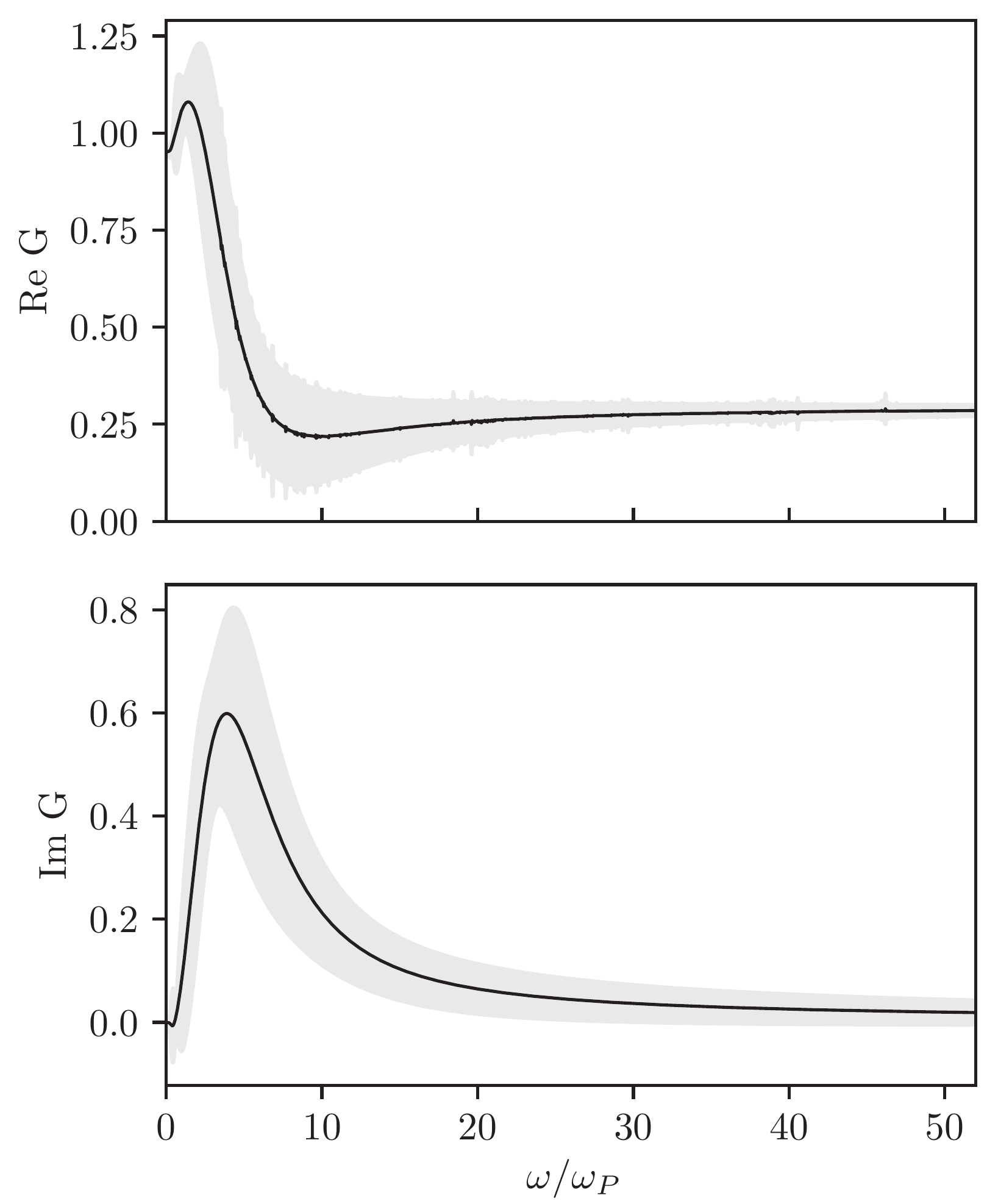}
\caption{\label{fig:Dynamic_LFC} \emph{Ab initio} PIMC results for the frequency-dependence of the dynamic local field correction $G(q,\omega)$ of the UEG with $\theta=1$ and $r_s=6$ for $q=1.88q_\textnormal{F}$. The shaded grey areas indicate the respective uncertainty intervals of the real [top] and imaginary [bottom] parts.
Reprinted with permission from Hamann \emph{et al.}~\cite{Hamann_PRB_2020}, \textit{Phys.~Rev.~B}~\textbf{102}, 125150 (2020). Copyright 2020 by the
American Physical Society.
}
\end{figure} 

The first highly accurate results for the dynamic electronic density response of the warm dense UEG have been presented by Dornheim \emph{et al.}~\cite{dornheim_dynamic} based on exact PIMC results for the ITCF $F(\mathbf{q},\tau)$. To render the numerical inversion of Eq.~(\ref{eq:Laplace}) tractable, the original problem has been re-cast into the reconstruction of the dynamic local field correction $G(\mathbf{q},\omega)$. More specifically, trial solutions for $G(\mathbf{q},\omega)$ give straightforward access to $\chi(\mathbf{q},\omega)$ [Eq.~(\ref{eq:LFC})], and via the fluctuation-dissipation theorem [Eq.~(\ref{eq:FDT})] also to the corresponding DSF $S(\mathbf{q},\omega)$; finally, the latter is inserted into 
Eq.~(\ref{eq:Laplace}) and compared to the PIMC reference data for $F(\mathbf{q},\tau)$. While formally being equivalent to the original problem of the analytic continuation~\cite{JARRELL1996133}, this approach allows one to incorporate a number of exact properties of $G(\mathbf{q},\omega)$~\cite{dornheim_dynamic,dynamic_folgepaper,Dabrowski_PRB_1986,dynamic1}, which sufficiently constrains the space of possible trial solutions for $S(\mathbf{q},\omega)$.

In Fig.~\ref{fig:Dynamic_LFC}, we show corresponding results for both the real (top panel) and the imaginary part (bottom panel) of $G(\mathbf{q},\omega)$ for the UEG at $\theta=1$, $r_s=6$~\cite{Hamann_PRB_2020}, with the shaded grey area indicating the given uncertainty interval. For the real part, the static $\omega\to0$ limit can be inferred from $\chi(\mathbf{q})$, while the high frequency $\omega\to\infty$ limit follows from the DSF cubic sum-rule, which allows expressing $M_3^{S}$ [cf.~Eq.~(\ref{eq:moments})] in terms of static properties such as the $S(\mathbf{q})$~\cite{Mihara_Puff_PR_1968,Iwamoto_PRB_1984,quantum_theory}. It is interesting that the real part exhibits a nontrivial and, in fact, non-monotonic behavior between these limits, with a pronounced maximum around $\omega\approx2\omega_\textnormal{p}$, with $\omega_\textnormal{p}=\sqrt{3/r_s^3}$ being the usual plasma frequency~\cite{quantum_theory}. The corresponding imaginary part of the dynamic local field correction vanishes in both the limits of $\omega\to0$ and $\omega\to\infty$, and exhibits a single broad peak around $\omega\approx5\omega_\textnormal{p}$.

From a theoretical perspective, accurate knowledge of $G(\mathbf{q},\omega)$ gives one direct access to the dynamic density response function $\chi(\mathbf{q},\omega)$ [cf.~Eq.~(\ref{eq:LFC})] and, in this way, a gamut of other properties such as the dynamic dielectric function $\epsilon(\mathbf{q},\omega)$ or the electric conductivity $\sigma(\mathbf{q},\omega)$ at finite wave numbers. This has been explored in detail in recent works by Hamann \emph{et al.}~\cite{Hamann_PRB_2020,Hamann_CPP_2020}.

\begin{figure}\centering
\includegraphics[width=0.5\textwidth]{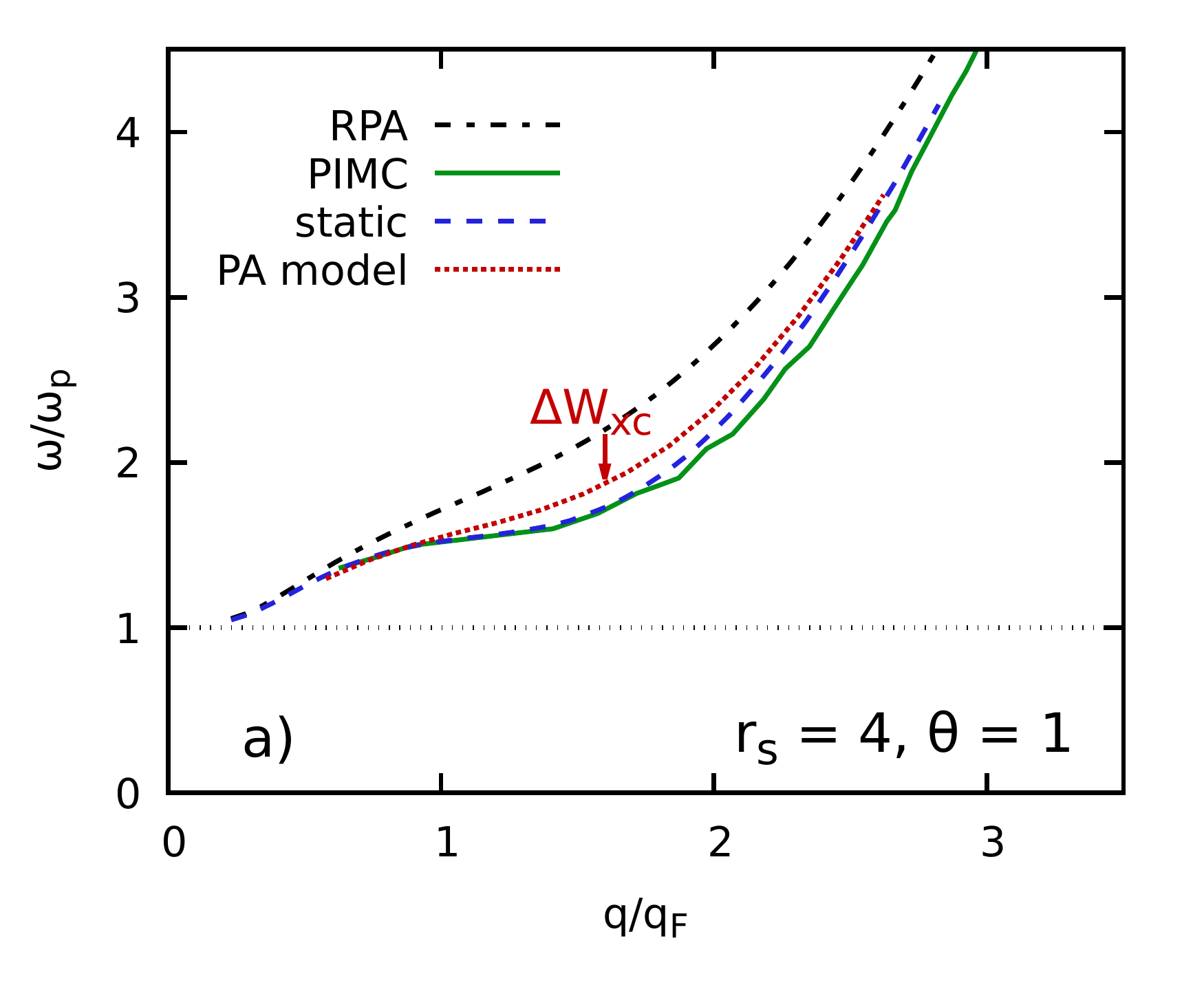}\\\vspace*{-1.425cm}\includegraphics[width=0.5\textwidth]{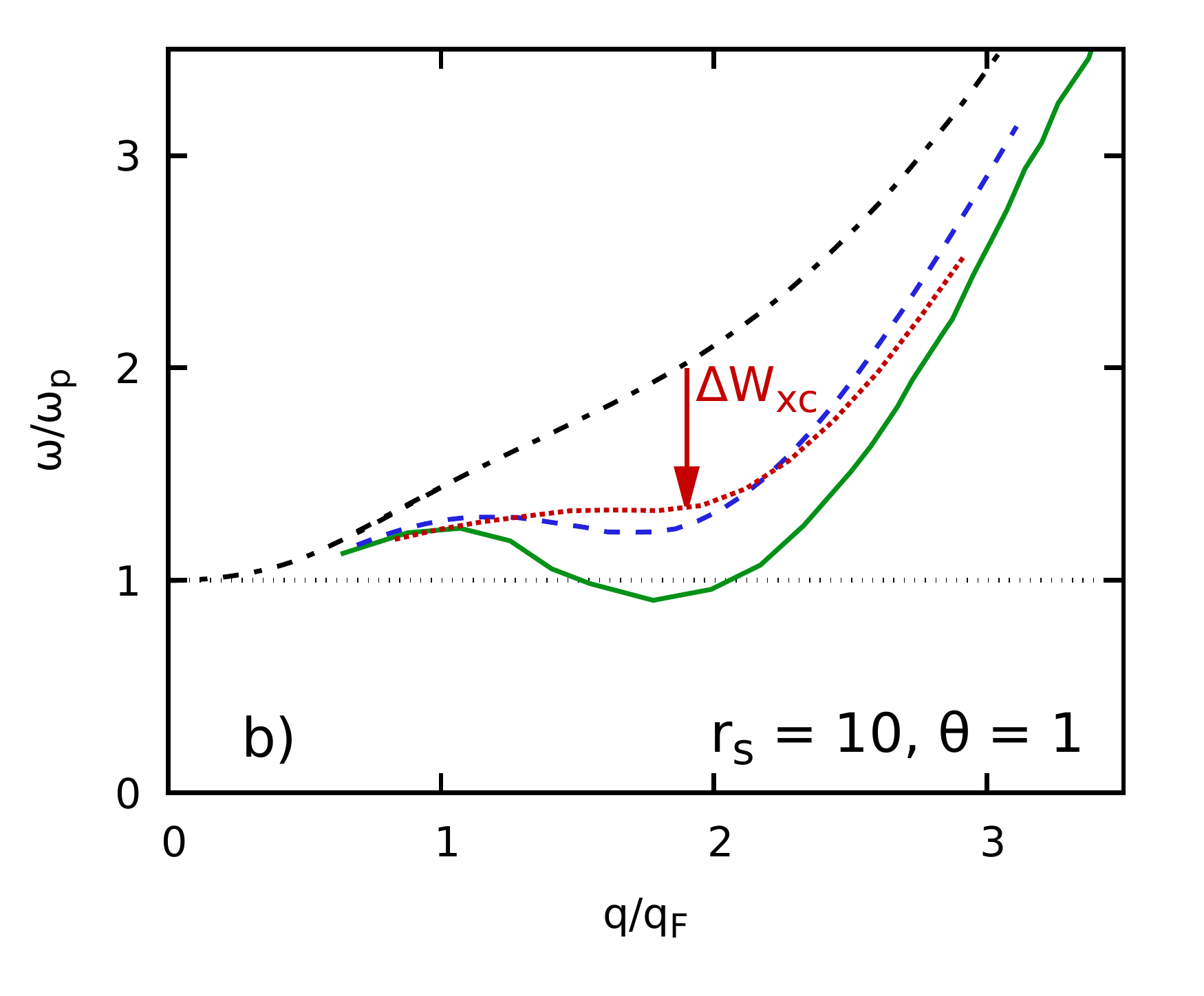}
\caption{\label{fig:roton} Position of the maximum of the DSF as a function of the wave number, $\omega(q)$, of the UEG at the electronic Fermi temperature, $\theta=1$, at a) $r_s=4$ and b) $r_s=10$. Solid green: exact PIMC results taken from Ref.~\cite{dornheim_dynamic}; dashed blue: \emph{static approximation}, i.e., setting $G(\mathbf{q},\omega)\equiv G(\mathbf{q},0)$ in Eq.~(\ref{eq:LFC}); dash-dotted black: RPA; dotted red: pair alignment model introduced in Ref.~\cite{Dornheim_Nature_2022}. The vertical red arrows indicate the XC induced down-shift $\Delta W_\textnormal{xc}$ of the true $\omega(q)$ compared to the mean-field description within RPA.
}
\end{figure} 

Due to its central role in the interpretation and modelling of XRTS experiments~\cite{siegfried_review,sheffield2010plasma} (cf.~Eq.~(\ref{eq:convolution}) above), one of the most interesting properties in this regard is given by $S(\mathbf{q},\omega)$, which has been studied extensively in Refs.~\cite{dornheim_dynamic,dynamic_folgepaper,Dornheim_Nature_2022,Dornheim_PRE_2020}. In particular, Dornheim \emph{et al.}~\cite{dornheim_dynamic} have found an XC induced red shift in the position of the maximum of the true $S(\mathbf{q},\omega)$, $\omega(q)$, compared to the RPA for metallic densities.
This is shown in Fig.~\ref{fig:roton}a) for $r_s=4$ at the electronic Fermi temperature, $\theta=1$. Specifically, the solid green curve depicts the exact $\omega(q)$
based on the full PIMC results for the dynamic local field correction $G(\mathbf{q},\omega)$, which are down-shifted compared to the RPA mean-field description that is depicted as the dash-dotted black curve.
In addition, the dashed blue curve in Fig.~\ref{fig:roton}a) shows the \emph{static approximation}~\cite{dornheim_dynamic}, where the dynamic local field correction in Eq.~(\ref{eq:LFC}) is approximated by its exact static limit, i.e., $G(\mathbf{q},\omega)\equiv G(\mathbf{q})$. The results are in excellent agreement with the full reference PIMC data. In other words, the combination of a dynamic description on the level of the RPA [where the frequency-dependence comes from the dynamic Lindhard function $\chi_0(\mathbf{q},\omega)$] with a static (frequency averaged) treatment of electron-electron XC effects via $G(\mathbf{q})$ provides a highly accurate description of the DSF of the UEG at metallic densities. Note that this is consistent with previous observations in the ground-state limit~\cite{fiolhais2008primer,dynamic1}.

Let us postpone the discussion of the dotted red curve for now and proceed with Fig.~\ref{fig:roton}b), where we show the same analysis for a lower density with $r_s=10$. Owing to the role of the Wigner-Seitz radius as the quantum coupling parameter of the UEG~\cite{quantum_theory,Ott2018}, these conditions are located at the margin of the strongly coupled electron liquid regime where electron-electron correlation effects play the dominant role~\cite{Tolias_JCP_2021,dornheim_electron_liquid}. In this case, the exact PIMC reference data for $\omega(q)$ exhibit a nonmonotonous behaviour, with a minimum at intermediate wave numbers, $q\sim2q_\textnormal{F}$. In fact, this phenomenologically resembles the \emph{roton feature} that is well known from quantum liquids such as ultracold helium~\cite{Ferre_PRB_2016,Godfrin2012,Dornheim_SciRep_2022,Nava_PRB_2013,Skold_PRL_1976,Skold_Trans_1980,griffin1993excitations,Trigger}. It is worth noting that a phonon-roton spectrum is an unambiguous feature of collective excitations in classical liquids and supercritical fluids\,\cite{Trigger,Kalman_2010}. This  includes not only model liquids (hard-sphere~\cite{Bryk_HS_2017}, Lennard-Jones~\cite{Kryuchkov_SciRep_2019}, Yukawa~\cite{Khrapak_PRE_2020}, inverse power law~\cite{Khrapak_SciRep_2017}) but also real liquids (water~\cite{Pontecorvo_PRE_2005,Cunsolo_PRB_2012}, elemental noble liquids~\cite{Cunsolo_PRL_1998}, liquid alkali metals~\cite{Giordano_PRB_2009}), for which the phonon-roton structure has been predicted by \emph{ab initio} molecular dynamics simulations and observed by inelastic X-ray and neutron scattering. Nevertheless, despite some numerical hints~\cite{Donko_Review_2008}, a plasmon-roton spectrum has not been univocally observed in MD simulation studies of the classical OCP~\cite{Hamaguchi_Ohta_2001,Choi_PRE_2019}, which is characterized by the onset of a negative long-wavelength limit dispersion, $d\omega(q)/dq<0$ as $q\to0$, at intermediate coupling $\Gamma\sim10$~\cite{Hansen_PRL_1974,Hansen_Letter_1981,Mithen_AIP_2012,Korolov_CPP_2015,Khrapak_POP_2016}. It is interesting that both quantum and classical cases have been elucidated by the onset of spatial structure~\cite{Kalman_2010}: through the form of the Feynman ansatz~\cite{Feynman_ansatz} [$S(\mathbf{q})$ connection] for quantum liquids and through the quasi-localized charge approximation~\cite{Kalman_Golden_QLCA_1990} [$g(\mathbf{r})$ connection] or the static local field corrected dielectric formalism~\cite{Tolias_PoPbrief_2021} [$S(\mathbf{q})$ connection] for classical liquids. Yet, no signatures of spatial structure can be found in either the static structure factor $S(\mathbf{q})$ or the pair correlation function $g(\mathbf{r})$ for the UEG at the conditions of Fig.~\ref{fig:roton}b).

To explain the mysterious roton feature in the UEG, Dornheim \emph{et al.}~\cite{Dornheim_Nature_2022} have recently introduced the \emph{pair alignment model} (PA) that interprets the minimum in $\omega(q)$ in terms of the relative length scales upon applying an external harmonic perturbation. In particular, the fluctuation-dissipation theorem [Eq.~(\ref{eq:FDT})] directly implies that the behaviour of $S(\mathbf{q},\omega)$ can be fully explained by understanding the response of any given system to a monochromatic external perturbation of the same wave vector $\mathbf{q}$ and frequency $\omega$. In the vicinity of the \emph{roton minimum} in $\omega(q)$ around $q\sim2q_\textnormal{F}$, the wavelength of the cosinuoidal perturbation is comparable to the average inter-particle distance, $\lambda_q=2\pi/q\sim d$. Therefore, the alignment of the electrons to the respective potential minima is associated with a reduction in the interaction energy $W$. 

The PA model introduced in Ref.~\cite{Dornheim_Nature_2022} postulates that the XC induced down-shift of the true $\omega(q)$ compared to the RPA is due to a deficiency of the latter to capture the true change in the interaction energy $W$. The latter can be accurately quantified using a suitable effective potential~\cite{Kukkonen_PRB_1979,Kukkonen_PRB_2021,Dornheim_JCP_2022}, and the difference between the true change in the interaction and the respective RPA result is indicated by the vertical red arrows labeled as $\Delta W_\textnormal{xc}$ in Fig.~\ref{fig:roton}; the corresponding dotted red curves have been computed as $\omega_\textnormal{PA}(q)=\omega_\textnormal{RPA}(q)+\Delta W_\textnormal{xc}(q)$. For the metallic density of $r_s=4$, the PA model nicely captures the true downshift of $\omega(q)$, and, therefore, is directly relevant for the modelling of XRTS experiments of WDM~\cite{Fortmann_PRE_2010}.
For the strongly coupled case of $r_s=10$, the PA model follows the dashed blue \emph{static approximation} curve rather than the actual PIMC results and, consequently, underestimates the true depth of the \emph{roton minimum} at these conditions.
This is not a coincidence, and it can be attributed to the construction of the PA model as an average energy shift. Empirically, the \emph{static approximation} exhibits the same behavior and merges the sharp roton peak at low $\omega$ with the shoulder at $\omega_\textnormal{RPA}(q)$; see Ref.~\cite{Dornheim_Nature_2022} for a more detailed explanation.

We note that the pair alignment that causes the XC induced down-shift of $\omega(q)$ is the same mechanism that has been evoked to explain the peak in the static linear density response function $\chi(\mathbf{q})$ in Fig.~\ref{fig:LRT_N34_rs4_theta1_3D} above. In fact, $S(\mathbf{q},\omega)$ and $\chi(\mathbf{q})$ are related via the inverse-frequency sum-rule, which states that 
\begin{eqnarray}\label{eq:inverse}
M_{-1}^{S}= - \frac{\chi(\mathbf{q})}{2n}\ ;
\end{eqnarray}
here $M_{-1}^{S}$ denotes the inverse frequency moment of the DSF defined in Eq.~(\ref{eq:moments}).
It is evident from Eq.~(\ref{eq:inverse}) that a shift of spectral weight in $S(\mathbf{q},\omega)$ to smaller frequencies $\omega$ is associated with an increase of the magnitude of $\chi(\mathbf{q})$.
This explains the emergence of an increasingly sharp peak in $\chi(\mathbf{q})$ around $q=2q_\textnormal{F}$ that has been reported in previous studies of the electron liquid both at finite temperature~\cite{dornheim_electron_liquid,Tolias_JCP_2021} and in the ground state~\cite{Tozzini_1996}.

\begin{figure}\centering
\includegraphics[width=0.5\textwidth]{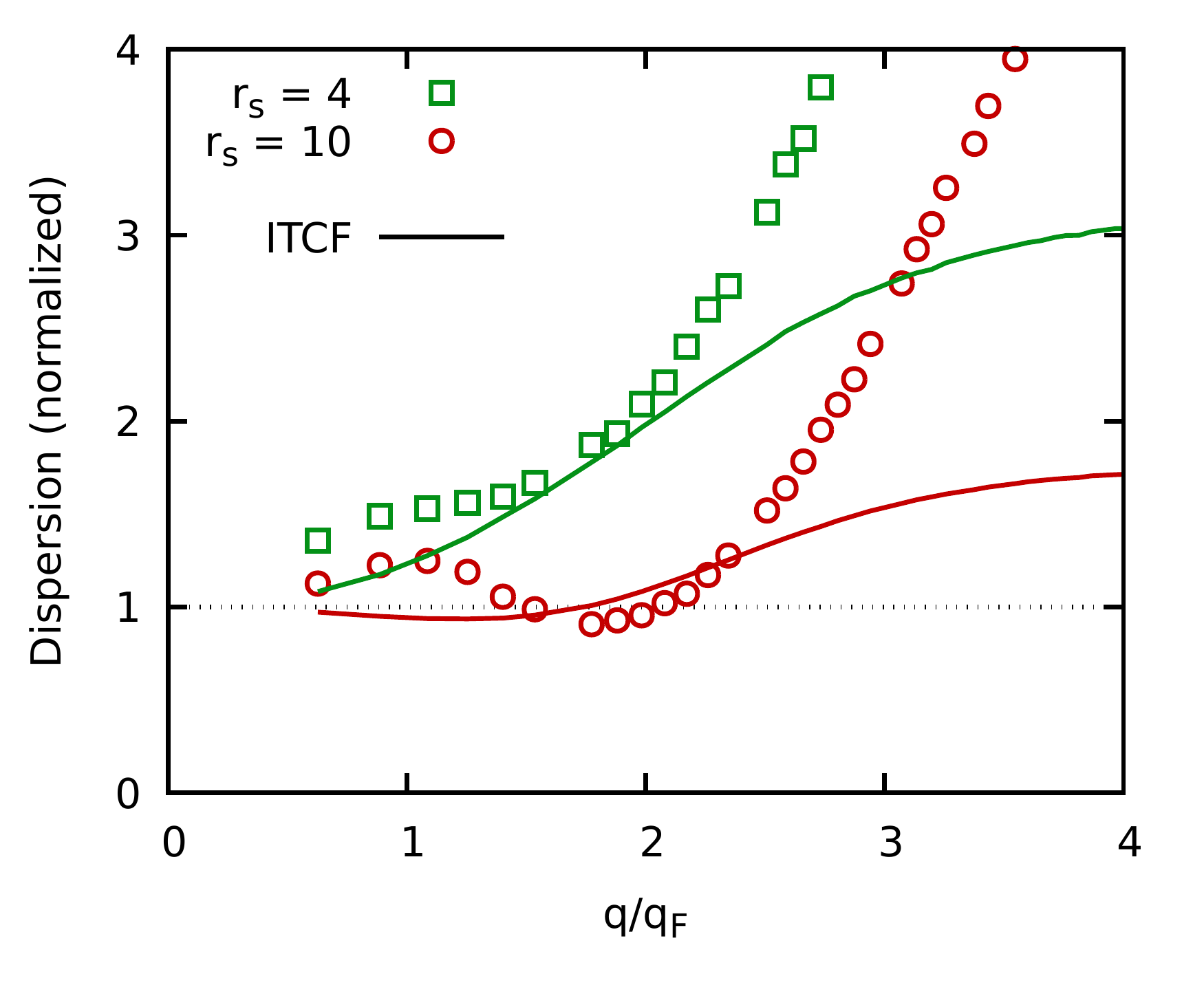}
\caption{\label{fig:dispersion}
Dispersion of dynamic density fluctuations in the UEG at $\theta=1$ for $r_s=4$ (green) and $r_s=10$ (red). Symbols: PIMC-based results for the position of the maximum in the DSF $\omega(q)$, cf.~Fig.~\ref{fig:roton}; the corresponding curves depict the relative-decay measure of the ITCF $\Delta F_{\beta/2}(\mathbf{q})$ defined in Eq.~(\ref{eq:decay_measure}). All data sets have been re-normalized with respect to the plasmon limit for $q\to0$.
Data taken from Refs.~\cite{dornheim_dynamic,Dornheim_PTR_2022}.
}
\end{figure} 

While having exact reference results for $S(\mathbf{q},\omega)$ gives one valuable insights into a number of physical effects, the required analytic continuation~\cite{JARRELL1996133} is not expected to be feasible for arbitrarily complex systems due to the ill-posed nature of the inverse Laplace transform. On the other hand, it is clear that, from a mathematical perspective, the ITCF $F(\mathbf{q},\tau)$ contains exactly the same information as the DSF, although in an unfamiliar representation. The task at hand is thus to understand the manifestation of a given physical effect of interest in the imaginary-time domain. 
For example, it is obvious from Eq.~(\ref{eq:symmetry}) that $F(\mathbf{q},\tau)$ is symmetric around $\tau=\beta/2$, and constitutes a direct and highly sensitive measure for the temperature of any system in thermodynamic equilibrium~\cite{Dornheim_T_2022,Dornheim_insight_2022}. This has important implications for the interpretation of XRTS experiments, and is discussed in Sec.~\ref{sec:experiment} below. 

In addition, understanding the physics encoded into $F(\mathbf{q},\tau)$~\cite{Dornheim_PTR_2022} is also directly helpful for the interpretation of simulation results. Very recently, Dornheim \emph{et al.}~\cite{Dornheim_insight_2022} have demonstrated that the \emph{roton feature} in the dispersion $\omega(q)$ of the DSF of the UEG manifests as a reduced decay with respect to $\tau$ of the ITCF for a given wave vector $\mathbf{q}$; see the spectral representation of $F(\mathbf{q},\tau)$ in Eq.~(\ref{eq:spectral_F}). In other words, the existence of energetically low-lying density excitations manifest as the stability of electron-electron correlations along the diffusion through the imaginary time $\tau$; this can be quantified using the relative decay measure $\Delta F_{\beta/2}(\mathbf{q},\tau)$ defined in Eq.~(\ref{eq:decay_measure}).

The results are shown in Fig.~\ref{fig:dispersion} for $\theta=1$ with green and red data sets corresponding to $r_s=4$ and $r_s=10$, respectively. More specifically, the symbols show the PIMC reference data for $\omega(q)$ that are also depicted in Fig.~\ref{fig:roton}, and the corresponding solid lines the relative decay measure $\Delta F_{\beta/2}(\mathbf{q},\tau)$ that has been directly computed from the ITCF without the need for an analytic continuation. We note that all curves have been re-normalized with respect to their respective plasmon limit for $q\to0$, which is given by the usual plasma frequency $\omega_\textnormal{p}$ in the case of $\omega(q)$. In addition, all curves exhibit a characteristic single-particle limit for $q\gg q_\textnormal{F}$. For the DSF, it is given by the well-known single-particle dispersion $\omega(q)\sim q^2$. For the ITCF, one finds $\Delta F_{\beta/2}(q\to\infty,\tau)=1$, which means that the corresponding re-normalized curves  attain a constant value in the short wavelength limit. This is due to the increasingly steep decay of $F(\mathbf{q},\tau)$ with $0\leq\tau\leq\beta/2$ in this regime, which can be explained with a simple semi-analytical Gaussian imaginary-time diffusion model; see Ref.~\cite{Dornheim_PTR_2022} for an extensive discussion of the physical origin of the $\tau$-dependence of the ITCF.

For intermediate $q$, we find a close correspondence between the respective $\omega(q)$ and $\Delta F_{\beta/2}(\mathbf{q},\tau)$ for the same value of the density parameter $r_s$. Indeed, the relative decay measure even becomes slightly negative for $r_s=10$. We note that the phenomenological similarity between the \emph{roton minimum} in the DSF and the reduced $\tau$-decay of the ITCF becomes even more apparent at $r_s=20$, where the alignment of pairs of electrons~\cite{Dornheim_Nature_2022} has a more pronounced impact on the physical observables of the system.

Despite these works, we emphasize that the thorough development of a dynamic quantum many-body theory in the imaginary-time domain as of yet remains in its infancy. Future works may start with the investigation of other systems such as quantum liquids~\cite{Dornheim_SciRep_2022,Ferre_PRB_2016,Nava_PRB_2013,Godfrin2012}, and of course real WDM applications that include both electrons and ions. It is evident from Eq.~(\ref{eq:Laplace}) that every theoretical model such as the Chihara decomposition~\cite{Chihara_1987,Gregori_PRE_2003,kraus_xrts}, but also TDDFT simulations can be straightforwardly translated into a description of the ITCF. At the same time, we also stress that working in the $\tau$-domain might allow for the development of entirely new concepts that naturally emerge from Feynman's imaginary-time path-integral representation of statistical mechanics, but may not have an obvious analogue in the traditional frequency-domain.

\subsubsection{Warm dense matter\label{sec:dynamic_WDM}}

\begin{figure}\centering
\includegraphics[width=0.5\textwidth]{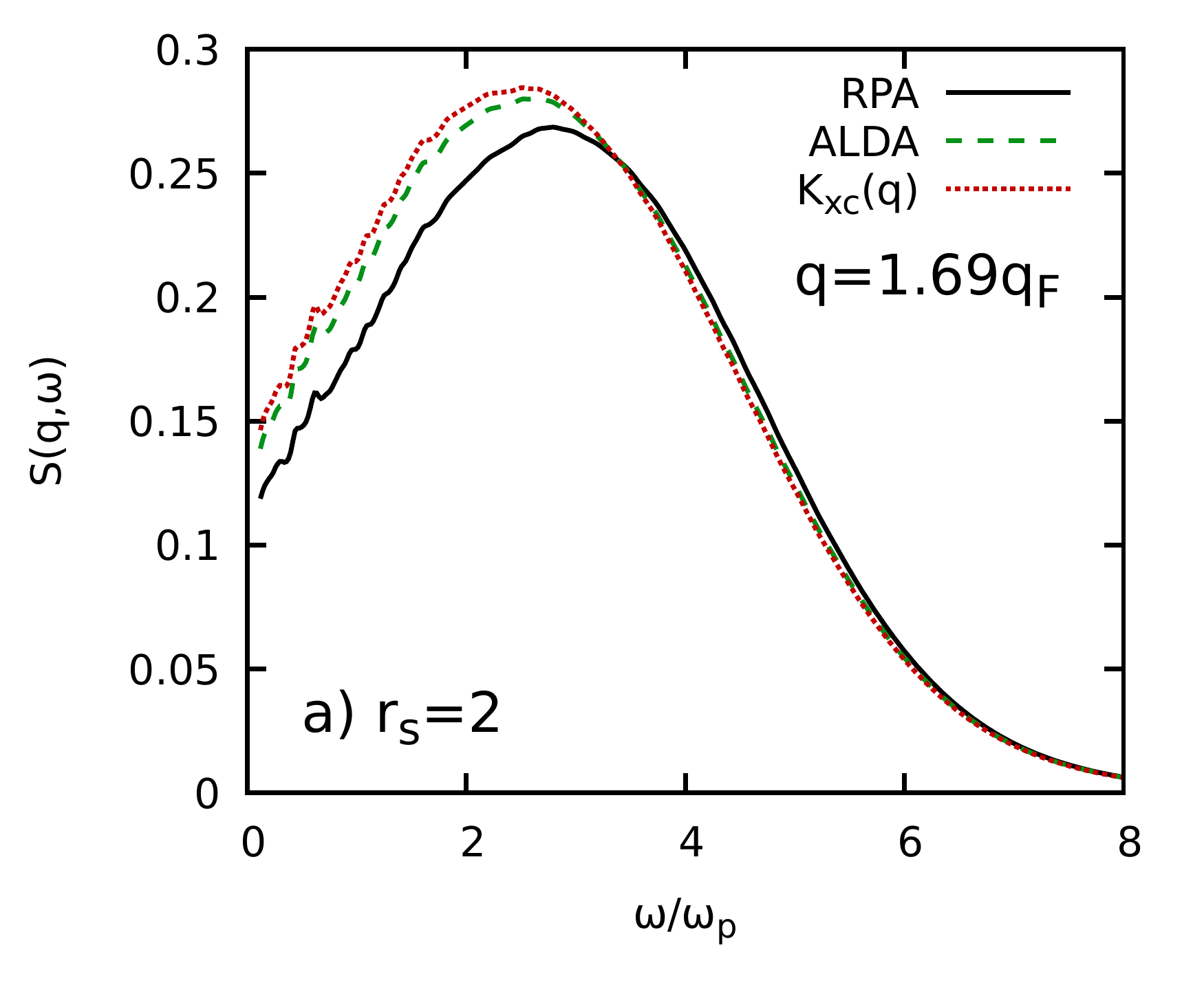}\\\vspace*{-1cm}\hspace*{0.02\textwidth}\includegraphics[width=0.475\textwidth]{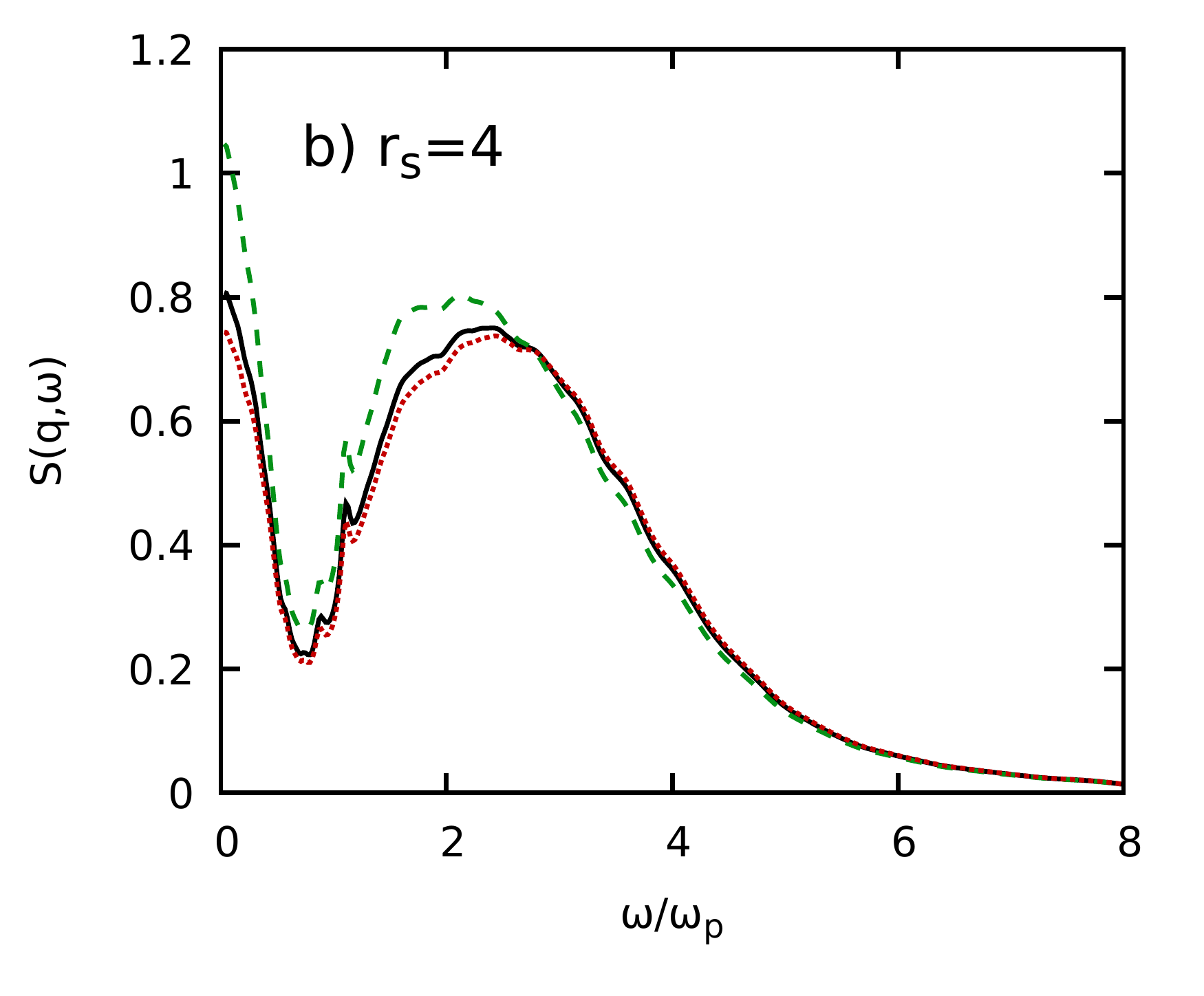}
\caption{\label{fig:DSF_hydrogen}
LR-TDDFT results for hydrogen at the electronic Fermi temperature $\theta=1$ for a) $r_s=2$ and b) $r_s=4$ for a wave vector of $q=1.69q_\textnormal{F}$. Solid black: RPA, i.e., setting $K_\textnormal{xc}(\mathbf{q},\omega)\equiv 0$ in Eq.~(\ref{eq:kernel}); dashed green: adiabatic LDA (ALDA), cf.~Eq.~(\ref{eq:CSR}); dotted red: \emph{static approximation}, $K_\textnormal{xc}(\mathbf{q},\omega)\equiv K_\textnormal{xc}(\mathbf{q})$ using exact PIMC data for the XC kernel in the static limit. Adapted from Ref.~\cite{Bohme_PRL_2022}.
}
\end{figure}

Let us next extend our analysis of the dynamic electronic density response of WDM beyond the UEG model. As we have elaborated above, no exact method for the simulation of frequency-dependent properties of real WDM systems is currently available to our knowledge. Being motivated by the remarkable performance of the \emph{static approximation} $K_\textnormal{xc}(\mathbf{q},\omega)\equiv K_\textnormal{xc}(\mathbf{q})$ for the UEG for metallic densities $r_s\lesssim4$, we here explore the combination of LR-TDDFT (cf.~Sec.~\ref{sec:DFT} above) with a consistent static XC kernel for the example of warm dense hydrogen. The results are shown in Fig.~\ref{fig:DSF_hydrogen} for a) $r_s=2$ and b) $r_s=4$ at the electronic Fermi temperature, $\theta=1$. More specifically, all curves have been computed via Eq.~(\ref{eq:kernel}) using the KS-response function $\chi_\textnormal{S}(\mathbf{q},\omega)$ based on a set of KS-orbitals from an equilibrium DFT simulation with the LDA functional.

Let us start our analysis for $r_s=2$, where hydrogen is mostly ionized, as it has been explained in Sec.~\ref{sec:hydrogen_results} above. In this case, the depicted LR-TDDFT results closely resemble the DSF of the UEG~\cite{dornheim_dynamic}. In particular, the solid black curve has been computed by setting $K_\textnormal{xc}(\mathbf{q},\omega)\equiv0$ in Eq.~(\ref{eq:kernel}), which is commonly being referred to as RPA in the DFT literature~\cite{marques2012fundamentals}.  
Including the appropriate PIMC results for the static XC kernel $K_\textnormal{xc}(\mathbf{q})$ by B\"ohme \emph{et al.}~\cite{Bohme_PRL_2022} leads to the dotted red curve; it exhibits a similar XC induced red-shift as we have observed in the case of the UEG in the previous section. Finally, the dashed green curve has been obtained by replacing the true static kernel with the ALDA, which is based on the $q\to0$ limit of the UEG at the same parameters, cf.~Eq.~(\ref{eq:CSR}). Evidently, the ALDA is in good agreement with the dotted red curve. This is expected, as the electrons of hydrogen basically behave like a UEG at these conditions due to the high degree of ionization~\cite{Bohme_PRL_2022,Militzer_PRE_2001}.

In stark contrast, the electrons are strongly influenced by the ions at $r_s=4$, which is shown in Fig.~\ref{fig:DSF_hydrogen}b). In particular, we find a diffusive peak around $\omega=0$, which is a direct consequence of the partial localization of the electrons around the protons. For these parameters, the RPA and the \emph{static approximation} are in close agreement, as the exact PIMC results for the static XC kernel nearly vanish, cf.~Fig.~\ref{fig:H_kernel}. The corresponding ALDA kernel fails to capture this trend, and the dashed green line substantially deviates from the dotted red reference curve. In other words, using the ALDA kernel that is based on the UEG in combination with the dynamic KS-response function $\chi_\textnormal{S}(\mathbf{q},\omega)$ that has been computed from the KS-orbitals for a specific model system leads to an actual deterioration of the quality of the results compared to the RPA. Therefore, we view the construction or utilization of universal model kernels as more or less futile, since a good kernel must 1) be consistent with the employed XC functional that has been used to compute $\chi_\textnormal{S}(\mathbf{q},\omega)$ and 2) depend on the physical behavior of the given system of interest.

A promising route is given by the recent framework by Moldabekov \emph{et al.}~\cite{Moldabekov_PRL_2022}, which allows one to compute the appropriate static XC kernel for any given system, and for arbitrary XC functionals across Jacob's Ladder~\cite{Perdew_AIP_2001} without the need for an explicit evaluation of functional derivatives. The accuracy of a corresponding LR-TDDFT calculation will then hinge on the \emph{static approximation}, which has to be studied in more detail in future works. In this regard, we mention the advent of direct PIMC simulations of real WDM systems without the fixed-node approximation~\cite{Bohme_PRL_2022,Bohme_PRE_2022}, which will give us the first exact results for the ITCF $F(\mathbf{q},\tau)$ of hydrogen in the near future. In addition to being interesting in their own right, such results will constitute an unassailable benchmark for the development of TDDFT simulations of WDM, and will guide the development of improved approximations that can be applied to more complex systems that are beyond the scope of PIMC.

\begin{figure}\centering
\includegraphics[width=0.5\textwidth]{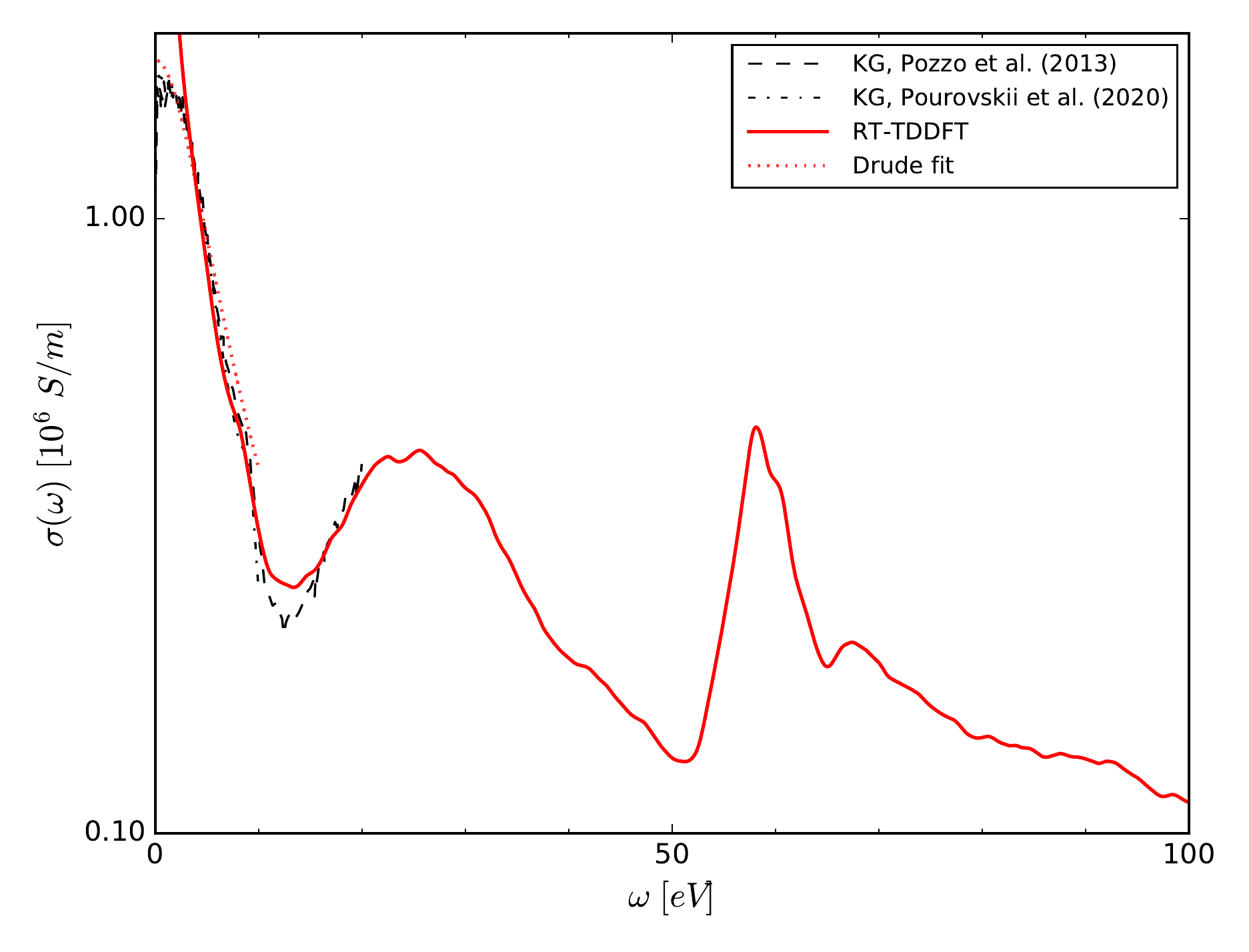}
\caption{\label{fig:iron}
Frequency-dependent electrical conductivity of iron under Earth-core conditions ($T\sim0.55$~eV, $r_{s} \sim 2.2$) from RT-TDDFT calculations (red curve). This is compared with previous works (black) using the KG formula~\cite{PhysRevB.87.014110,pourovskii2020electronic}. The Drude fit to RT-TDDFT data is indicated by the red dotted line. }
\end{figure}

Based on the methodology introduced in section \ref{sec:RTTDDFT}, RT-TDDFT is a powerful method for evaluating experimentally relevant features such as the DSF, and the electrical conductivity (in the optical limit $\bf q \rightarrow 0$). 
Fig.~\ref{fig:iron} shows the frequency-dependent electrical conductivity~\cite{ramakrishna2022electrical} of iron at earth-core conditions ($P\sim320$~GPa or $r_{s} \sim 2.2$, $T\sim0.55$~eV).  The red curve shows the tensor averaged result of several ionic configurations of size N=16 with an energy resolution $\Delta \omega \sim$0.11~eV which is proportional to the inverse of the total propagation time ($t\sim1500 $~a.u.). 
We compare our calculations with prior results (black dashed and dotted curves) obtained  using the KG formula based on static DFT (up to 20~eV)~\cite{PhysRevB.87.014110,pourovskii2020electronic}. In general, KG results are affected by finite-size effects and are highly dependent on the density and location of the KS eigenvalues~\cite{PhysRevB.84.054203,PhysRevB.85.024102}. The red dotted line shows the Drude fit to RT-TDDFT data to obtain the DC conductivity which is in the range observed in previous \emph{ab-initio}~\cite{PhysRevB.87.014110,pourovskii2020electronic,de2012electrical} methods.

\begin{figure}\centering
    \includegraphics{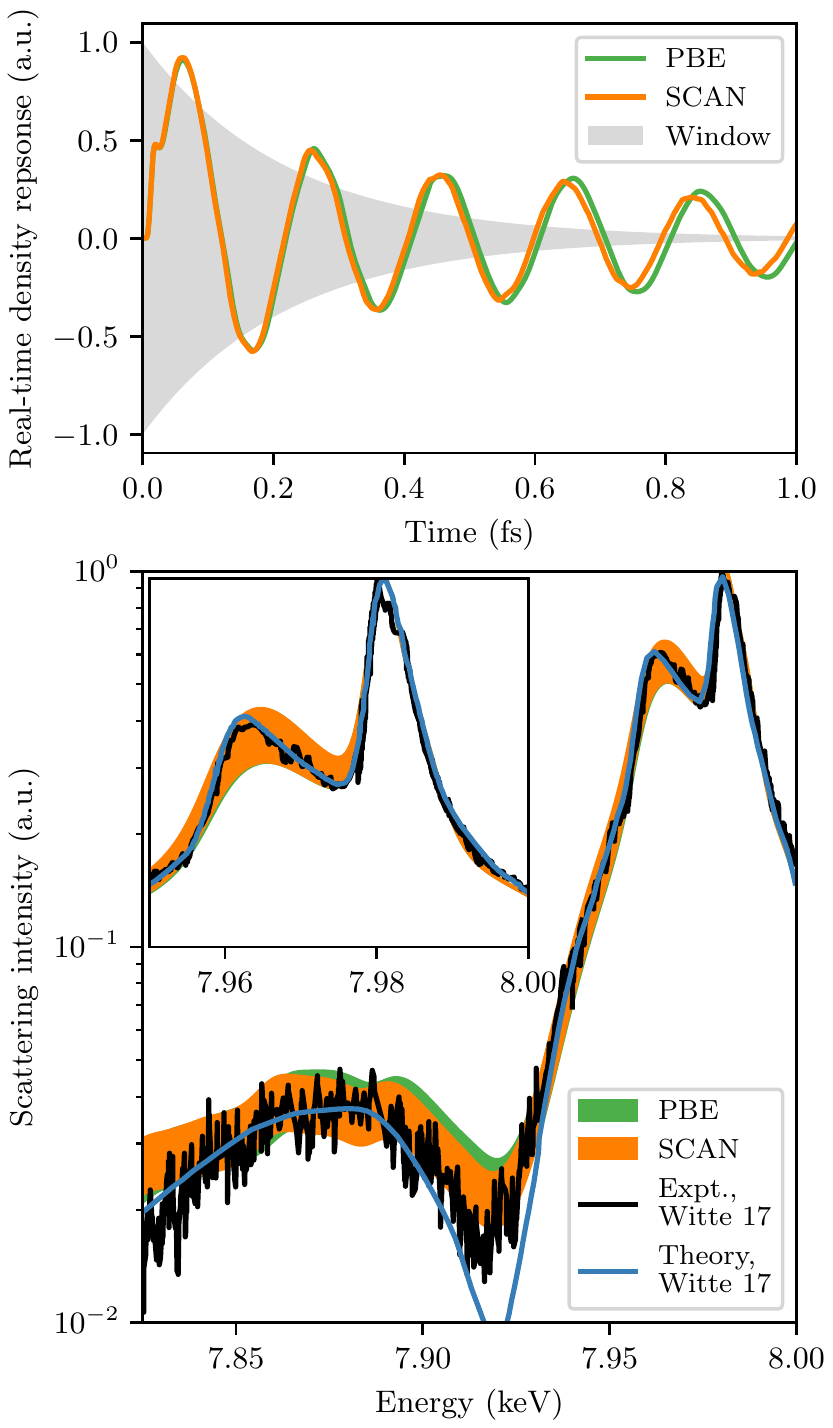}
    \caption{\label{fig:al_dsf_tddft} (Top) The RT-TDDFT density response to a perturbation of the form $v_0\exp(i{\bf q}\cdot {\bf r}) f(t)$, where ${\bf q}=0.67$ a.u. and $f(t)$ is a Gaussian envelope. Results for adiabatic PBE and SCAN are included and the envelope for a postprocessing window with a $3$-eV Lorentzian broadening is indicated in the background. (Bottom) The RT-TDDFT results postprocessed into simulated XRTS spectra through convolution with a 4-Gaussian least-squares fit to the instrument function in Ref.~\cite{Witte_PRL_2017}. We simulate a range of ionic structures, spanned by the shaded regions, by subtracting the elastic peak from the RT-TDDFT and adding a range of model ionic peaks. We compare to digitizations of both the experimental spectra and a subset of the theoretical data in Ref.~\cite{Witte_PRL_2017}. The theoretical data from Ref.~\cite{Witte_PRL_2017} are the bound-free edge from the Kubo-Greenwood optical conductivity and the rest of the spectrum from the Mermin approximation with collision frequencies determined using Kubo-Greenwood - both of which used HSE. Note that the SCAN results are on top of the PBE results, with the exception of a slight difference in the bound-free edge.}
\end{figure}

Next, in Fig.~\ref{fig:al_dsf_tddft} we consider a RT-TDDFT calculation of the DSF of isochorically heated aluminum ($T_e=0.3$ eV) at a momentum transfer of ${\bf q}=0.67$ a.u., compared to both experiment and a DFT-based theory for these same conditions first studied in Ref.~\cite{Witte_PRL_2017}.
The RT-TDDFT calculations make use of in-house extensions to the Vienna \emph{ab-initio} simulation package (VASP)~\cite{kresse1996efficient,kresse1996efficiency,kresse1999ultrasoft} first reported in Ref.~\cite{dynamic2} and we consider adiabatic PBE and SCAN exchange-correlation potentials.
The plasmon for both RT-TDDFT functionals are notably red-shifted relative to the experiment, similar to the LR-TDDFT ALDA results reported in Ref.~\cite{Dornheim_PRL_2020}, but inconsistent with LR-TDDFT results for cold solid aluminum reported in Ref.~\cite{cazzaniga2011dynamical}, which also indicated better agreement (and a small blue shift) between ALDA and experiment.
The DFT-based theory to which we compare from Ref.~\cite{Witte_PRL_2017} uses separate models for the bound-free edge (the ${\bf q}\rightarrow 0$ limit of the optical conductivity) and the plasmon (the Mermin approximation, with collision frequencies taken from Kubo-Greenwood calculations~\cite{plagemann2012dynamic}), but both are based on thermal DFT in a limit that is equivalent to an LR-TDDFT calculation in which $f_{Hxc}=0$ and both make use of the HSE exchange-correlation functional in the attendant ground-state calculations~\cite{krukau2006influence}.
To capture the bound-free edge we used an 11-electron projector augmented-wave pseudization~\cite{blochl1994projector} of the electron-ion interaction. Relative to a 3-electron pseudization, the additional spectral width adversely impacts the condition number of the linear system of equations in the time propagation scheme, making HSE impractical relative to simply verifying expected systematics for much cheaper PBE and SCAN calculations.
We expect that the only impact of using HSE would be to more strongly bind the 2p states and to shift the bound-free edge a bit further than SCAN, consistent with prior DFT-MD calculations~\cite{witte2018observations}. 
This is consistent with the systematics that are observable in switching between PBE and SCAN in RT-TDDFT, in which no change in the shape of the plasmon is observable.

The RT-TDDFT DSF for a range of plausible ionic structures are convolved with a fit to the reported instrument function and while the simulated scattering intensity is in reasonable agreement with experiment, the discrepancy between RT-TDDFT and the methods based strictly on DFT is notable.
The bound-free edge is approximately consistent with Kubo-Greenwood at ${\bf q} \rightarrow 0$, which suggests that this region of the DSF should be well treated within a ground-state framework.
But the plasmon is notably distorted relative to the \emph{ab-initio} formulation of the Mermin approximation.
We note that this has previously been observed~\cite{dynamic2} and more severe discrepancies between Kubo-Greenwood and RT-TDDFT have been observed in the context of simulations of bound-bound features in XRTS~\cite{baczewski2021predictions}.
In as far as RT-TDDFT is ostensibly a higher level of theory, this warrants further investigation to reach a better understanding of the relative strengths/weaknesses of the two approaches, as well as ultimate consistency.

\subsection{Nonlinear density response\label{sec:nonlinear_results}}

The concept of linear density response is ubiquitous throughout physics~\cite{nolting}, in general, and WDM description, in particular. From a theoretical perspective, neglecting all nonlinear terms e.g.~in Eq.~(\ref{eq:n_ind_re}) leads to substantial simplifications, which often render calculations feasible in the first place. Indeed, the assumption of a linear response is well justified for many applications~\cite{Dornheim_PRL_2020}, such as XRTS experiments with low to moderate intensities as they are employed for WDM diagnostics~\cite{siegfried_review}. At the same time, non-linear effects are known to play an important role in high-Z stopping power calculations~\cite{PhysRevB.37.9268,Echenique_PRB_1986,Nagy_PRA_1989} and have been suggested to be important in the construction of effective pair potentials~\cite{Rasolt_PRB_1975,Gravel,zhandos_cpp21}. To be more specific, the Bethe stopping power formula, together with its many corrections, remains the cornerstone of our understanding of electron ionization-excitation energy losses of charged heavy particles in matter~\cite{sigmund_stopping}. One of the primary limitations of the bare (uncorrected) Bethe formula is reflected on its origin from first-order quantum mechanical perturbation theory~\cite{Fano_review}. In fact, for high-Z particles, it can be expected that higher order terms of the Born series become significant~\cite{Ashley_PRB,Hill_PRA}. Translating from the language of quantum scattering theory to the language of density response theory, the stopping power of the high-Z particle is the force that it experiences from its own induced field in the medium~\cite{Akhiezer2}. Thus, it is straightforward to deduce that the next high-order correction will roughly be $\propto{Z}^3$ and will emerge from the quadratic density response~\cite{Echenique_PRB_1986,MeyerterVehn_PRA,PitarkeEPL}. This is the famous Barkas effect and corresponding Barkas correction that also accounts for the stopping power differences between equal mass projectiles of opposite charge sign~\cite{BarkasPRL,AndersenPRL}. 

In addition, it can be shown~\cite{Dornheim_JPSJ_2021} that nonlinear response functions are directly connected to higher-order correlation functions. Therefore, the experimental investigation of nonlinear observables might allow one to gain new insights into many-body effects in a given system. Finally, we note that the nonlinear density response has been shown~\cite{Dornheim_PRL_2020,Dornheim_PRR_2021}
to depend much more sensitively on system parameters like the temperature compared to the usually considered linear response, which might make them useful as a new tool for WDM diagnostics~\cite{Moldabekov_SciRep_2022}.

To our knowledge, the first rigorous analysis of the static nonlinear density response of WDM has been presented in Ref.~\cite{Dornheim_PRL_2020} based on exact PIMC simulations of a harmonically perturbed UEG, cf.~Eq.~(\ref{eq:Hamiltonian_modified}). The basic idea can be seen in Fig.~\ref{fig:LRT_N34_rs4_theta1_3D}d) above, where we show the induced density $\braket{\hat\rho_\mathbf{k}}_{q,A}$ [Eq.~(\ref{eq:rho})] as a function of the perturbation amplitude $A$. More specifically, the symbols show the raw PIMC results, and the dotted curves fit functions based on the expansion given in Eq.~(\ref{eq:rho1_fit}).
Note that the induced density has been divided by $A$, so that the data sets attain a constant value in the linear-response limit.
For the smallest depicted value of the wave number $q$ (blue diamonds), the density response of the UEG is comparably small, and nonlinear effects are basically negligible over the entire depicted $A$-range.
In stark contrast, the induced density for $q=3q_\textnormal{min}$ is substantially larger in magnitude, and visibly deviates from the constant LRT limit for $A\gtrsim 0.02$. This is reflected by a large value in the cubic coefficient in Eq.~(\ref{eq:rho1_fit}), which directly corresponds to the cubic density response at the first harmonic, i.e., at the third power of the perturbation amplitude but exactly at the perturbation wave vector (and, in general, frequency). Detailed investigations of the latter have been presented in Refs.~\cite{Dornheim_PRL_2020,Dornheim_PRR_2021}.

While being formally exact, the approach based on the direct PIMC simulation of the harmonically perturbed system is computationally very demanding as it requires one to carry out a set of independent simulations over a sufficient interval of perturbation amplitudes $A$ to extract the nonlinear density response for a single combination of $r_s$, $\theta$, and $q$. A promising route to circumvent this issue has been suggested by Moldabekov \emph{et al.}~\cite{Moldabekov_JCTC_2022}, who have shown that the nonlinear density response can be studied very accurately based on KS-DFT simulations. In particular, this will allow one to transcend the current limitations of PIMC, and to study nonlinear effects in real materials.
From the perspective of PIMC itself, Dornheim \emph{et al.}~\cite{Dornheim_JCP_ITCF_2021} have presented relations between generalized nonlinear response functions and high-order imaginary-time correlation functions, which can be viewed as generalizations of the imaginary-time version of the fluctuation-dissipation theorem given in Eq.~(\ref{eq:chi_static}). In the present work, we restrict ourselves to quadratic orders in the perturbation strength $A$, and the corresponding relation between the quadratic density response function at the second harmonic and the respective imaginary-time three-body correlation function $F^{(2)}(\mathbf{q};\tau_1,\tau_2)$ is given by Eq.~(\ref{eq:Y_imaginary}).

\begin{figure}\centering\vspace*{-1cm}
\includegraphics[width=0.5\textwidth]{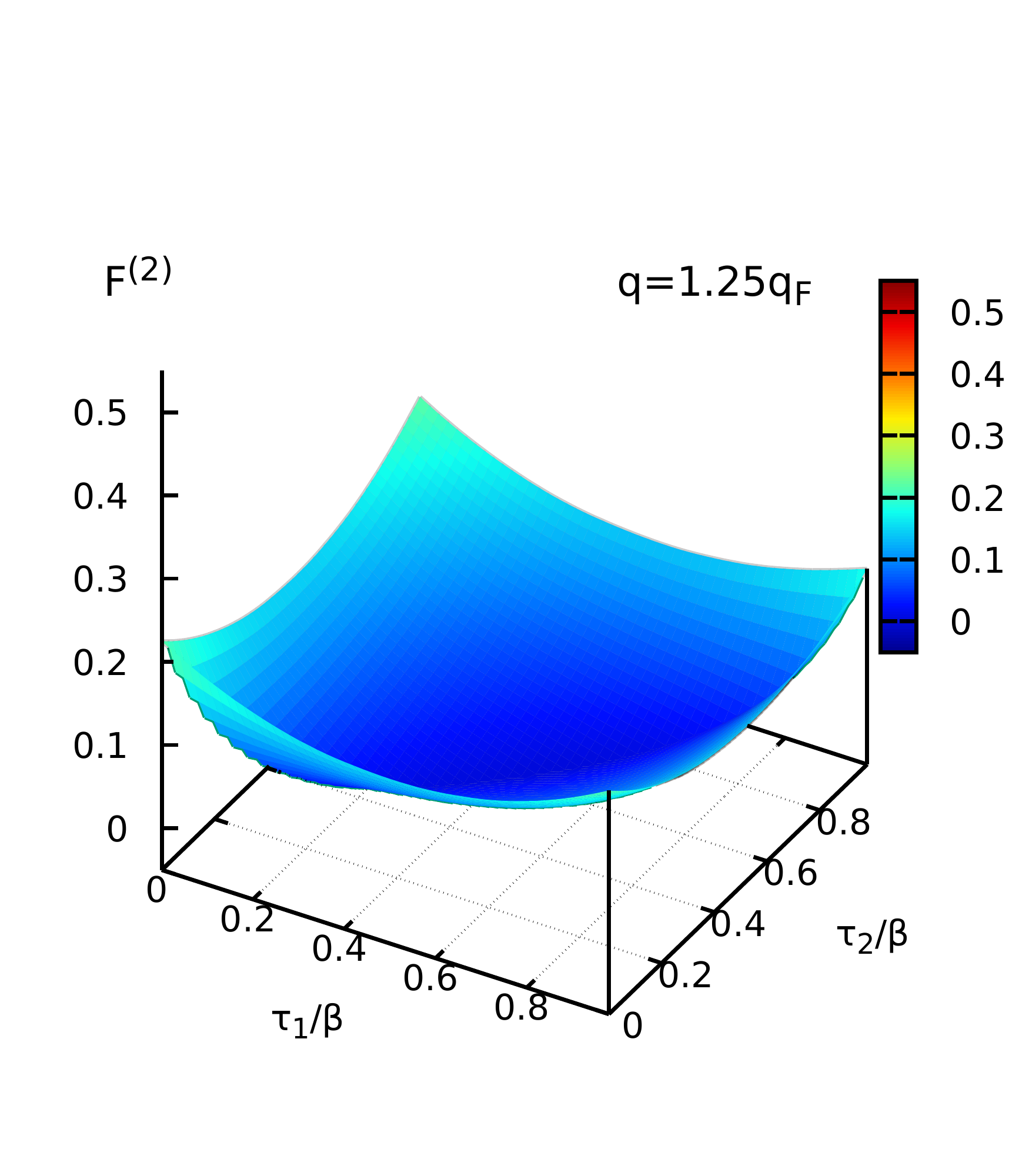}\\\vspace*{-2.5cm}\includegraphics[width=0.5\textwidth]{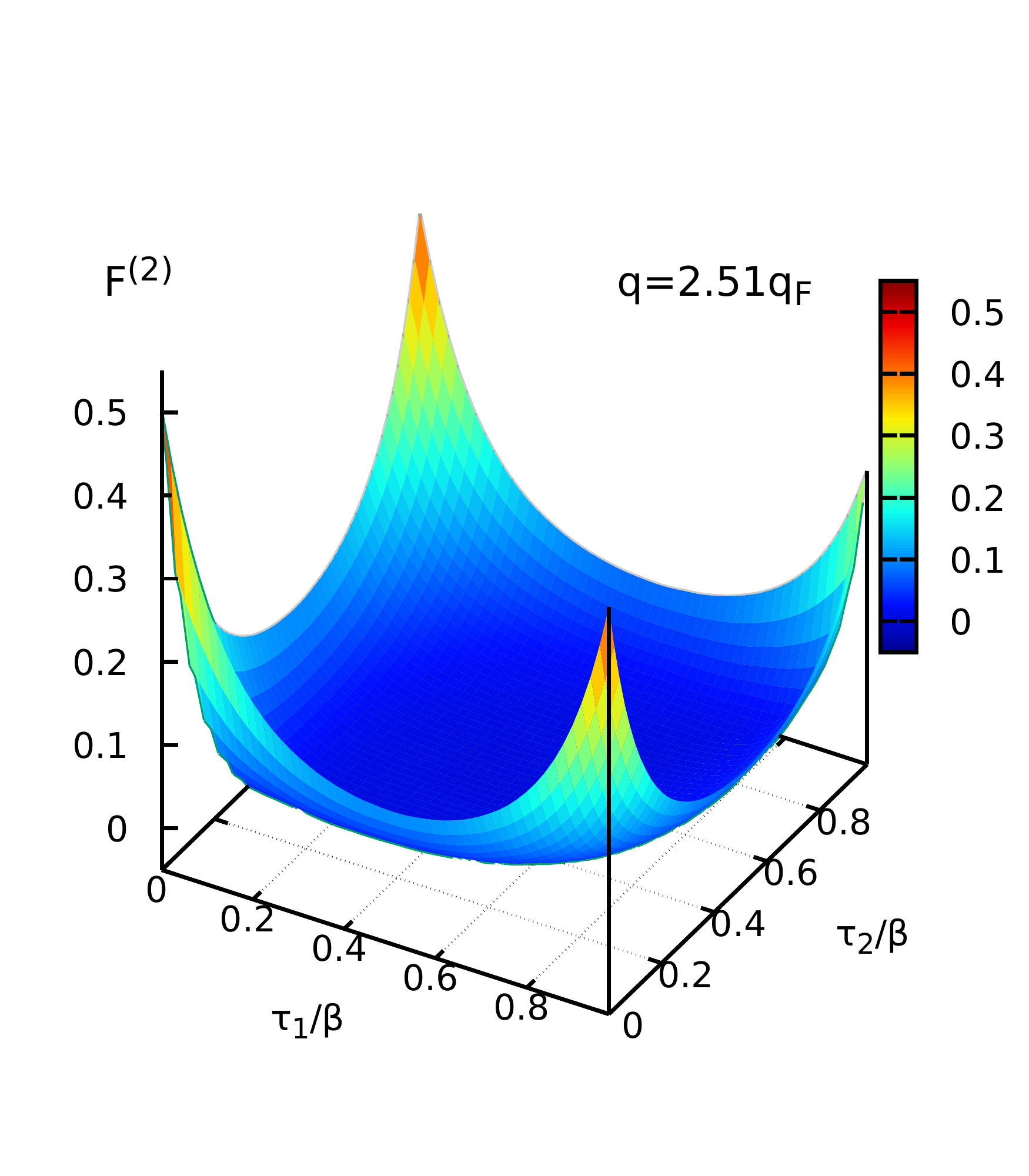}\\\vspace*{-1cm}
\caption{\label{fig:ITCF2}
\emph{Ab initio} PIMC results for the three-body ITCF $F^{(2)}(\mathbf{q};\tau_1,\tau_2)$ [Eq.~(\ref{eq:F2})] for $N=34$, $r_s=4$, and $\theta=1$ in the $\tau_1$-$\tau_2$-plane; a) $q=1.25q_\textnormal{F}$ and b) $q=2.51q_\textnormal{F}$.
}
\end{figure}

In Fig.~\ref{fig:ITCF2}, we show new \emph{ab initio} PIMC results for $F^{(2)}(\mathbf{q};\tau_1,\tau_2)$ in the $\tau_1$-$\tau_2$-plane for two different values of the wave number $q$ for the UEG at $r_s=4$ and $\theta=1$, i.e., for the same conditions as in Fig.~\ref{fig:LRT_N34_rs4_theta1_3D} above.
Overall, we find that $F^{(2)}(\mathbf{q};\tau_1,\tau_2)$ exhibits a qualitatively similar behavior to the two-body ITCF $F(\mathbf{q},\tau)$. Specifically, we observe an increasingly steep decay with respect to both $\tau_1$ and $\tau_2$ with increasing $q$, whereas the three-body ITCF becomes flatter in the long-wavelength limit. At the same time, we note that the physical behavior of $F^{(2)}(\mathbf{q};\tau_1,\tau_2)$ remains very poorly understood, and its improved understanding along similar lines as it has been achieved for $F(\mathbf{q},\tau)$ in Refs.~\cite{Dornheim_insight_2022,Dornheim_PTR_2022} remains an important task for future works.

\begin{figure}\centering
\includegraphics[width=0.5\textwidth]{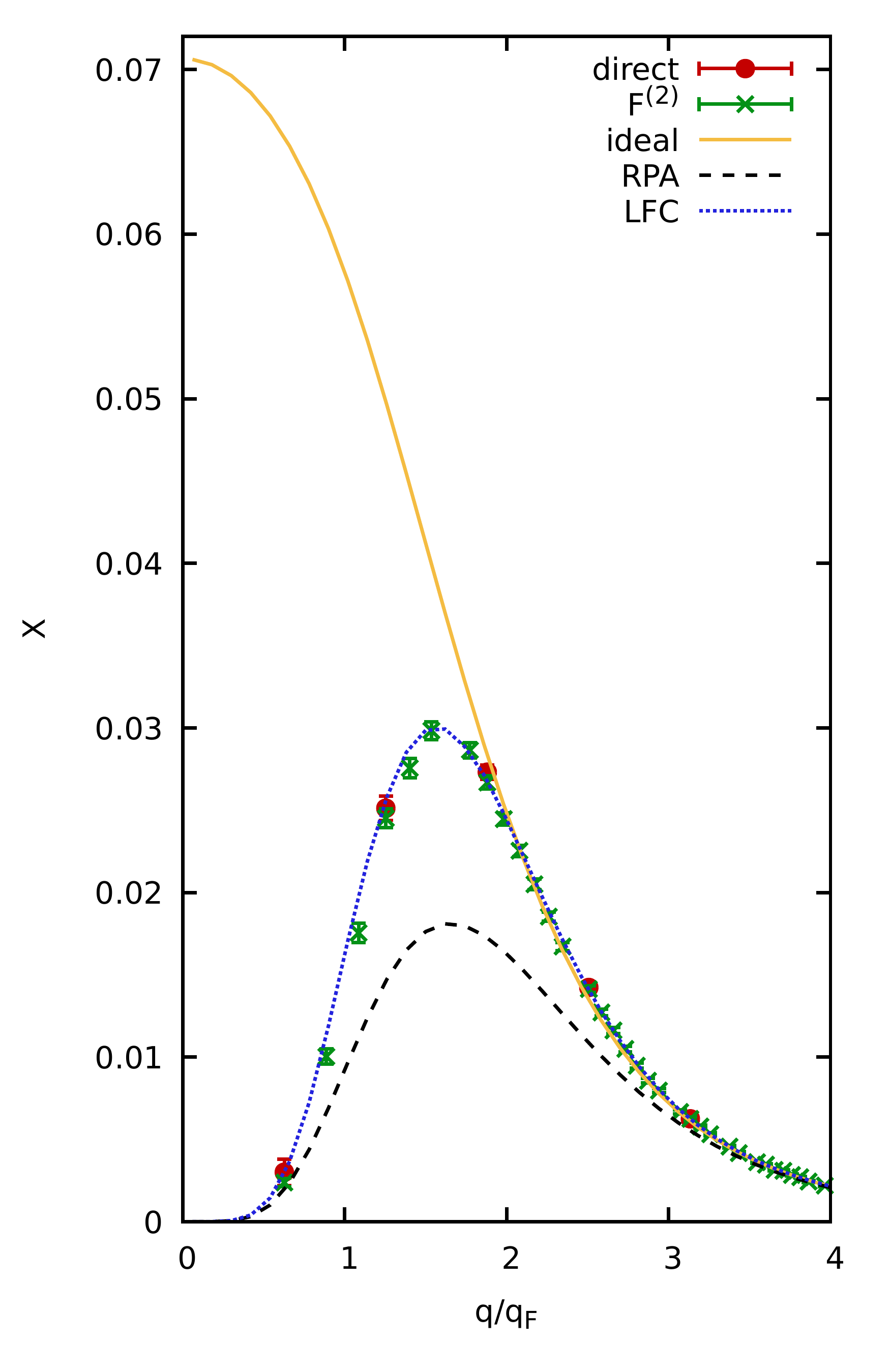}
\caption{\label{fig:quadratic}
Quadratic density response at the second harmonic $\chi^{(2,2)}(\mathbf{q},0)$ for $N=34$, $r_s=4$, and $\theta=1$. Red circles: estimation from PIMC simulations of the harmonically perturbed system [cf.~Eq.~(\ref{eq:Hamiltonian_modified})] via Eq.~(\ref{eq:rho2_fit}); green crosses: integration of the three-body ITCF $F^{(2)}(\mathbf{q},\tau_1,\tau_2)$ shown in Fig.~\ref{fig:ITCF2} via Eq.~(\ref{eq:Y_imaginary}); solid yellow: quadratic density response of the ideal Fermi gas, Eq.~(\ref{eq:Mikhailov2}); dashed black: RPA expression, Eq.~(\ref{eq:chi2_RPA}); dotted blue: incorporation of XC effects via the static local field correction $G(q)$ taken from Ref.~\cite{dornheim_ML} via Eq.~(\ref{eq:quadratic_LFC}).
}
\end{figure} 

In the present context, the main utility of $F^{(2)}(\mathbf{q};\tau_1,\tau_2)$ is given by its direct connection to the quadratic static density response function of the second harmonic, $\chi^{(2,2)}(\mathbf{q})$, via Eq.~(\ref{eq:Y_imaginary}), which is explored in Fig.~\ref{fig:quadratic} for the same conditions as in Fig.~\ref{fig:ITCF2}. In particular, the green crosses have been obtained from a single simulation of the unperturbed UEG in this way.
Evidently, $\chi^{(2,2)}(\mathbf{q})$ exhibits the opposite sign compared to the linear static density response function $\chi(\mathbf{q})$ shown in Fig.~\ref{fig:LRT_N34_rs4_theta1_3D}, which, together with the positive sign of the cubic response at the first harmonic, leads to a saturation of the true response compared to LRT for medium to large perturbation amplitudes $A$.
In addition, we find that the PIMC based evaluation of Eq.~(\ref{eq:Y_imaginary}) is in excellent agreement to the five red dots, which have been obtained from a multitude of PIMC simulations of the harmonically perturbed system by fitting Eq.~(\ref{eq:rho2_fit}) to the induced density at the second harmonic, i.e., $\braket{\hat\rho_{2\mathbf{q}}}_{q,A}$. In practice, the imaginary-time framework introduced in Ref.~\cite{Dornheim_JCP_ITCF_2021} thus gives superior results for the quadratic density response of the second harmonic for all relevant wave numbers with a fraction of the computation cost. For completeness, we note that similar relations also exist for the nonlinear interaction between multiple external perturbations, which has been explored in more detail in Ref.~\cite{Dornheim_CPP_2022}.

In addition to their direct value for the understanding of nonlinear effects in WDM, the availability of exact PIMC reference data is also pivotal to guide the development of new theoretical frameworks. 
In this regard, a highly useful result is given by the recursion formula for the quadratic density response of the ideal Fermi gas by Mikhailov~\cite{Mikhailov_Annalen,Mikhailov_PRL}, cf.~Eq.~(\ref{eq:Mikhailov2}). The results are shown as the solid yellow curve in Fig.~\ref{fig:quadratic} and exhibit the correct asymptotic behaviour for $q\gtrsim2q_\textnormal{F}$. Screening effects for small $q$, on the other hand, are not included.
In Ref.~\cite{Dornheim_PRR_2021}, five of us have introduced an RPA-like expression for $\chi^{(2,2)}(\mathbf{q})$--- and also for high-order response functions which are beyond the scope of the present overview --- that is truncated at the linear level with regard to screening effects, see Eq.~(\ref{eq:chi2_RPA}) above. The results have been included as the dashed black curve. They exhibit the correct trends in the limits of $q\to0$ and $q\to\infty$, but substantially underestimate the true quadratic response around its maximum at $q\sim2q_\textnormal{F}$. Including the static local field correction $G(q;r_s,\theta)$ evaluated from the neural-net representation from Ref.~\cite{dornheim_ML} via Eq.~(\ref{eq:chi2_LFC}) gives the dotted blue curve, which is in perfect agreement to the PIMC data sets over the entire $q$-range. This is a strong empirical verification for the small effect of the screening truncation employed in Eqs.~(\ref{eq:chi2_RPA}) and (\ref{eq:chi2_LFC}). Moreover, it confirms the more pronounced importance of electronic XC effects for the description of nonlinear effects, since the RPA curve is less accurate for the description of $\chi^{(2,2)}(\mathbf{q})$ compared to $\chi(\mathbf{q})$; this is consistent with previous investigations presented in Refs.~\cite{Dornheim_PRR_2021,Dornheim_JPSJ_2021}.

Future investigations of the nonlinear density response of the UEG will include the study of dynamic effects, which is possible in multiple ways. For example, the RPA and LFC-based expressions for the quadratic density response Eqs.~(\ref{eq:chi2_RPA}) and (\ref{eq:chi2_LFC}) remain the same for all frequencies. As a first step, one might therefore study dynamic nonlinear effects based on the \emph{static approximation}, which is straightforward given the easy availability of $G(\mathbf{q};r_s,\theta)$ based on different representations~\cite{dornheim_ML,Dornheim_PRB_nk_2021}. Moreover, these efforts can be improved by including the full, frequency-dependent results for $G(\mathbf{q},\omega)$ that are available for certain parameters of the UEG based on the analytic continuation~\cite{dornheim_dynamic} discussed in Sec.~\ref{sec:dynamic_UEG} above. Such an approach will allow one to gain both qualitative and quantitative insights into a number of interesting phenomena, such as double plasmons~\cite{huotari2008electron,Panholzer_PRL_2018,PhysRevLett.95.157401} and dynamic three-body correlations~\cite{SEDRAKIAN1998,Hata2021,Schmiedinghoff2022}.

In addition, upcoming direct PIMC simulations (i.e., without the fixed-node approximation) of real WDM systems such as hydrogen~\cite{Bohme_PRL_2022,Bohme_PRE_2022} will allow one to directly estimate $F^{(2)}(\mathbf{q};\tau_1,\tau_2)$, as well higher-order ITCFs that are connected to the cubic density response and so on~\cite{Dornheim_JCP_ITCF_2021}. First and foremost, this will facilitate investigations of the static nonlinear density response of WDM, without any assumptions or approximations. In particular, the presence of both electrons and ions leads to various cross terms, which deserve a closer examination. Finally, we envision the systematic development of the study of dynamic many-body effects in the imaginary-time domain similar to the efforts that have been developed in Refs.~\cite{Dornheim_T_2022,Dornheim_insight_2022,Dornheim_PTR_2022} for $F(\mathbf{q},\tau)$, and which have been touched upon throughout the present work.

\section{Interpretation of XRTS experiments\label{sec:experiment}}

\begin{figure*}\centering
\includegraphics[width=0.995\textwidth]{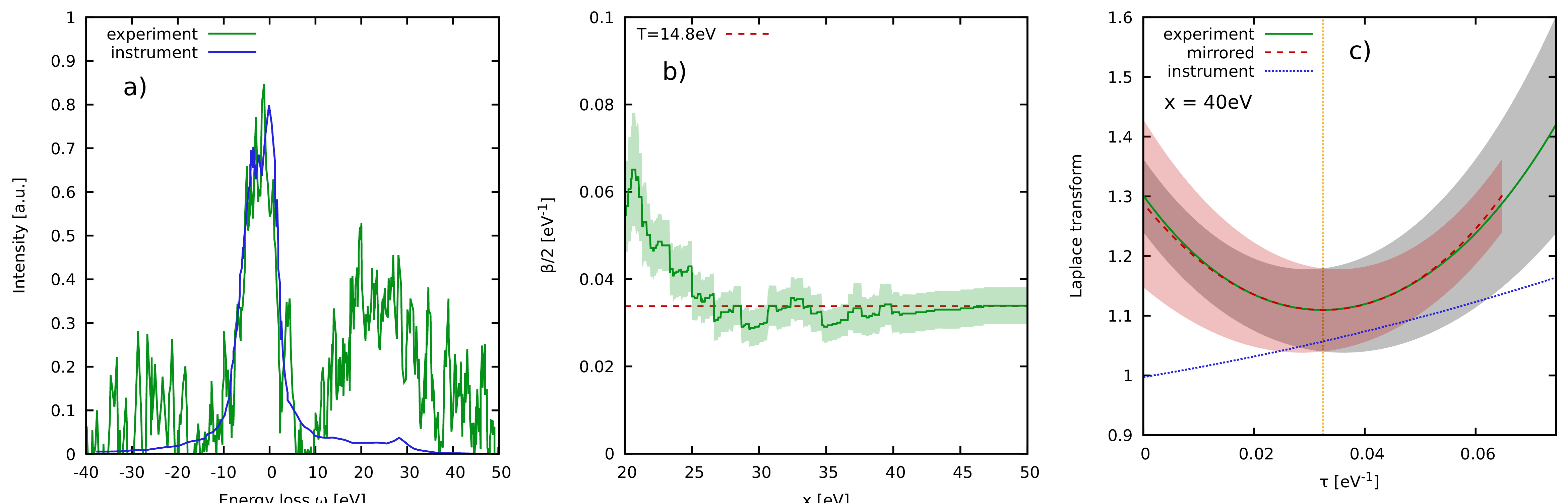}
\caption{\label{fig:Glenzer} Temperature diagnosis of the XRTS measurement of warm dense beryllium by Glenzer \emph{et al.}~\cite{Glenzer_PRL_2007}. a) XRTS signal (green) and combined source and instrument function (blue); b) convergence of the inferred temperature with the integration boundary $x$ [Eq.~(\ref{eq:F_truncated})] and the corresponding uncertainty interval due to the noise in the intensity $I(\mathbf{q},\omega)$; c) extracted ITCF for $x=40\,$eV (solid green); the shaded grey area illustrates the uncertainty in $F(\mathbf{q},\tau)$, and the dotted blue line shows the Laplace transform of the instrument function; the dotted red curve has been mirrored around $\tau=\beta/2$, which illustrates that the symmetry relation Eq.~(\ref{eq:symmetry}) is fulfilled to a high degree.
Taken from Ref.~\cite{Dornheim_T_follow_up}.
}
\end{figure*}

One of the most important practical applications of the electronic density response of WDM is given by the interpretation of XRTS experiments. In the traditional way, one constructs a theoretical model for the DSF $S(\mathbf{q},\omega)$ with a-priori unknown system variables such as the temperature $T$, the density $n$, or the charge state $Z$ being the free parameters~\cite{siegfried_review}. Convolving the DSF with the combined source and instrument function $R(\omega)$, cf.~Eq.~(\ref{eq:convolution}) above, then allows one to compare these results with the experimentally measured scattering intensity $I(\mathbf{q},\omega)$, and the best fit between model and experiment then gives one access to the system parameters of interest.  We note that the numerical deconvolution of Eq.~(\ref{eq:convolution}) is generally rendered highly unstable by the inevitable experimental noise. Therefore, XRTS does not give one direct access to the DSF, which could have been used e.g.~to extract the temperature via the detailed balance relation Eq.~(\ref{eq:detailed_balance})~\cite{DOPPNER2009182}.

To our knowledge, the most widely used theoretical model for the interpretation of XRTS experiments with WDM is given by the Chihara decomposition~\cite{Chihara_1987,Gregori_PRE_2003,siegfried_review,kraus_xrts}, which is based on a separation into \emph{bound} and \emph{free} electrons. Yet, such a chemical picture is expected to break down for significant parts of the WDM regime, where the electrons are partially localized around the ions; in this case, they are neither \emph{bound} nor \emph{free}, see also the discussion of warm dense hydrogen in Sec.~\ref{sec:hydrogen_results} above.

We note that overcoming these difficulties with more advanced simulation methods such as LR-TDDFT is also no trivial matter for numerous reasons. First and foremost, such \emph{ab initio} simulations are computationally very costly, and is unclear if it will be feasible to carry them out on a sufficient grid of free parameters to infer thermodynamic variables such as the temperature from XRTS measurements. In this regard, modern machine learning based interpolation methods~\cite{Fiedler_PRM_2022} will likely be helpful, but their practical application to this problem remains an important task for future research. Second, the accuracy of LR-TDDFT simulations is determined by the XC kernel, and the development of consistent useful approximations for real WDM systems is still in its infancy.

In any case, it is safe to say that previous approaches for the interpretation of XRTS measurements are based on approximations, and their accuracy remains largely unclear. Therefore, EOS measurements~\cite{Falk_HEDP_2012,Falk_PRL_2014} are actually model-dependent and do not necessarily constitute a reliable baseline either for practical applications or to guide the development of improved theoretical methods.

Very recently, Dornheim \emph{et al.}~\cite{Dornheim_T_2022,Dornheim_insight_2022,Dornheim_T_follow_up} have suggested that this problem can be circumvented by switching from the usual frequency-domain that is centered around $S(\mathbf{q},\omega)$ to the imaginary time $\tau$; here, the main property is the ITCF $F(\mathbf{q},\tau)$, which contains exactly the same information as the DSF. In particular, the convolution theorem Eq.~(\ref{eq:convolution_theorem}) implies that the deconvolution is trivial in the Laplace domain. In other words, XRTS gives us direct access to $F(\mathbf{q},\tau)$, which can subsequently be used to infer information about the given system without any intermediate models or simulations. For example, the symmetry relation Eq.~(\ref{eq:symmetry}) indicates that the ITCF is symmetric around $\tau=\beta/2=1/2T$. In this sense, one could say that XRTS measurements effectively function as a thermometer as the relation between $F(\mathbf{q},\tau)$ and $T$ is trivial.

In Fig.~\ref{fig:Glenzer}, we demonstrate this idea for the pioneering observation of plasmons in warm dense beryllium by Glenzer \emph{et al.}~\cite{Glenzer_PRL_2007}. 
Panel a) shows the actual experimental scattering intensity (green) and the corresponding combined source and instrument function (blue). In particular, the elastic feature around $\omega=0$ is dominated by the latter, and both the up- and down-shifted plasmons are nicely visible around $\omega\approx\pm25\,$eV. 
From a practical perspective, it is clear that the evaluation of the Laplace transform Eq.~(\ref{eq:Laplace}) --- and also the proof of the convolution theorem, Eq.~(\ref{eq:convolution_theorem}) --- requires an integration over an infinite frequency-range, whereas the available $\omega$-range
is clearly limited in any given experiment.
In practice, we thus define a symmetrically truncated Laplace transform,
\begin{eqnarray}
\mathcal{L}_x\left[S(\mathbf{q},\omega)\circledast R(\omega)\right] &=& \int_{-x}^x \textnormal{d}\omega\ e^{-\tau\omega} \left\{S(\mathbf{q},\omega)\circledast R(\omega)\right\}\ , \nonumber \\ & & \label{eq:Laplace_truncated}
\end{eqnarray}
with the corresponding truncated ITCF
\begin{eqnarray}\label{eq:F_truncated}
F_x(\mathbf{q},\tau) = \frac{\mathcal{L}_x\left[S(\mathbf{q},\omega)\circledast R(\omega)\right] }{\mathcal{L}\left[R(\omega)\right]}\ .
\end{eqnarray}
It is easy to see that the truncation error vanishes in the limit of large $x$
\begin{eqnarray}\label{eq:lim1}
\lim_{x\to\infty} F_x(\mathbf{q},\tau) = F(\mathbf{q},\tau)\ ,
\end{eqnarray}
and the convergence with $x$ has to be carefully checked.

This is demonstrated in Fig.~\ref{fig:Glenzer}b), where we plot the position of the minimum in $F_x(\mathbf{q},\tau)$ as a function of the integration boundary. We note that the shaded green area constitutes a measure of the respective uncertainty due to the experimental noise, and the utilized procedure is explained in detail in Ref.~\cite{Dornheim_T_follow_up}.
Evidently, we observe a convergence of the thus inferred temperature around $\omega\gtrsim30\,$eV, i.e., upon capturing the main plasmon peak in the integration range. As our final estimate for the temperature, we find $T=14.8\pm2\,$eV. This is comparably close to the nominal temperature of $T=12\,$eV, which was found in the original publication~\cite{Glenzer_PRL_2007} based on a simple Mermin model~\cite{Holl_EPJ_2004}.

Let us conclude this reassessment of the beryllium XRTS data by examining the actual result for $F(\mathbf{q},\tau)$; shown in Fig.~\ref{fig:Glenzer} for the converged value of $x=40\,$eV. Specifically, the green line shows the deconvolved ITCF (up to the unknown normalization factor), and the shaded grey area is the corresponding uncertainty interval due to the experimental noise. The vertical yellow line indicates the location of the minimum, which has been used to infer the temperature via Eq.~(\ref{eq:symmetry}). Moreover, the dotted blue line shows the Laplace transform of the instrument function $R(\omega)$, i.e., the denominator of the RHS of Eq.~(\ref{eq:convolution_theorem}). Without this correction, the minimum of the ITCF would be shifted to smaller values of $\tau$, leading to an overestimation of the temperature $T$~\cite{Dornheim_T_2022,Dornheim_T_follow_up}.
Finally, the dashed red curve has been obtained by mirroring the ITCF around $\tau=\beta/2$. Evidently, the red curve is in excellent agreement to the green curve, which means that the ITCF computed from the XRTS data shown in Fig.~\ref{fig:Glenzer}a) is really symmetric to a remarkable degree given the comparably high noise level in the latter. 
The stability of the method with respect to experimental noise is analyzed in detail in Ref.~\cite{Dornheim_T_follow_up}.

In summary, the formally exact method proposed in Ref.~\cite{Dornheim_T_2022} allows one to extract very accurate results for the temperature of arbitrarily complex systems in thermodynamic equilibrium from an XRTS measurement without any models, approximations or simulations. From a practical perspective, a decisive factor is given by the accurate quantification of the combined source and instrument function $R(\omega)$. This is relatively straightforward using source monitors at modern XFEL facilities~\cite{Tschentscher_2017,Glenzer_2016}, whereas the characterization of backlighter emission spectra~\cite{MacDonald_POP_2022} at facilities like NIF is more challenging. On the other hand, we note that the high temperatures typical for NIF implosion experiments make the influence of $R(\omega)$ on the extracted temperature less pronounced in any case.
Furthermore, it has been shown~\cite{Dornheim_T_follow_up} that the characteristic width of the instrument function also determines the minimum temperature that can be inferred from the ITCF method, which becomes apparent both from considerations of the respective DSF and the corresponding ITCF. Specifically, the detailed balance relation Eq.~(\ref{eq:detailed_balance}) of the DSF directly implies that the signal at negative frequencies is exponentially damped with increasing $\beta$, i.e., upon lowering the temperature. At some point, the signal for $\omega<0$ will vanish within the given experimental noise, and no inference of a temperature will be possible. In the $\tau$-domain, large $\beta$ means that one has to resolve the ITCF over a correspondingly large imaginary-time interval. With increasing $\tau$, the effect of the convolution becomes exponentially more pronounced. In this case, every small uncertainty in the determination of $R(\omega)$ will have increasing consequences for the evaluation of Eq.~(\ref{eq:convolution_theorem}) and thus make the localization of the minimum around $\tau=\beta/2$ less accurate.

Additional challenges for future works include the investigation of the effect of inhomogeneity onto the ITCF, that is particularly important for spatially extended systems such as the fuel capsule in implosion experiments at the NIF~\cite{Chapman_POP_2014,Poole_POP_2022}. Moreover, the symmetry relation Eq.~(\ref{eq:symmetry}) naturally only holds in thermodynamic equilibrium, and the influence even of small nonequilibrium effects~\cite{Chapman_HEDP_2012,Vorberger_PRE_2018} should be quantified in the future. As a final point, we mention the potential impact of the (small) frequency-dependence of the wave vector, $\mathbf{q}=\mathbf{q}(\omega)$, even though a first investigation in Ref.~\cite{Dornheim_T_follow_up} has led to the conclusion that the effect can likely be neglected for XRTS experiments, and is of no consequence for the interpretation of the beryllium data shown in Fig.~\ref{fig:Glenzer}.

\section{Summary and Outlook\label{sec:summary}}

In this work, we have given an overview of our current understanding of the electronic density response of WDM. More specifically, we have attempted to form a coherent picture that elucidates the connections between various theoretical approaches and simulation methods such as PIMC and different flavors of DFT, and between different representations such as the DSF $S(\mathbf{q},\omega)$ and the ITCF $F(\mathbf{q},\tau)$. In addition, we have discussed both the ubiquitous weak-perturbation limit described by linear-response theory and its extension to include nonlinear effects with respect to the perturbation amplitude.

From a theoretical perspective, density response theory constitutes a powerful framework for the investigation of a host of observables. Indeed, it forms the basis for the widely used LR-TDDFT method, and helps to connect approaches such as dielectric theories with PIMC and DFT.
In addition, density response theory is pivotal for the understanding and modelling of experiments, with XRTS measurements of WDM being a prime example. In this regard, special attention has been given to the imaginary-time domain, which combines a number of key advantages. First, obtaining the ITCF $F(\mathbf{q},\tau)$ from XRTS experiments is straightforward; this allows for exact and model-free diagnostics of the temperature~\cite{Dornheim_T_2022} and can likely be extended to other observables in future works~\cite{Dornheim_insight_2022}. Second, the ITCF contains the same information as the DSF, which makes it a useful tool for the investigation of dynamic effects from exact PIMC simulations~\cite{Dornheim_insight_2022}. Finally, we argue that the ITCF is at the heart of density response theory as it is directly connected to $\chi(\mathbf{q})$ and $S(\mathbf{q},\omega)$ in different ways. In fact, we view the imaginary-time domain as a complementary representation of dynamic quantum many-body theory. As such, it is reasonable to consider it even in cases where frequency-dependent properties such as the DSF are exactly known, since different representations tend to emphasize different aspects of the same information.

Let us conclude this work by outlining three particularly promising routes for future research.

\textbf{Improved real-time TDDFT and NEGF simulations.}
As we discussed in Secs.~\ref{sec:RTTDDFT} and \ref{sec:NEGF}, real-time simulations offer a number of attractive features: direct access to nonlinear response properties, to non-adiabatic effects, as well as to high-level electronic correlations.
The observation that real-time NEGF simulations with rather simple self-energies allow one to reproduce linear response results that otherwise require to solve complicated Bethe-Salpeter-type equations indicates a promising route for future developments in NEGF and real-time TDDFT. This can benefit from recent developments in NEGF theory that allowed to dramatically speed up the simulations achieving time-linear scaling \cite{Schluenzen_PRL_2020, Joost_PRB_2020} also for advanced self-energies \cite{Joost_PRB_2022}. These simulations can be used, among other things, to benchmark existing and derive improved exchange-correlation potentials for RT-TDDFT.
In this context, we also mention the recently developed 
 quantum fluctuations approach~\cite{schroedter_cmp_22}, which constitutes another promising concept for the real-time computation of the density response.

\begin{figure}\centering
\includegraphics[width=0.45\textwidth]{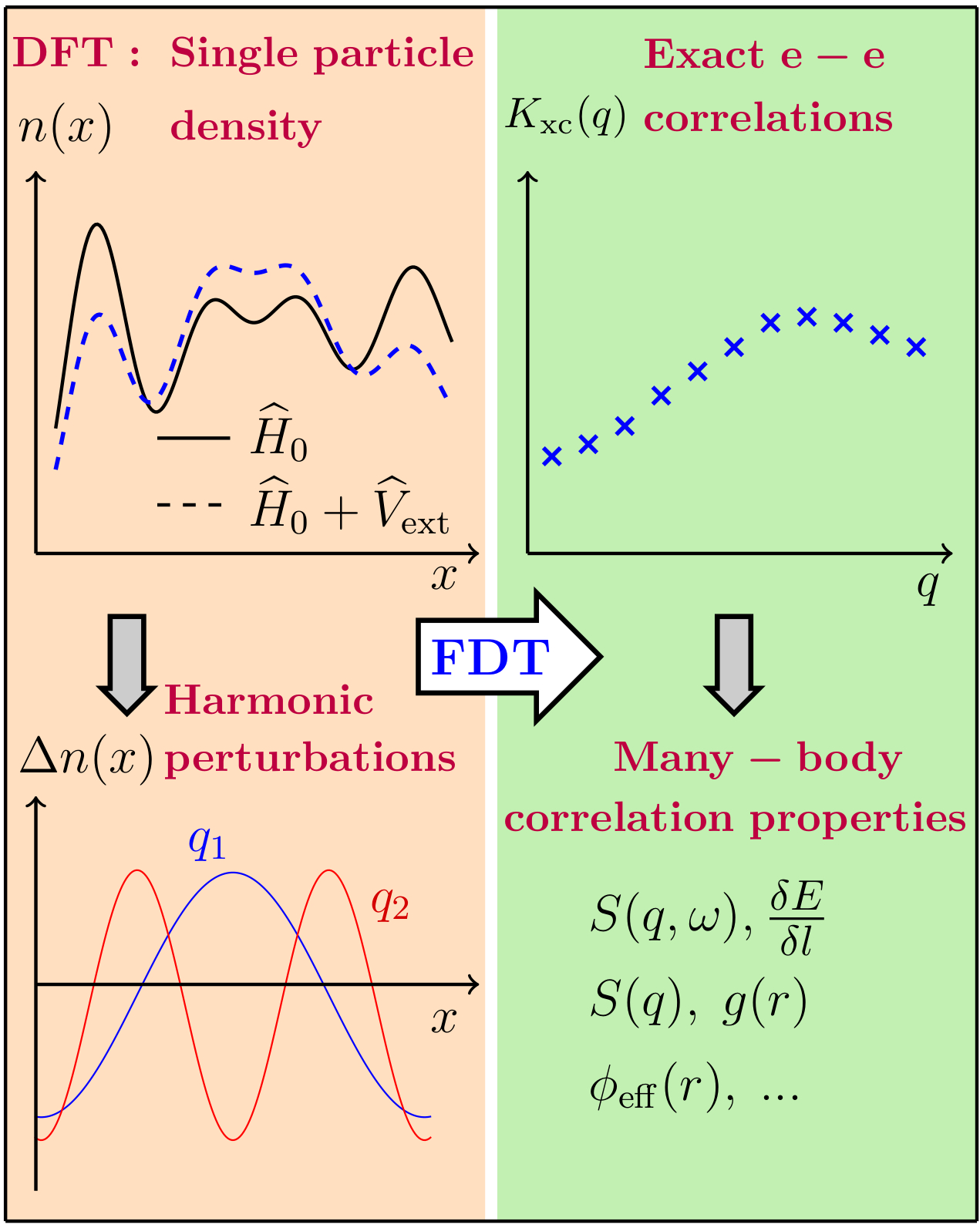}
\caption{\label{fig:Zhandos}
Schematic illustration of the DFT setup to the static XC kernel by Moldabekov \emph{et al.}~\cite{Moldabekov_PRL_2022}. Top left: Standard KS-DFT is used to compute the single-electron density $n(\mathbf{r})$ a) with respect to the original electronic Hamiltonian of interest $\hat{H}_0$ (solid black) and b) with respect to a modified Hamiltonian subject to an external monochromatic perturbation $\hat{V}_\textnormal{ext}$ (dashed blue). Bottom left: This gives access to the corresponding density modulation $\Delta n(\mathbf{r})$ for different wave vectors $\mathbf{q}$. Top right: In the linear-response regime ($\hat{V}_\textnormal{ext}\ll\hat{H}_0$), the induced density gives access to the static density response function $\chi(\mathbf{q})$ and the corresponding XC kernel $K_\textnormal{xc}(\mathbf{q})$ (blue crosses). Bottom right: The FDT [Eq.~(\ref{eq:FDT})] provides a direct connection to electron-electron correlation functions such as the static structure factor $S_{ee}(\mathbf{q})$, that cannot be readily computed within standard DFT.
}
\end{figure}

\textbf{Systematic improvement of LR-TDDFT.} In Fig.~\ref{fig:Zhandos}, we illustrate the recent framework by Moldabekov \emph{et al.}~\cite{Moldabekov_PRL_2022} for the DFT-based estimation of the static XC kernel $K_\textnormal{xc}(\mathbf{q})$. As the first step (top left), KS-DFT calculations are carried out with respect to the original electronic Hamiltonian $\hat{H}_0$ and with respect to a modified Hamiltonian $\hat{H}=\hat{H}_0+\hat{V}_\textnormal{ext}$ that is perturbed by a monochromatic static potential $\hat{V}_\textnormal{ext}$, cf.~Eq.~(\ref{eq:Hamiltonian_modified}).
The difference between these two calculations gives one the induced density shown in the bottom left panel. In the limit of small $A$, one gets direct access to the physical static density response $\chi(\mathbf{q})$ and, in this way, to the static XC kernel $K_\textnormal{xc}(\mathbf{q})$ of real materials (top right).

As the first future improvement, it will be indispensable to compare these DFT results for different conditions and for a variety of XC functionals across Jacob's Ladder to upcoming exact PIMC simulations of real WDM, starting with hydrogen. It can be expected that a comparison of the static XC kernel by itself will give invaluable insights into the performance of different functionals, and may guide the development of improved approximations. In this context, we also mention the enticing possibility to actually measure the exact $K_\textnormal{xc}(\mathbf{q})$ in XRTS experiments, which is discussed in more detail below. Second, we propose to utilize the DFT results for $K_\textnormal{xc}(\mathbf{q})$ as input for LR-TDDFT simulations within the \emph{static approximation}~\cite{dornheim_dynamic}. This can be motivated by the high accuracy of this approach for the UEG at metallic densities, and should work particularly well for highly ionized matter. The corresponding results for $S(\mathbf{q},\omega)$ can then a) be used to interpret and model XRTS experiments [Eq.~(\ref{eq:convolution})] and, after performing the Laplace transform Eq.~(\ref{eq:Laplace}), compared to exact PIMC reference data for $F(\mathbf{q},\tau)$ e.g.~for hydrogen.

In fact, the last step with the DFT results can be viewed as a linear-response imaginary-time-dependent DFT (LR-iTDDFT) calculation within the respective \emph{static approximation}. From a theoretical perspective, iTDDFT has a crucial advantage over usual TDDFT methods that operate in frequency-space: exact PIMC calculations can be used to extract an exact dynamic XC kernel that contains the full dependence on the complex frequency $z$, $K_\textnormal{xc}(\mathbf{q},z)$. Such information can give key insights into the limits of the \emph{static approximation}. Moreover, it can form the basis for the construction of novel dynamic XC kernels for LR-iTDDFT calculations of real materials that transcend the current limitations of TDDFT. We re-iterate that such calculations will allow one to accurately predict the ITCF $F(\mathbf{q},\tau)$, which is directly accessible from XRTS experiments.

Finally, we propose to use either the aforementioned LR-TDDFT calculations within the \emph{static approximation} or improved LR-iTDDFT calculations with a dynamic, $z$-dependent kernel as the basis for the estimation electron-electron correlation functions such as the static structure factor $S(\mathbf{q})$ or the electron-electron pair correlation function $g(\mathbf{r})$. From a philosophical perspective, the key ingredient to this idea is given by the fluctuation-dissipation theorem [Eq.~(\ref{eq:FDT})] that provides a straightforward relation between the induced single-particle density --- a property that is easily accessed within standard KS-DFT --- and an electron-electron correlation function such as the DSF $S(\mathbf{q},\omega)$. In this way, it will be possible to reconstruct information about electronic correlations; we also note that this idea is not limited to pair correlations and can be extended to higher-order correlation functions~\cite{Dornheim_JPSJ_2021} by using KS-DFT to study the nonlinear density response of a given system of interest~\cite{Moldabekov_JCTC_2022}.

\textbf{Improved XRTS measurements at modern XFEL facilities.}
The proposed systematic improvements of our theoretical understanding of the electronic density response of WDM can be decisively aided by new emerging experimental capabilities. Indeed, the new framework for the interpretation of XRTS experiments with WDM in the imaginary-time domain allows us to suggest new experimental setups that exploit these ideas in an optimal way. While this applies to XRTS experiments in general, we find that modern XFEL facilities such as the European XFEL~\cite{Tschentscher_2017} are particularly promising in this regard, as they offer a high degree of flexibility for new developments, and for the exploration of novel concepts and ideas. In practice, a key advantage of the European XFEL is given by the rep-rated high-power drive laser systems ReLaX and DiPOLE100X, which are made available by the HIBEF user consortium and match the macro-bunch frequency of the X-ray laser of 10\,Hz~\cite{Zastrau2021}.  This increase in the data rate of up to several thousand times in comparison to similar installations at LCLS and SACLA leads to a revolutionary improvement in photon statistics~\cite{Voigt_POP_2021}, which can result in high-quality scattering spectra with a dynamic range over four orders of magnitude or more. This accuracy is particularly important for clear measurements of the blue-shifted part of the scattering spectrum, which is exponentially damped by the Boltzmann factor in detailed balance. Furthermore, the use of self-seeding, both with and without a monochromator, can substantially reduce the bandwidth of the instrument function while keeping the photon flux in a comparable realm as for pure SASE radiation, or indeed even better~\cite{Liu2019}. This reduces the influence of uncertainties of the instrument function for further analysis. 

A clear example of a first experiment to test the proposed imaginary-time framework would be the measurement of XRTS of high energy density matter with unprecedented quality for an entire set of scattering angles. This would give us access to $I(\mathbf{q},\omega)$ over the full relevant $q$-range and will be beneficial for two reasons. First, this would allow the ITCF-based temperature diagnostics developed in Ref.~\cite{Dornheim_T_2022} to be applied for all wave vectors. Since the extracted temperature has to be the same independent of a particular $q$, such an experiment can be used to further demonstrate the consistency of the idea, and to quantify the underlying uncertainty. Second, recording the XTRS spectrum for multiple $q$ will be indispensable to test theoretical models, and to guide the development of methodological improvements. In Sec.~\ref{sec:UEG_results}, we have made extensive use of the imaginary-time version of the fluctuation-dissipation theorem, Eq.~(\ref{eq:chi_static}), which gives a straightforward relation between $F(\mathbf{q},\tau)$ and the static density response function $\chi(\mathbf{q})$.
The proposed set of XRTS measurements will thus make it possible to measure the static density response function of WDM systems. In addition to being very interesting in their own right, these data can be used to invert Eq.~(\ref{eq:kernel}) and, in this way, to obtain unassailable experimental reference data for the XC kernel of real materials.

\section*{Acknowledgments}
This work was partially supported by the Center for Advanced Systems Understanding (CASUS) which is financed by Germany’s Federal Ministry of Education and Research (BMBF) and by the Saxon state government out of the State budget approved by the Saxon State Parliament. M.B.~acknowledges funding by the Deutsche Forschungsgemeinschaft via project BO1366-15.
 A.D.B.~acknowledges support from Sandia's Laboratory Directed Research and Development Program and US Department of Energy Science Campaign 1. Sandia National Laboratories is a multimission laboratory managed and operated by National Technology and Engineering Solutions of Sandia, LLC, a wholly-owned subsidiary of Honeywell International Inc., for the U.S. Department of Energy’s National Nuclear Security Administration under contract DE-NA0003525.  The work of Ti.~D.~and F.G.~was performed under the auspices of the U.S. Department of Energy by Lawrence Livermore National Laboratory under Contract No. DE-AC52-07NA27344.

The PIMC and DFT calculations were partly carried out at the Norddeutscher Verbund f\"ur Hoch- und H\"ochstleistungsrechnen (HLRN) under grant shp00026, on a Bull Cluster at the Center for Information Services and High Performance Computing (ZIH) at Technische Universit\"at Dresden, and on the Hemera cluster at Helmholtz-Zentrum Dresden-Rossendorf (HZDR).

\bibliography{bibliography}
\end{document}